\pdfoutput=1
\documentclass[acmtog]{acmart}
\acmSubmissionID{789}
\usepackage{diagbox}
\usepackage{verbatim}
\usepackage{amssymb}
\usepackage{latexsym}
\usepackage{lineno}
\usepackage{xcolor}
\usepackage{tabularx}
\usepackage{xspace}
\usepackage{color}
\usepackage{amsmath,bm}
\usepackage{psfrag}
\usepackage{mathtools}
\usepackage{hyperref}

\usepackage{multirow, tabularx}
\newcolumntype{Y}{>{\centering\arraybackslash}X}
\usepackage{capt-of}%
\usepackage{array, boldline, makecell, booktabs}

\usepackage{colortbl}

\usepackage{array}
\usepackage{amsthm}

\usepackage{pifont}

\usepackage{psfrag}
\usepackage{forest}
\usepackage{makecell}
\usepackage{physics}
\usepackage{algorithm}
\usepackage{threeparttable}
\usepackage{graphicx}
\usepackage{overpic}
\usepackage{subcaption}
\usepackage{flushend}
\usepackage{wrapfig}
\usepackage{algpseudocode}
\usepackage{fancyhdr}
\usepackage{algpseudocode}

\newtheorem{remark}{Remark}

\usepackage{xcolor}
\usepackage{graphicx}

\newif\ifverbose
\verbosetrue

\ifverbose

\else

\newcommand{\vb}[1]{}
\fi

\usepackage{url}
\usepackage{xcolor}
\definecolor{newcolor}{rgb}{.8,.349,.1}

\usepackage{enumitem}
\usepackage{svg}
\svgpath{{./img/}} 

\usepackage[normalem]{ulem}

\newcommand{\gradF}[1]{\nabla \mathcal{F}}
\newcommand{\surfmask}[1]{\chi_S^\text{surf}}
\newcommand{\inmask}[1]{\chi_S^\text{in}}
\newcommand{\solidmask}[1]{\chi_S}
\newcommand{\xp}[1]{\bm x^p}
\newcommand{\midvortp}[1]{\bm \omega_b^p}
\newcommand{\initvortp}[1]{\bm \omega_a^p}
\newcommand{\currvortp}[1]{\bm \omega_c^p}
\newcommand{\currvortg}[1]{\bm \omega_c^g}
\newcommand{\currvelg}[1]{\bm u_c^g}
\newcommand{\Tab}[1]{\mathcal{T}^p_{[a, b]}}
\newcommand{\Tbc}[1]{\mathcal{T}^p_{[b, c]}}
\newcommand{\Fab}[1]{\mathcal{F}^p_{[a, b]}}
\newcommand{\Fbc}[1]{\mathcal{F}^p_{[b, c]}}
\newcommand{\Fac}[1]{\mathcal{F}^p_{[a, c]}}
\newcommand{\Tac}[1]{\mathcal{T}^p_{[a, c]}}
\newcommand{\T}[1]{\mathcal{T}}
\newcommand{\F}[1]{\mathcal{F}}
\newcommand{\ce}[1]{\centering}
\newcommand{\gradFbc}[1]{\gradF{}_{[b, c]}^p}
\newcommand{\gradFac}[1]{\gradF{}_{[a, c]}^p}
\newcommand{\omegavisc}[1]{\nu \Delta \bm \omega}
\newcommand{\normv}[1]{\bm u_{Sn}^g}
\newcommand{\tanv}[1]{\bm u_{St}^g}
\newcommand{\currdeltagammag}[1]{\delta \Gamma^g}
\newcommand{\gammapac}[1]{\gamma^p_{a \rightarrow c}}
\newcommand{\gammapbc}[1]{\gamma^p_{b \rightarrow c}}

\newcommand{\rev}[1]{#1}
\newcommand{\revv}[1]{#1}

\newcommand{\revvvv}[1]{}
\newcommand{\revvvvv}[1]{}

\begin{document}

\author{Sinan Wang}
\authornote{Both authors contributed equally to this research.}
\email{swang3081@gatech.edu}
\affiliation{
\institution{Georgia Institute of Technology}
\country{USA}
}

\author{Jinjin He}
\authornotemark[1]
\email{jhe433@gatech.edu}
\affiliation{
\institution{Georgia Institute of Technology}
\country{USA}
}

\author{Shenyifan Lu}
\email{slu361@gatech.edu}
\affiliation{
\institution{Georgia Institute of Technology}
\country{USA}
}

\author{Ruicheng Wang}
\email{wrc0326@outlook.com}
\affiliation{
\institution{Georgia Institute of Technology}
\country{USA}
}

\author{Greg Turk}
\email{turk@cc.gatech.edu}
\affiliation{
\institution{Georgia Institute of Technology}
\country{USA}
}

\author{Bo Zhu}
\email{bo.zhu@gatech.edu}
\affiliation{
\institution{Georgia Institute of Technology}
\country{USA}
}

\title{Generative Modeling with Orbit-Space Particle Flow Matching}

\begin{teaserfigure}
    \centering
    \includegraphics[width=\textwidth]{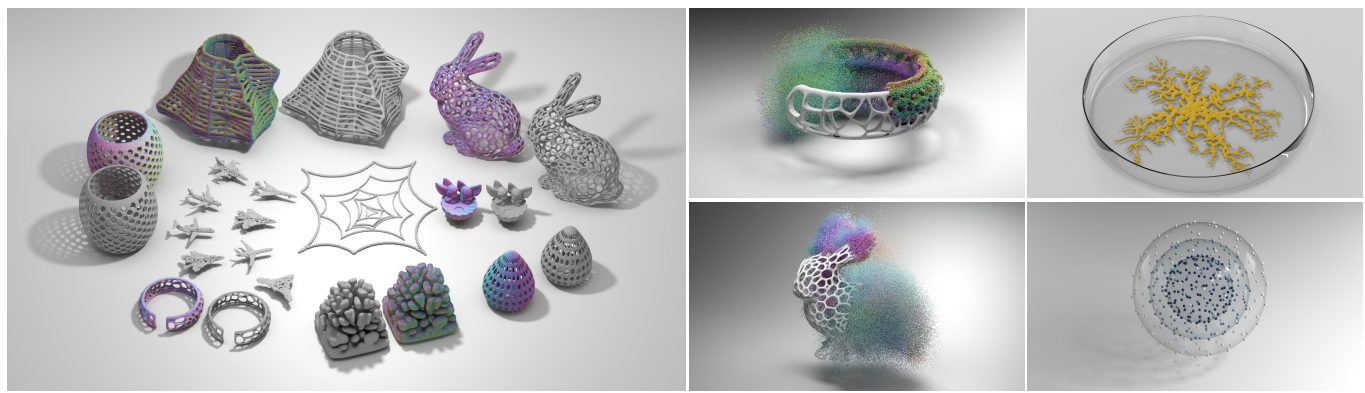}
    \caption{\textit{Left:} ShapeNet point cloud generation, single-shape encoding on complex Thingi10k meshes with Poisson-reconstructed surfaces, and minimal surface generation. \textit{Middle:} Generation process visualization showing geometric probability paths transporting noise to surface points with encoded normals. \textit{Right:} Energy-driven particle generation: diffusion-limited aggregation (top) and multilayer Thomson problem with electrons on concentric shells (bottom).}
    \label{fig:teaser}
\end{teaserfigure}

\begin{abstract}
We present \textbf{Orbit-Space Geometric Probability Paths (OGPP)}, a particle-native flow-matching framework for generative modeling of particle systems. OGPP is motivated by two insights: (i) particles are defined up to \emph{permutation symmetries}, so anonymous indexing inflates per-index target variance and yields curved, hard-to-learn flows; (ii) particles live in physical space, so the flow's \emph{terminal velocity} has physical meaning and can encode geometric attributes (e.g., surface normals). OGPP instantiates three key components: (1) orbit-space canonicalization of the probability-path terminal endpoint, (2) particle index embeddings for role specialization, and (3) geometric probability paths with arc-length-aware terminal velocities that generate normals as a byproduct of the flow. \rev{We evaluate OGPP on minimal-surface benchmarks, where it reduces metric error by up to two orders of magnitude in a single inference step; on ShapeNet, where it matches the state-of-the-art with $5\times$ fewer steps and reaches airplane EMD comparable to DiT-3D with $26\times$ fewer parameters and $5\times$ fewer steps; and on single-shape encoding, where it produces normals and reconstructions competitive with 6D generators while operating entirely in 3D.}

\end{abstract}

\keywords{Generative Modeling, Flow Matching, Particle Systems}

\setcopyright{cc}
\setcctype{by}
\acmJournal{TOG}
\acmYear{2026} \acmVolume{45} \acmNumber{4} \acmArticle{117}
\acmMonth{7} \acmDOI{10.1145/3811342}

\ccsdesc[500]{Computing methodologies~Point-based models}
\maketitle

\section{Introduction}

Particles constitute a central representation in computer graphics, where sampling, geometry, appearance, and physics are often modeled as structured sets of particles embedded in 2D or 3D physical space. Such particle-based representations arise across graphics pipelines for different purposes, from Poisson sampling for ray tracing \cite{ahmed2020screen,ahmed2021optimizing}, to point-set surfaces and clouds for geometric modeling \cite{alexa2001point,guennebaud2007algebraic,peng2021shape,kerbl20233d}, to Lagrangian particles for solid and fluid simulations \cite{muller2003particle,stomakhin2013material,muller2007position, zhou2024eulerian}, and to agent-based animation such as crowd and flock simulation \cite{reynolds1987flocks,narain2009aggregate,thalmann2012crowd, guy2010pledestrians}. Therefore, a generative model natively defined on particles and leveraging their connectivity-free structure and physical-space dynamics \rev{is well-motivated for} graphics generation tasks.

However, modern generative models are built on grids rather than particles (e.g., diffusion \cite{ho2020denoising,song2020score,rombach2022high,blattmann2023stable} and flow matching \cite{lipman2022flow}). In these settings, the representation lives on a fixed grid (e.g., a 2D lattice of pixels), and generation amounts to mapping noise to data distributions on the grid. Despite their successes in generating images or videos, these models do not transfer efficiently to particle generation because they ignore two fundamental differences.
\emph{First}, \textbf{particles exhibit pronounced \textit{symmetries}}: permuting particle indices leaves the underlying configuration unchanged, yet can arbitrarily alter its vectorized representation in high-dimensional space. In group-theoretic terms, the collection of all such symmetry-related configurations forms the \emph{orbit} of a particle state. As a result, naively applying grid-based generative frameworks (e.g., flow matching~\cite{lipman2022flow}) to particle data by flattening particles into long vectors leads to fundamental difficulties: For images, a pixel at a fixed coordinate exhibits consistent statistics across samples. In contrast, particle systems are defined only up to permutation symmetry: a particle at a fixed index does not correspond to any consistent spatial or statistical role across the dataset. Consequently, probability-path endpoints associated with a given index are dispersed throughout space, forcing velocity predictors to average over incompatible targets during training and yielding noisy, poorly structured per-particle flows. State-of-the-art particle generators such as Equivariant Flow Matching~\cite{klein2023equivariant,song2023equivariant} mitigate permutation ambiguity via optimal transport couplings. However, these methods incur high computational cost and still operate on flattened, anonymous particle representations, so individual indices must aggregate over many symmetry-induced roles, leading to increased target variance and highly curved flows.
%
\emph{Second}, \textbf{particles live in \emph{physical space}}. 
Generating a set of particles can be viewed as simulating their spatiotemporal evolution under a learned physical velocity field. This differs fundamentally from image generation, where the velocity field in flow matching merely transports pixel values and carries no intrinsic physical meaning. In particle-based settings, however, the velocity field is defined in physical space, and the terminal velocity at $t=1$ represents a well-defined geometric quantity. For example, when particles sample a surface, this terminal velocity can encode meaningful local geometric information, such as surface normals or orientation. Standard linear paths place particles at the correct locations but do not exploit this geometric degree of freedom.

\begin{figure}[t]
    \centering
    \includegraphics[width=0.5\textwidth]{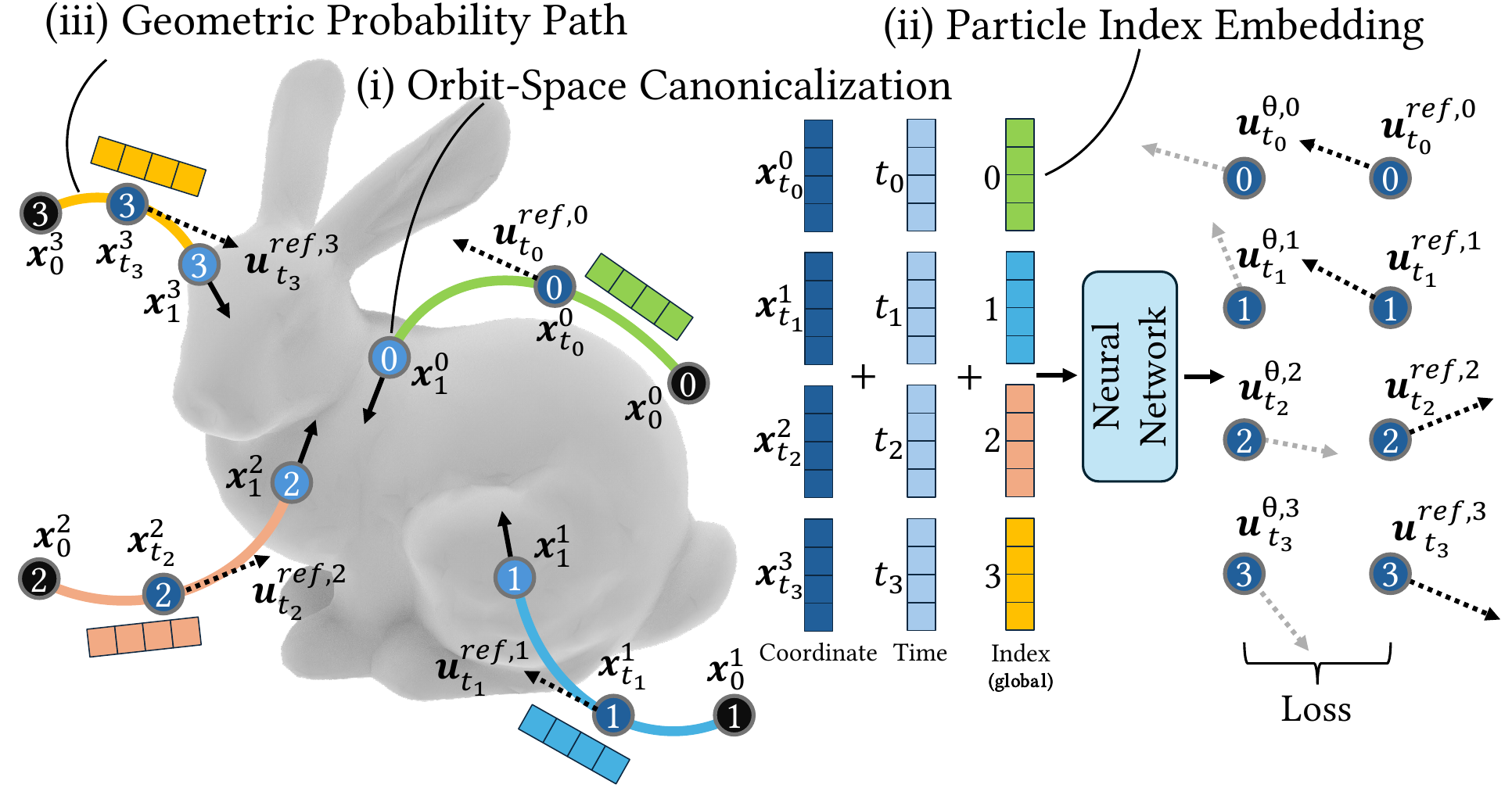}
    \caption{\textbf{OGPP.} Our framework integrates three key components: (i) \textit{orbit-space canonicalization} assigns canonical indices (0,1,2,3) to $X_1$ while keeping $X_0$ uncanonicalized, (ii) \textit{particle index embeddings} (colored blocks) allow each index to specialize to its canonical role, and (iii) \textit{geometric probability paths} encode surface normals via arc-length-aware terminal velocities. Per-particle coordinates $\bm{x}_{t_i}^i$ and learnable per-index embeddings are fed into a NN, predicting velocities $\bm{u}_{t_i}^{\theta,i}$ supervised by reference velocities $\bm{u}_{t_i}^{\text{ref},i}$.}
    \label{fig:workflow}
\end{figure}

Motivated by these two insights, we propose a generative framework for particles that both respects orbit structure and exploits geometric path. Our key idea is to untangle mixed particle roles by combining orbit-space canonicalization with identity-aware particle index embeddings. On top of this, we design geometric probability paths whose terminal tangents encode per-particle normals, so a single flow jointly generates particle positions and attributes. Our framework consists of three key components (See Figure~\ref{fig:workflow}): (i) \textit{orbit-space canonicalization}, (ii) \textit{particle index embeddings}, and (iii) \textit{geometric probability paths}. All three are expressed as choices of \emph{conditional probability path}, which we collectively call \textbf{\emph{Orbit-Space Geometric Probability Paths (OGPP)}}.

For \textit{orbit-space canonicalization}, we perform symmetry reduction at the terminal endpoint \(X_1\): for each particle configuration, we sort indices according to a geometric criterion (e.g., a space-filling curve) and select a single representative from the orbit. This enforces that particle index \(i\) consistently lands in a localized and stable spatial region, \revvvv{significantly} reducing variability in the training targets seen by each index. 
Next, for \textit{particle index embeddings}, we attach a learnable identity embedding to each particle index and provide it to the velocity network. This allows the model to condition on particle identity, enabling different indices to specialize to distinct velocity-field roles, analogous to class-conditional generation. Together, canonicalization and identity embeddings convert noisy mixtures of regression targets into well-separated, easier-to-learn trajectory families, yielding \revvvv{substantially} straighter flows.
Finally, for \textit{geometric probability paths}, we replace linear interpolation with geometry-aware paths that exploit the structure of particle systems. Specifically, we construct Hermite-type probability paths whose terminal tangents align with per-particle normals: the endpoint specifies particle position, while the terminal velocity encodes local surface orientation. As a result, the learned flow simultaneously transports particles from noise to data and produces accurate surface normals as an intrinsic byproduct.

\rev{We evaluate \textbf{OGPP} on a range} of  graphics-oriented generative tasks, including geometric reconstruction, shape generation, and physics simulation. \revvvv{OGPP consistently outperforms baseline methods across all evaluated benchmarks.} For minimal-surface generation, \rev{OGPP reduces metric error by up to two orders of magnitude in a single inference step.} On ShapeNet, it improves 1-NNA, matches the particle-generator SOTA NSOT~\cite{hui2025not} with \textbf{5}$\times$ fewer inference steps, \rev{reaching} airplane EMD comparable to DiT-3D~\cite{mo2023dit} using \textbf{26}$\times$ fewer parameters and \textbf{5}$\times$ fewer steps. On single-shape encoding benchmarks~\cite{zhang2025geometry}, it yields \revvvv{substantially} better normal estimation and reconstruction quality than generalized VP-based paths~\cite{albergo2022building,ma2024sit,chang20243d}, while remaining comparable to state-of-the-art 6D generators.

\paragraph{Contributions.}
Our main contributions are:

\begin{enumerate}[leftmargin=*]
  \item \textbf{Orbit-space particle flow matching.} We introduce \emph{particle flow matching} as a Lagrangian formulation of flow matching for particle systems, in contrast to the Eulerian image models, combining identity embeddings on individual particles with an orbit-space canonicalization so that each particle can learn its own consistent velocity field while simplifying the learning.

  \item \textbf{Geometric probability paths.} We construct Hermite-type geometric probability paths whose terminal tangent encodes per-particle attributes such as surface normals, enabling surface normal generation as a byproduct of the learned flow.

  \item \textbf{Energy-driven evaluation for particle generators.}
  We propose self-referential, physics- and geometry-based metrics that directly assess the quality of generated particle sets (e.g., blue-noise spectra, fractal dimension, residual Coulomb forces, minimal-surface deviation), together with matching benchmark datasets.
\end{enumerate}

\rev{
\paragraph{Paper outline.}
Section~2 reviews related work, and Section~3 recalls the background on flow matching and group theories. Sections~4 and~5 present our two main theoretical contributions, orbit-space probability paths and geometric probability paths for attribute encoding, respectively, and are \emph{essential} reading. Section~6 summarizes the overall algorithm. Section~7 reports experimental results, before Section~8 offers further discussion and Section~9 concludes.
}

\section{Related Work}

\subsection{Generative Models}

\paragraph{Continuous-Time Generative Models}
From a modern continuous-time perspective \cite{lipman2024flow,holderrieth2025introduction}, generative models can be formulated as stochastic differential equations (SDEs), as in denoising diffusion models \cite{ho2020denoising,song2020score}, or ordinary differential equations (ODEs), as in flow models \cite{chen2018neural,grathwohl2018ffjord,lipman2022flow,albergo2023stochastic,liu2022flow}. This unified view has enabled broad progress across image and video synthesis \cite{ramesh2022hierarchical,rombach2022high,blattmann2023stable,brooks2024video}, 3D shape generation \cite{zhou20213d,mo2023dit,vahdat2022lion,hui2022neural}, point cloud modeling \cite{luo2021diffusion,yang2019pointflow}, and neural rendering \cite{poole2022dreamfusion,wang2023prolificdreamer}. Recently, IADB \cite{heitz2023iterative} reinterprets DDIM as an ODE-based deterministic diffusion process. Within the ODE family, unlike CNFs \cite{chen2018neural,grathwohl2018ffjord}, flow matching \cite{lipman2022flow} enables simulation-free vector field learning and the use of Optimal Transport (OT) \cite{lipman2022flow} paths. Techniques like Minibatch OT \cite{tong2023improving,pooladian2023multisample} and Rectified Flow \cite{liu2022flow} further straighten trajectories for efficiency. However, these models are primarily tailored for objects in Euclidean space $\mathbb{R}^d$ (e.g., images), and do not naturally accommodate the unique quotient geometry of particle systems. While Riemannian Flow Matching \cite{chen2023flow} extends flow matching to non-Euclidean manifolds and SE(3) flow matching \cite{yim2023fast} applies it to structured proteins, both remain Eulerian and do not treat particles individually.


\paragraph{Alternative Probability Path Designs}
Beyond linear interpolation, recent works explore alternative path designs. BNDM \cite{huang2024blue}, motivated by the spectral bias of diffusion models, injects time-dependent blue noise into deterministic diffusion to modify the probability path. Generalized VP interpolants \cite{albergo2022building,ma2024sit}, building on VP and VE SDEs \cite{song2020score}, enable flexible nonlinear schedules in flow matching. Recent 3D shape tokenization work \cite{chang20243d} adopts gVP paths for latent flow matching and zero-shot normal estimation.


\subsection{Point Cloud Generation}

\paragraph{Point Cloud Generative Models}
Early point cloud generation relied on GANs \cite{achlioptas2018learning,xie2021generative,shu20193d,li2021sp} and set-structured VAEs such as SetVAE \cite{kim2021setvae}, alongside CNF-based models like PointFlow \cite{yang2019pointflow} and SoftFlow \cite{kim2020softflow}, which offer exact likelihoods. To achieve scalable high-fidelity synthesis, recent works adopt a two-stage strategy: compressing high dimensional vectors into a compact latent space via VAEs \cite{kingma2013auto} before training generative models. Early latent representations used voxel grids (e.g., ConvOccNet \cite{peng2020convolutional}), suffering from cubic memory costs, while later works explored more efficient structures such as irregular grids \cite{zhang20223dilg}, hierarchical point-based latents (LION \cite{vahdat2022lion}), or latent sets without explicit spatial structure (3DShape2VecSet \cite{zhang20233dshape2vecset}). Despite improved scalability, these frameworks typically rely on category-specific autoencoders. In contrast, generative models like PVD \cite{zhou20213d}, DiT-3D \cite{mo2023dit}, DPM \cite{luo2021diffusion} apply diffusion directly in data space. PSF \cite{wu2023fast} accelerates sampling via Reflow \cite{liu2022flow}. While effective, these works primarily advance model architectures or data representations and do not explicitly model orbit-space symmetries, retaining an Eulerian viewpoint.

\paragraph{Canonicalization for Permutation Handling}
Permutation ambiguity in particle systems is commonly addressed via canonicalization by deterministic ordering, such as Z-order (Morton order)~\cite{morton1966computer} or Hilbert curves~\cite{hilbert1935stetige}, which map spatial coordinates to one-dimensional sequences while preserving locality. Recent Transformer-based models adopt similar strategies to stabilize attention and improve scalability, e.g., Point Transformer v3~\cite{wu2024point}, OctFormer~\cite{wang2023octformer}, and FlatFormer~\cite{liu2023flatformer}. These methods canonicalize the \emph{Transformer input representation}, primarily to improve computational efficiency and architectural stability. 
\paragraph{Symmetry Modeling}
The recent frontier focuses on enforcing symmetries. Equivariant Flow Matching~\cite{klein2023equivariant,song2023equivariant} achieves this via optimal transport (OT) couplings but with a training-step complexity of $O(B^2 N^3)$~\cite{hui2025not}, making it unscalable. NSOT~\cite{hui2025not} improves scalability by offline OT precomputation and hybrid coupling, and SGFM~\cite{puny2025space} extends such constraints to enforce complex space-group symmetries inherent to crystalline structures. From an architectural perspective, these permutation-equivariant models~\cite{hui2025not,klein2023equivariant,song2023equivariant}, and more broadly, mainstream point-cloud architectures~\cite{liu2019point,zhou20213d} treat particles as anonymous coordinates, so the network is not allowed to distinguish particle indices. Diffusion Transformers such as DiT-3D~\cite{mo2023dit} do employ learned positional embeddings, but operates on voxel grids without orbit-space canonicalization. Consequently, these methods still adopt an Eulerian view, which makes the regression problem ill-conditioned and the flow highly curved.



\subsection{Energy-Driven Particle Systems}
\paragraph{Physical Particle Systems}
Particle systems are a ubiquitous representation across physics, graphics, and vision, used to model phenomena ranging from N-body simulations \cite{barnes1986hierarchical}, to molecular dynamics \cite{frenkel2023understanding}, fluids via SPH \cite{muller2003particle} and vortex particles \cite{park2005vortex}, and flocking or crowd behavior \cite{reynolds1987flocks,thalmann2012crowd}. A fundamental subclass involves systems governed by energy functionals, where equilibrium states correspond to stationary points of pairwise or global potentials: blue-noise sampling seeks point sets with suppressed low-frequency spectra and isotropy \cite{yellott1983spectral,ulichney1988dithering,cook1986stochastic}, computed via Lloyd relaxation \cite{lloyd1982least}, capacity-constrained Voronoi tessellations \cite{balzer2009capacity}, optimal transport \cite{de2012blue,qin2017wasserstein}, or kernel-based methods \cite{fattal2011blue,ahmed2022gaussian}; the Thomson problem \cite{thomson1904xxiv,bowick2009two,smale1998mathematical} seeks minimum-energy configurations of repelling charges, tackled via basin-hopping \cite{wales1997global}, genetic algorithms \cite{morris1996genetic}, or simulated annealing \cite{erber1991equilibrium}; diffusion-limited aggregation \cite{witten1981diffusion} produces fractal clusters through Brownian-motion attachment \cite{meakin1983formation,s1992fractal,halsey2000diffusion}; and minimal surfaces \cite{plateau1873statique} minimize area under boundary constraints via variational methods \cite{brakke1992surface,pinkall1993computing,dziuk1990algorithm,wang2021computing}.  Recently Geometry Distributions \cite{zhang2025geometry,tang2025generative,tang2025human} represent single surfaces as infinite point distribution via diffusion models.

\paragraph{Physics-Aware Evaluation for Generative Models}
Despite their scientific importance, these physically grounded particle systems have received limited attention from the generative modeling community, which prioritizes shape-level point cloud generation evaluated by distribution-matching metrics such as 1-NNA \cite{lopez2016revisiting}. \revv{We observe that energy-driven particle systems offer intrinsic evaluation criteria (e.g., spectral characteristics, fractal dimensions, residual forces, surface deviations) that can complement distributional metrics by more directly measuring physical fidelity.} To this end, \rev{for energy-driven tasks such as blue noise, minimal surfaces, DLA, and the Thomson problem, we first use classical solvers to produce large datasets of equilibrium configurations, and then train a generative model on these datasets}. 

\section{Background}

\newcolumntype{z}{X}
\newcolumntype{s}{>{\hsize=.25\hsize}X}
\begin{table}[h]
\caption{Summary of the main symbols and notations.}
\centering
\small
\begin{tabularx}{0.47\textwidth}{scz}
\hlineB{2.5}
Notation & Type & Definition\\
\hlineB{2.5}
\multicolumn{3}{c}{\textit{General}} \\
\hline
$t$ & scalar & time $\in [0,1]$\\
\hline
$d, D$ & scalar & dimension\\
\hline
$N$ & scalar & number of particles\\
\hlineB{2.5}
\multicolumn{3}{c}{\textit{Flow Matching}} \\
\hline
$\bm{u}_t$ & vector field & velocity field at time $t$\\
\hline
$\bm{u}_t^\theta$ & vector field & neural network velocity field\\
\hline
$\bm{u}_t^\mathrm{ref}$ & vector field & reference (target) velocity field\\
\hline
$X_0$ & random var. & initial point (noise)\\
\hline
$X_1$ & random var. & terminal point (data)\\
\hline
$X_t$ & random var. & interpolated point at time $t$\\
\hline
$\bm{x}_0, \bm{x}_1, \bm{x}_t$ & vector & realizations of $X_0, X_1, X_t$\\
\hline
$p_\mathrm{init}$ & distribution & initial (noise) distribution\\
\hline
$p_\mathrm{data}$ & distribution & data distribution\\
\hline
$p_t$ & distribution & marginal probability path\\
\hline
$p_t(\cdot|\bm{x}_1)$ & distribution & conditional probability path\\
\hline
$Z$ & random var. & joint variable: $(X_1, N_1)$\\
\hlineB{2.5}
\multicolumn{3}{c}{\textit{Canonicalization}} \\
\hline
$C(\cdot)$ & map & orbit-space canonicalization map\\
\hline
$G$ & group & symmetry group (e.g., permutation)\\
\hline
$\rho(g)$ & matrix & orthogonal representation of $g$\\
\hline
$\mathrm{Orb}(\bm{x})$ & set & orbit of $\bm{x}$ under the group action\\
\hline
$\zeta_{\bm{x}}$ & random var. & canonical representative of \(\mathrm{Orb}(X_1)\)\\
\hlineB{2.5}
\multicolumn{3}{c}{\textit{Geometric Path}} \\
\hline
$\bm{n}$ & vector & per-particle attribute (surface normal)\\
\hline
$\bm{v}_1$ & vector & terminal tangent velocity\\
\hline
$\alpha(t), \beta(t)$ & scalar & Hermite basis functions\\
\hline
$\gamma(t)$ & curve & conditional probability path curve\\
\hlineB{2.5}
\end{tabularx}
\label{tab:notation_table}
\end{table}

\subsection{Naming Conventions}
We adopt the following conventions throughout this paper.
Bold symbols (e.g., $\bm{u}$, $\bm{x}$, $\bm{n}$) denote vector fields or vectors, while regular symbols denote scalars.
Capital letters (e.g., $X$, $N$) represent random variables, and lowercase bold letters (e.g., $\bm{x}$, $\bm{n}$) denote their realizations or fixed values.
Specifically, $X$ denotes position random variables, $N$ denotes attribute random variables, and $Z = (X_1, N_1)$ denotes the joint random variable of position and attribute.
Superscripts without parentheses (e.g., $\bm{x}_t^i$) denote particle indices,
while superscripts in parentheses (e.g., $\bm{x}_0^{(i)}$) denote sample indices.
We summarize the main symbols and notations in Table~\ref{tab:notation_table}.

\subsection{Flow Matching}
\rev{Flow matching~\cite{lipman2022flow,lipman2024flow} trains a velocity field that transports a noise distribution $p_\mathrm{init}$ to $p_\mathrm{data}$ by integrating an ODE.
A flow model generates samples by solving}
\begin{equation}
\label{eq:flow_ode}
\frac{\mathrm{d} X_t}{\mathrm{d} t} = \bm{u}_t^\theta(X_t), \quad X_0 \sim p_\mathrm{init},
\end{equation}
\rev{where $\bm{u}_t^\theta : \mathbb{R}^d \times [0,1] \to \mathbb{R}^d$ is a neural network chosen so that $X_1 \sim p_\mathrm{data}$.
For each data point $\bm{x}_1 \sim p_\mathrm{data}$, a \emph{conditional probability path} $p_t(\cdot \mid \bm{x}_1)$ interpolates from $p_\mathrm{init}$ at $t{=}0$ to a point mass at $\bm{x}_1$ at $t{=}1$; averaging over $\bm{x}_1$ yields the \emph{marginal probability path} $p_t$.
Since most of our constructions are defined at the conditional level, we refer to $p_t(\cdot \mid \bm{x}_1)$ simply as a \emph{probability path} and reserve \emph{marginal probability path} for $p_t$.}

\paragraph{Marginalization trick.}
\label{thm:marginalization}
\rev{The marginal velocity field can be expressed as a posterior-weighted average of conditional velocities (see Appendix~\ref{sec:appendix_flow_matching} for details):}
\begin{equation}
\label{eq:marginal_vf}
\bm{u}_t^\mathrm{ref}(\bm{x})
= \int \bm{u}_t^\mathrm{ref}(\bm{x} \mid \bm{x}_1)\,
\frac{p_t(\bm{x} \mid \bm{x}_1)\, p_\mathrm{data}(\bm{x}_1)}{p_t(\bm{x})} \, \mathrm{d}\bm{x}_1.
\end{equation}
\rev{In practice, the network is trained via the \emph{conditional flow matching loss}:}
\begin{equation}
\label{eq:cfm_loss}
\mathcal{L}_\mathrm{CFM}(\theta)
= \mathbb{E}_{t,\, \bm{x}_1,\, \bm{x} \sim p_t(\cdot \mid \bm{x}_1)}
\Big[ \big\| \bm{u}_t^\theta(\bm{x}) - \bm{u}_t^\mathrm{ref}(\bm{x} \mid \bm{x}_1) \big\|^2 \Big].
\end{equation}

\subsection{Group Theory}
\label{sec:group_theory}
\rev{We briefly review concepts needed for orbit-space probability paths; extended definitions are in Appendix~\ref{sec:appendix_group_theory}.
A \emph{group} $G$ acts on $\mathbb{R}^d$ via an orthogonal representation $\rho: G \to O(d)$, i.e., $g \cdot x = \rho(g) x$.
The \emph{orbit} of $x$ is $\mathrm{Orb}(x) := \{ \rho(g) x : g \in G \}$.}

\rev{In all our experiments we normalize away global pose by recentering and PCA alignment, so the remaining symmetry is $G = S_N$ acting on $(\mathbb{R}^D)^N$ by permuting particle indices.}

\paragraph{Canonicalization.}
\label{def:canonical_map}
\rev{A \emph{canonicalization map} $C: \mathbb{R}^d \to \mathbb{R}^d$ selects a representative from each orbit in a $G$-invariant way, requiring:
(1) $C(\rho(g) x) = C(x)$ for all $g \in G$ ($G$-invariance), and
(2) $C(x) \in \mathrm{Orb}(x)$ (the output lies in the orbit of $x$).
Together, these conditions imply $C$ induces a bijection between orbits and their canonical representatives.
Concretely, a $G$-canonicalization fixes a deterministic particle ordering (e.g., via a space-filling curve such as Morton or Hilbert).}

\section{Orbit-Space Probability Paths}
\label{sec:ogpp}

\begin{figure}[t]
    \centering
    \includegraphics[width=0.48\textwidth]{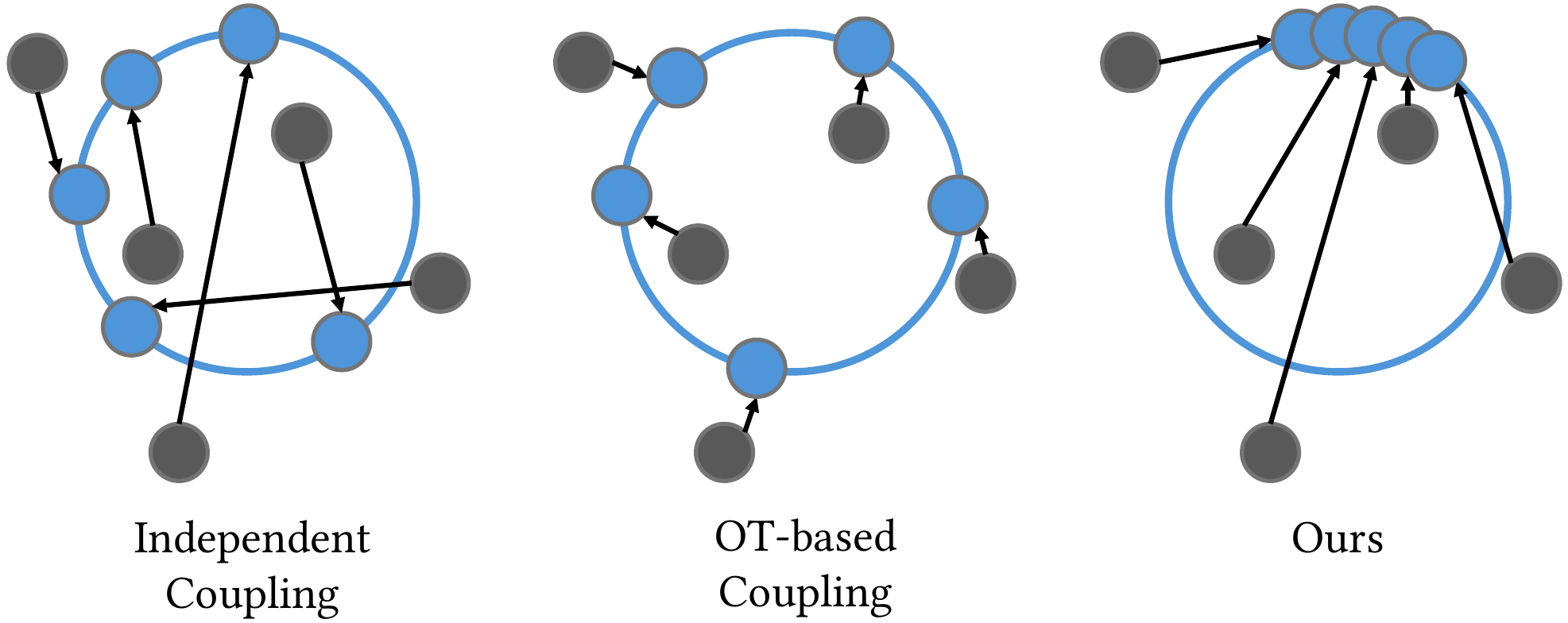}
    \caption{Conceptual illustration of the conditional distribution of the terminal endpoint $X_1^0$ for a fixed particle index $0$ under different coupling strategies. Each figure shows five sampled $(\bm{x}_0,\bm{x}_1)$ pairs for particles with index $0$: gray points denote the noise positions $\bm{x}_0^{(i),0}$, blue points denote the corresponding targets $\bm{x}_1^{(i),0}$ on the surface, and arrows indicate their displacements. Left: independent coupling~\cite{lipman2022flow}; middle: OT-based coupling~\cite{song2023equivariant,klein2023equivariant}; right: our orbit-space canonicalization. Independent and OT-based  couplings spread the possible endpoints $X_1^0$ around the surface, yielding a broad conditional distribution $p(X_1^0 \mid X_t=\bm{x}, I=0)$, \revv{whereas orbit-space canonicalization concentrates them into a smaller region, which is expected to reduce the conditional covariance and simplify the velocity regression.}}
    \label{fig:simple_illustration}
\end{figure}

In this section, we focus on the first two key components of OGPP, \textbf{orbit-space canonicalization} and \textbf{particle index embeddings}. These two mechanisms are designed to work in tandem: our ablation in Sec.~\ref{sec:ablation} shows that each alone brings only limited gains, while their combination is crucial. We discuss our third key component, \textbf{geometric probability paths}, in Sec.~\ref{sec:att_prob_path}.

In a traditional flow-matching framework, the neural network takes as input the particle
positions $\bm{x}_t$ at time $t$ (together with $t$ itself) and outputs a
velocity field $\bm{u}_\theta(\bm{x}_t, t)$.
Both the intermediate states $X_t$ and the reference targets $Y$ are determined
by the choice of conditional probability path $p_t(\cdot|\bm{x}_1)$.
We discuss the interior geometry of $p_t(\cdot|\bm{x}_1)$ in Section~\ref{sec:att_prob_path}
and focus on its endpoints in this section.
As outlined in the introduction, we pursue two objectives:
(i) \emph{make the regression task easier} by reducing the conditional covariance; and
(ii) \emph{encourage straight flows} by reducing Lipschitz ratios.
We begin by describing our model architecture with particle index embedding in Section~\ref{subsec:model_arch}.
We then show in Section~\ref{subsec:per_particle_cov} that orbit-space canonicalization on $X_1$ reduces the conditional covariance of the regression targets.
In Section~\ref{subsec:orbit_straightness} we formalize a requirement on orbit-space canonicalization maps: they must be orbit-continuous, thereby encouraging straighter flows.
Finally, in Section~\ref{subsec:lipschitz_view} we study nearest-neighbor Lipschitz ratios and show that further canonicalizing the noise endpoint $X_0$ inflates these ratios, leading \rev{to} less straight flows.

\begin{figure*}[!htb]
    \centering
    \includegraphics[width=1.0\textwidth]{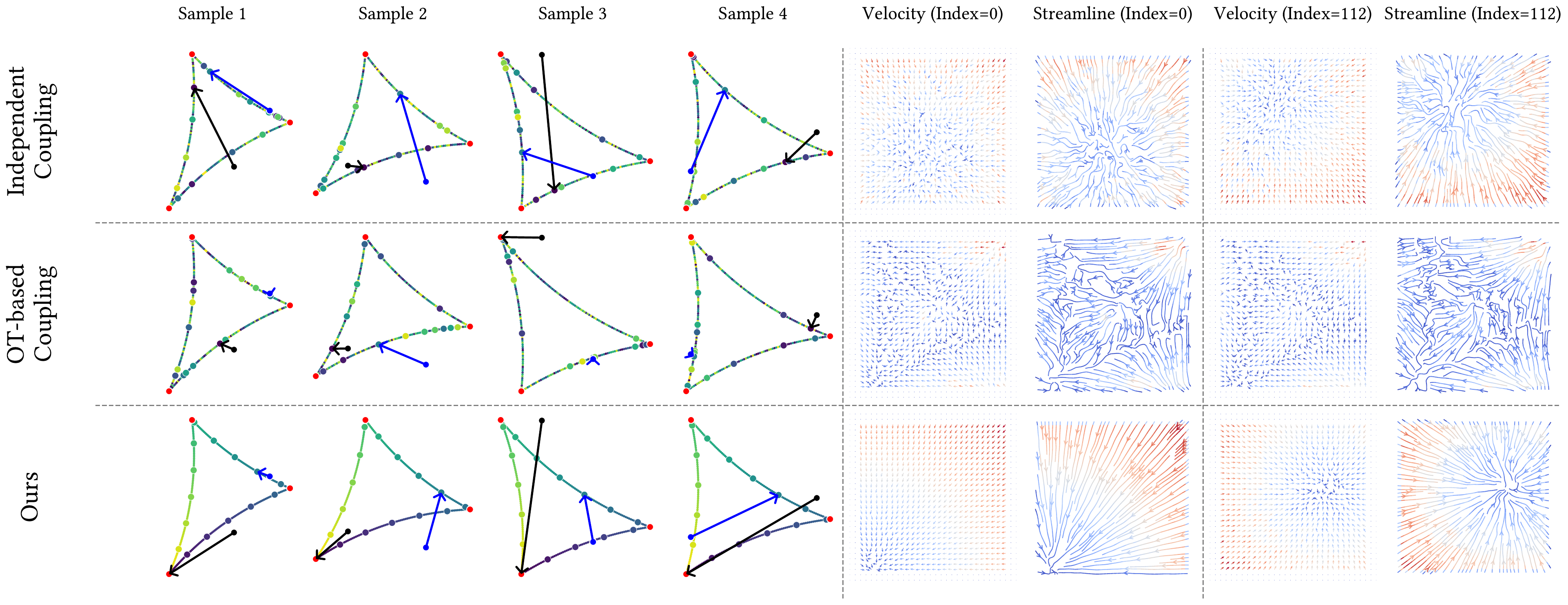}
    \caption{Visualization of index-conditioned velocity fields in a realistic minimal-surface configuration (area-constrained). For each strategy, we sample 1000 random noise configurations $\bm{x}_0$ for each minimal-surface target $\bm{x}_1$ and construct couplings using independent coupling~\cite{lipman2022flow}, OT-based coupling~\cite{klein2023equivariant,song2023equivariant}, and our orbit-space canonicalization. Left: four minimal-surface boundary point sets with particles colored by index; arrows show, for a single trial, the velocity $\bm{x}_1^i - \bm{x}_0^i$ of two highlighted particles (black: Index=0, blue: Index=112) from a shared initialization $\bm{x}_0$. Right: empirical per-particle velocity fields for the same indices, obtained by aggregating these velocities over the 1000 $(\bm{x}_0,\bm{x}_1)$ pairs and interpolating them onto a grid; streamlines visualize flow trajectories induced by these vector fields. As in the conceptual illustration Figure~\ref{fig:simple_illustration}, independent and OT-based couplings spread the possible endpoints for each index around the surface, whereas our orbit-space canonicalization concentrates them into a small, stable region and yields straighter \revvvv{, more coherent index-conditioned} flows.}
    \label{fig:cond_dist_minimal}
\end{figure*}

\subsection{Model Architecture with Particle Index Embeddings}
\label{subsec:model_arch}
We instantiate the particle-indexed velocity field
$\bm{u}_\theta$ with a plain Transformer encoder \cite{vaswani2017attention} that operates on sets of
particles.
Given an input configuration $\bm{x}_t = (\bm{x}_t^1,\dots,\bm{x}_t^N)$, each particle
is represented by a feature vector in $\mathbb{R}^{D_{\mathrm{in}}}$, consisting
primarily of spatial coordinates and, in a few experiments, an additional time
coordinate (see Section~\ref{sec:dla_validation}). We first apply a linear
projection to an embedding dimension $D_{\mathrm{emb}}$, add a \emph{particle index embedding} $e_i \in \mathbb{R}^{D_{\mathrm{emb}}}$, and add a
global time embedding $\phi_t(t) \in \mathbb{R}^{D_{\mathrm{emb}}}$:
\(
\bm{h}_i^{(0)}
=
W_{\mathrm{in}} \bm{x}_t^i
+
e_i
+
\phi_t(t),
\text{with } i = 1,\dots,N.
\)
The sequence $(\bm{h}_1^{(0)},\dots,\bm{h}_N^{(0)})$ is then processed by a
$L$-layer Transformer encoder with multi-head self-attention and GELU-activated
MLPs, yielding representations $\bm{h}_i^{(L)}$.
The final velocity prediction for particle $i$ is obtained by a shared linear
head
\(
\bm{u}_\theta^i(X_t, t)
=
W_{\mathrm{out}} \bm{h}_i^{(L)}.
\)

Architecturally, the particle index embedding $e_i$ is deliberately simple: it plays the same role as a standard positional embedding in Transformers, i.e., a learned token-wise bias that lets the network distinguish different positions. We refer to it as a \emph{particle index embedding} to emphasize that the “position” here is the canonical particle index rather than a grid coordinate. \revvvv{In contrast,}



\revv{This architecture is intentionally plain; we attribute the observed improvements primarily to the probability path design and the use of identity embeddings rather than to architectural sophistication.}

For conditional generation tasks such as minimal surface generation with anchor points, we extend this architecture with cross-attention layers interleaved every few self-attention blocks. Condition tokens are projected to the embedding dimension and serve as keys and values, with particle tokens as queries. To handle variable numbers of condition tokens (e.g., 3--8 anchors), we pad to a maximum count using learnable missing embeddings and apply attention masks to ignore padded positions.

\subsection{Orbit-Space Canonicalization on \(X_1\)}
\label{subsec:per_particle_cov}
In this subsection we analyze the regression problem and show that orbit-space canonicalization of the terminal endpoint \(X_1\)  reduces the conditional covariance \(\mathrm{Cov}(Y \mid X_t = \bm{x})\) of the regression target \(Y\), which measures how noisy the regression problem is for the velocity predictor. We illustrate the conditional distribution conceptually in Figure~\ref{fig:simple_illustration} and in a real training scenario in Figure~\ref{fig:cond_dist_minimal}. 

\begin{figure}[t]
    \centering
    \includegraphics[width=0.5\textwidth]{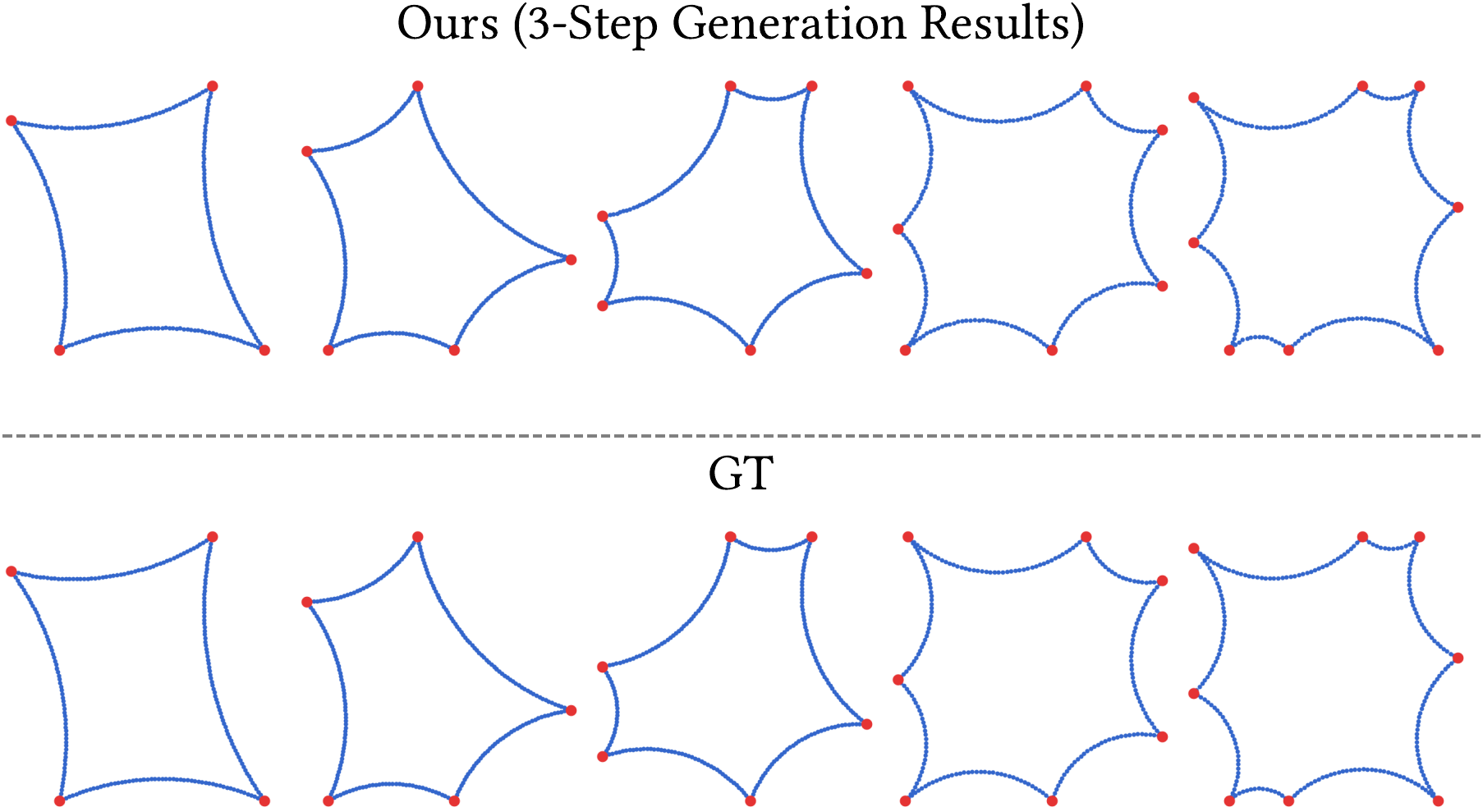}
    \caption{\textbf{Minimal surface generation with variable anchors (3--8).} 3-step generation results from a single conditional model trained on varying anchor counts. \revv{The generated boundaries appear smooth and accurate across diverse configurations.}}
    \label{fig:minimal_surface_multi}
\end{figure}

\rev{
\rev{We define the regression target (for the linear path; the geometric-path extension is in Section~\ref{subsec:joint_canon}) as}
\begin{equation}
Y
:=
\frac{\bm{X}_1 - \bm{X}_t}{1-t}
=
\bm{X}_1 - \bm{X}_0.
\label{eq:YI}
\end{equation}
\rev{The Bayes-optimal velocity is}
\label{lem:bayes_opt_per_particle}
\begin{equation}
u^*(\bm{x}, t)
=
\mathbb{E}\!\left[ Y \mid X_t = \bm{x} \right].
\label{eq:bayes_opt}
\end{equation}
\rev{A smaller conditional covariance $\mathrm{Cov}(Y \mid X_t = \bm{x})$ directly lowers the irreducible MSE of the Bayes-optimal predictor, making the velocity regression easier to learn.
We use the trace $\mathrm{tr}\,\mathrm{Cov}(\cdot)$ as a scalar measure of this covariance.}

\rev{\paragraph{Orbit symmetry and canonicalization.}
After pose normalization (Section~\ref{sec:group_theory}), we model the residual permutation symmetry by assuming that, for each fixed $X_t = \bm{x}$,}
\begin{equation}
\label{eq:orbit-factorization}
X_1 \mid (X_t = \bm{x})
\;\overset{d}{=}\;
\rho(G)\,\zeta_{\bm{x}},
\end{equation}
\rev{where $G$ is a random permutation in $S_N$ and $\zeta_{\bm{x}}$ is a canonical representative.}
\rev{Applying the law of total covariance to $X_1$ with respect to $G$ yields the decomposition (see Appendix~\ref{sec:appendix_cov_derivation} for the full derivation):}
\begin{equation}
\label{eq:per-particle-cov-decomp}
\rev{\begin{aligned}
\mathrm{Cov}( X_1 \mid X_t\!=\!\bm{x} )
&=
\underbrace{\mathbb{E}_{G}\bigl[
  \mathrm{Cov}(
    X_1 \mid X_t\!=\!\bm{x}, G
  )
\bigr]}_{\text{intrinsic variability}}
\\
&+
\underbrace{\mathrm{Cov} \bigl(
  \mathbb{E}[
    X_1 \mid X_t\!=\!\bm{x}, G
  ]
  \bigm|
  X_t\!=\!\bm{x}
\bigr)}_{\text{role-ambiguity term}\;\succeq\; 0}.
\end{aligned}}
\end{equation}
\rev{The first term is the intrinsic variability conditioned on a fixed permutation, averaged over $G$; the second term captures the additional variability from random $G$.
Exploiting the $G$-invariance of a canonicalization map $C$ (Section~\ref{def:canonical_map}), the second term vanishes for $\widetilde{X}_1 := C(X_1)$, giving for $\widetilde{Y} := (\widetilde{X}_1 - X_t)/(1-t)$:}
\begin{equation}
\label{eq:per-particle-trace-ineq-Y}
\mathrm{tr}\,
\mathrm{Cov}\!\left( Y \mid X_t = \bm{x} \right)
\;\ge\;
\mathrm{tr}\,
\mathrm{Cov}\!\left( \widetilde{Y} \mid X_t = \bm{x} \right).
\end{equation}
}
\begin{figure*}[!htb]
    \centering
    \includegraphics[width=1.0\textwidth]{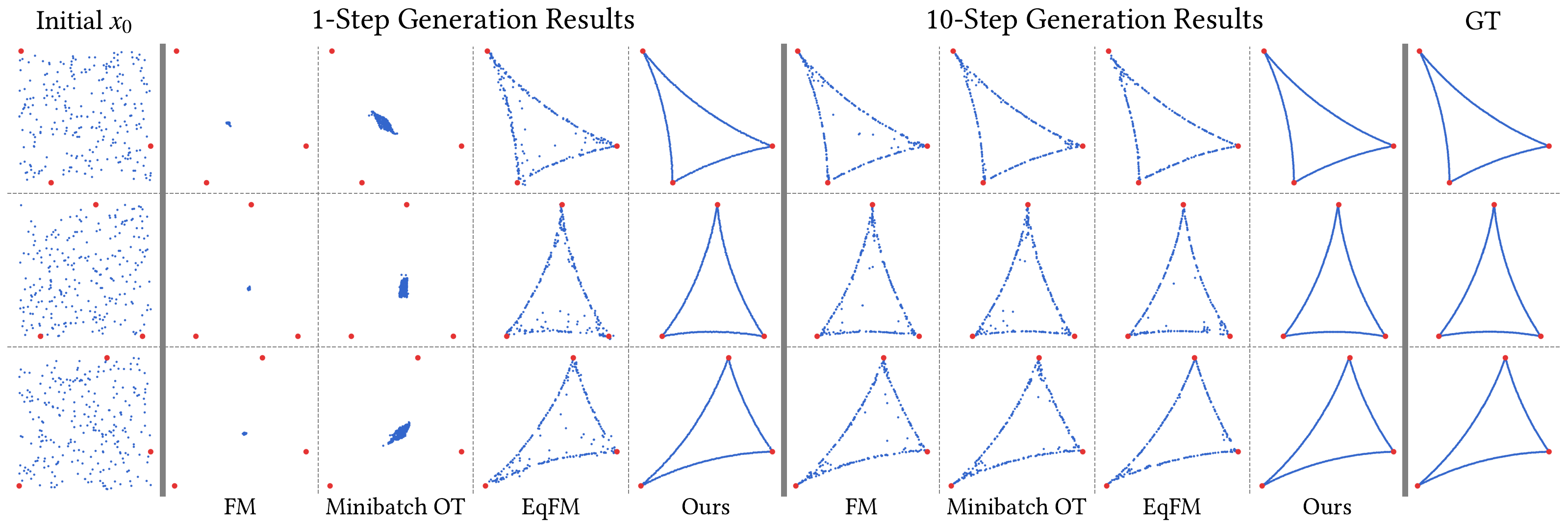}
    \caption{\textbf{Minimal surface generation (3 anchors).} We consider the 2D analog of minimal surfaces: soap film boundaries satisfying  area constraints. We compare 1-step and 10-step generation results with different methods. Red dots indicate anchor particles; blue dots show generated boundary particles. Our method produces accurate minimal surface boundaries in a single step, while baselines require multiple steps and \revvvv{still} exhibit artifacts. Ground truth (GT) shown on the right.}
    \label{fig:minimal_surface}
\end{figure*}


Therefore, without canonicalization, this covariance contains an extra component from random permutations, forcing identity embeddings to average over different roles.
Orbit-space canonicalization of $X_1$ removes exactly this role-ambiguity term,
so that each identity embedding can specialize to a well-defined canonical role and the targets become \revvvv{substantially} easier to learn.
Our ablations in Section~\ref{ablation:index} empirically confirm that the largest
gains arise when identity embeddings and one-sided canonicalization are used
together.

\subsection{Orbit-continuous canonicalization and straight flows}
\label{subsec:orbit_straightness}

We now turn to our second objective: \textbf{encouraging straight flows}.
Beyond reducing conditional variance at each fixed configuration $\bm{x}_t$, we
would like the Bayes-optimal velocity field
$\bm{u}^*(\bm{x},t)$ in Eq.~\eqref{eq:bayes_opt} to vary smoothly across nearby
configurations.
Since the full trajectory satisfies the ODE
\[
\frac{\mathrm{d}}{\mathrm{d}t}\bm{x}_t = \bm{u}^*(\bm{x}_t,t),
\]
a locally Lipschitz velocity field with a small Lipschitz constant encourages
nearby trajectories to remain coherent and to change direction smoothly over
time.
This motivates a canonicalization map $C$ that is well behaved on the orbit
space, so that nearby orbits are mapped to nearby canonical representatives.

To make this precise, we consider the orbit space
$\mathcal{O} = (\mathbb{R}^D)^N/G$ and equip it with a metric
$d_{\mathcal{O}}$ that is invariant under the group action.
We say that a canonicalization map
$C : (\mathbb{R}^D)^N \to (\mathbb{R}^D)^N$ is \emph{orbit-continuous} if it
maps nearby orbits to nearby canonical representatives in $(\mathbb{R}^D)^N$ in
a Lipschitz manner, i.e., if there exists a constant $L_{\mathrm{orb}}$ such that
for all $\bm{x},\bm{x}' \in (\mathbb{R}^D)^N$,
\begin{equation}
\label{eq:orbit-Lipschitz}
\|C(\bm{x}) - C(\bm{x}')\|
\;\le\;
L_{\mathrm{orb}}\, d_{\mathcal{O}}\bigl(\mathrm{Orb}(\bm{x}), \mathrm{Orb}(\bm{x}')\bigr).
\end{equation}

\rev{
The Bayes-optimal velocity can be written as
\begin{equation}
\label{eq:bayes-vel-canon}
\bm{u}^*(\bm{x},t)
=
\mathbb{E}\!\left[Y \mid X_t = \bm{x}\right]
=
\frac{1}{1-t}\,
\bigl(\bm{m}(\bm{x}) - \bm{x}\bigr),
\end{equation}
where the canonical mean is
\(\bm{m}(\bm{x}) := \mathbb{E}\!\left[\widetilde{X}_1 \mid X_t = \bm{x}\right].\)
}

\rev{Under natural smoothness assumptions on the endpoint distribution over the orbit space (see Appendix~\ref{sec:appendix_orbit_continuous} for details), the canonical means $\bm{m}(\bm{x})$ inherit orbit-Lipschitz regularity from the orbit-continuity of $C$.
Combining this with Eq.~\eqref{eq:bayes-vel-canon}, we obtain a local Lipschitz bound for the Bayes-optimal velocity field:}
\begin{equation}
\label{eq:u-star-Lip}
\rev{\|\bm{u}^*(\bm{x},t) - \bm{u}^*(\bm{x}',t)\|
\;\le\;
L_{\mathrm{vel}}(t)\,
d_{\mathcal{O}}\bigl(\mathrm{Orb}(\bm{x}), \mathrm{Orb}(\bm{x}')\bigr),}
\end{equation}
\rev{where $L_{\mathrm{vel}}(t)$ is a time-dependent constant controlled by $L_{\mathrm{orb}}$ and the intrinsic smoothness of the canonical means.
Thus, orbit-continuous canonicalization maps that align neighboring orbits with neighboring canonical representatives ensure a continuous velocity field and thereby encourage straight flows. Since each per-particle velocity is a component of the full vector field $\bm{u}^*(\bm{x},t)$, this regularity also transfers componentwise to the individual particle trajectories.}

\paragraph{Practical canonicalization maps.}
In practice, we first apply a simple pose-normalization step (recenter and
align a PCA frame), which removes translations and global rotations.
The main challenge lies in the remaining symmetry, the \emph{permutation} part $S_N$, whose
combinatorial complexity grows as $N!$.
We therefore focus our design effort on a robust permutation canonicalization.
We find that a Hilbert space-filling curve ordering provides a stable
permutation of particle indices under small perturbations.
Figure~\ref{fig:ablation_bn} compares several alternatives (e.g., Z-order,
Moore curve).
For joint canonicalization (See Section~\ref{subsec:joint_canon}) we analogously apply a \(n\)-dimensional Hilbert sort \cite{skilling2004programming}.
In the minimal-surface experiments (Fig.~\ref{fig:minimal_surface_multi} and Fig.~\ref{fig:minimal_surface}) we instead
use a simple rule: we pin left bottom anchor as index~0 and then
enumerate boundary particles in counterclockwise order along the curve.
All of these constructions are designed to satisfy the orbit-continuity
intuition that neighboring orbits should induce nearby canonical representatives and avoid abrupt role flips.

\subsection{Canonicalizing \(X_0\) increases Lipschitz ratios}
\label{subsec:lipschitz_view}

\begin{figure}
    \centering
    \includegraphics[width=0.5\textwidth]{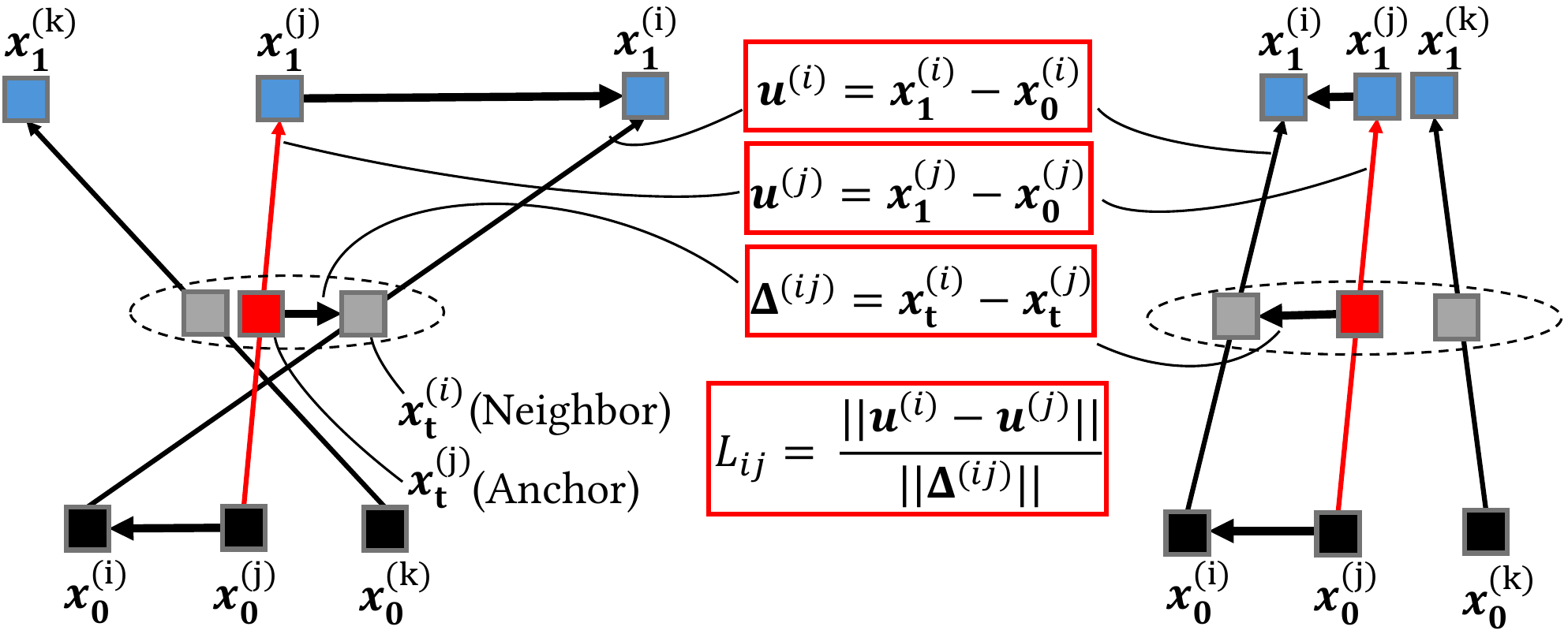}
\caption{\rev{Illustration of directional cancellation in the Lipschitz ratio. Squares represent stacked particle vectors in $\mathbb{R}^{DN}$.
\textbf{Left:} When $\Delta_0^{(ij)}$ and $\Delta_1^{(ij)}$ point in opposite directions, the denominator nearly vanishes while the numerator remains large, yielding a large Lipschitz ratio.
\textbf{Right:} Without such cancellation, the ratio stays moderate.}}
    \label{fig:Lij_illustration}
\end{figure}

\rev{A natural follow-up question is whether we should also canonicalize the noise endpoint $X_0$.
We show that further canonicalizing $X_0$ amplifies \emph{directional cancellation} events and inflates the local Lipschitz ratios of the velocity field; a detailed derivation is given in Appendix~\ref{sec:appendix_lipschitz}.}

\rev{We measure the smoothness of the velocity field via nearest-neighbor Lipschitz ratios.
For each $k$-NN edge $(i,j)$ built on the interpolants $\bm{x}_t^{(i)} = (1{-}t)\bm{x}_0^{(i)} + t\bm{x}_1^{(i)}$, the Lipschitz ratio is}
\begin{equation}
\label{eq:Lij-deltas}
\rev{L_{ij}(t)^2
=
\frac{\bigl\|\Delta_1^{(ij)} - \Delta_0^{(ij)}\bigr\|^2}
     {\bigl\|(1-t)\,\Delta_0^{(ij)} + t\,\Delta_1^{(ij)}\bigr\|^2},}
\end{equation}
\rev{where $\Delta_\ell^{(ij)} := \bm{x}_\ell^{(i)} - \bm{x}_\ell^{(j)}$.
Once $X_1$ is canonicalized, $\Delta_1^{(ij)}$ is typically small.
If $X_0$ is also canonicalized, the contracted $\widetilde{\Delta}_0^{(ij)}$ reaches a comparable scale to $\Delta_1^{(ij)}$, making it much easier for the two vectors to nearly cancel in the denominator while the numerator stays large (Figure~\ref{fig:Lij_illustration}).
By contrast, keeping $X_0$ uncanonicalized preserves a large spread in $\Delta_0^{(ij)}$, making such cancellation statistically unlikely.}

\rev{We empirically verify in Section~\ref{subsec:mid_time_analysis} (Figure~\ref{fig:midtime-analysis}) that our one-sided canonicalization strategy (canonicalizing $X_1$ only) achieves the smallest local Lipschitz ratios and the lowest prevalence of high-cancellation edges among all four canonicalization regimes (no canonicalization, $X_0$ only, $X_1$ only, and both).}


    


\section{Geometric Probability Paths for Attribute Encoding}
\label{sec:att_prob_path}

This section focuses on the third key component of OGPP: \textbf{geometric probability paths}.
The previous section focused on how to process the \emph{endpoints} of the probability path by performing symmetry reduction on the terminal distribution.
We now turn to the second design axis: the \emph{shape} of the probability path itself.

\begin{figure}
    \centering
    \includegraphics[width=0.42\textwidth]{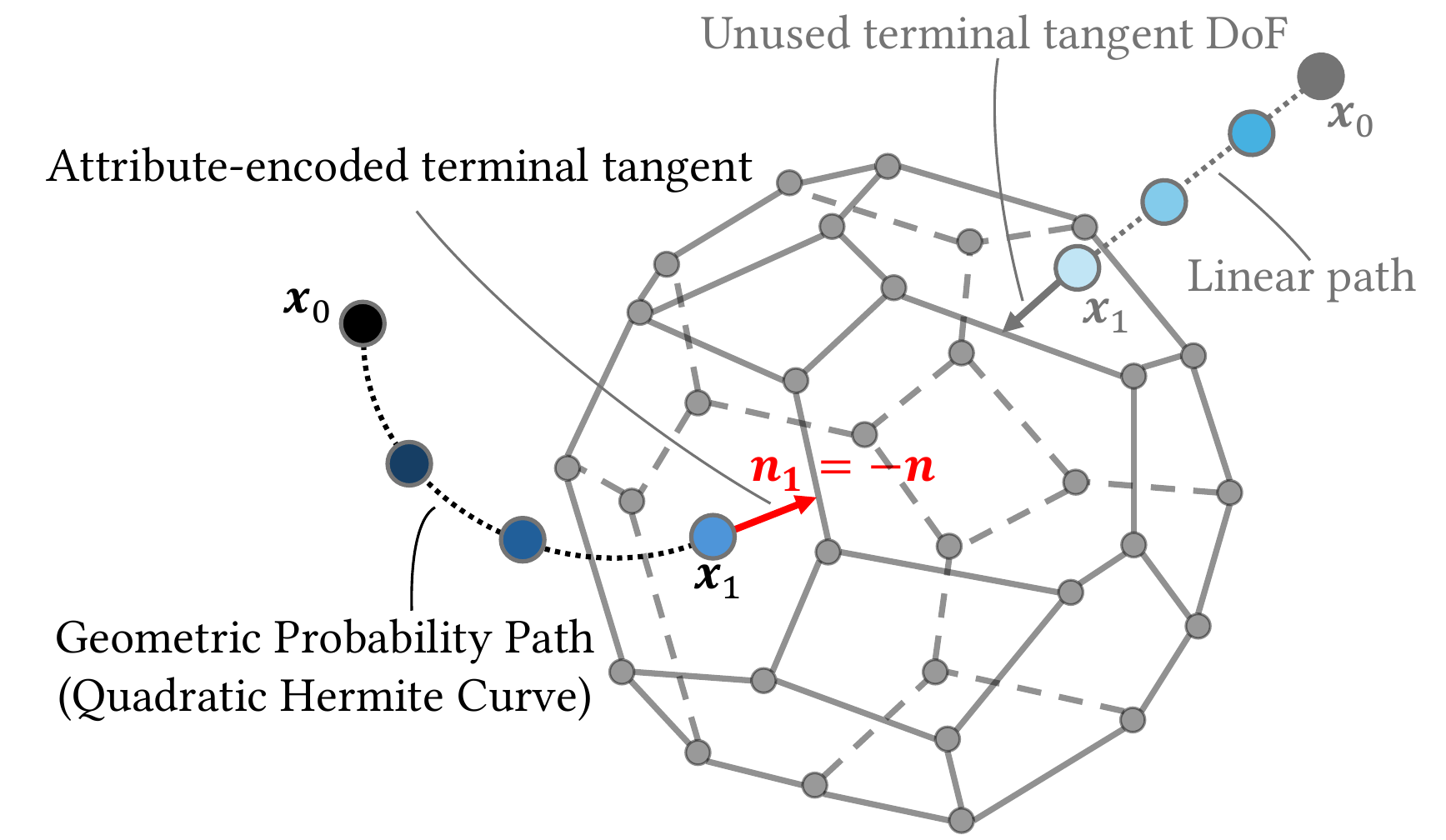}
    \caption{\textbf{Geometric probability paths for attribute encoding.} \textit{Left:} Our geometric probability path (quadratic Hermite curve) aligns the terminal tangent with the surface normal $\bm{n}_1$, encoding per-particle attributes into the path geometry. \textit{Right:} Standard linear interpolation leaves the terminal velocity as an unused degree of freedom.}
    \label{fig:att_illustration}
\end{figure}

In standard flow matching, the conditional path between noise and data is often taken to be a linear interpolation.
While this choice is simple and effective for transporting particle positions (or, more generally, distributions), it leaves the terminal velocity field at $t = 1$ geometrically under-utilized: the velocity at the endpoint does not carry any intrinsic meaning.
We exploit these unused degrees of freedom by constructing \emph{geometric probability paths} whose terminal tangent aligns with a per-particle attribute.
In this work, we instantiate this attribute as the surface normal of the shape.
Formally, let the conditioning variable be $\bm{z} = (\bm{x}_1, \bm{n}_1)$, where $\bm{x}_1 \in \mathbb{R}^3$ is the target position and $\bm{n}_1 \in \mathbb{R}^3$ is the associated surface normal.
We design conditional probability paths that satisfy three boundary conditions:
(i) the path starts at noise, $\bm{x}(0) = \bm{x}_0 \sim p_\mathrm{init}$;
(ii) the path ends at the target position, $\bm{x}(1) = \bm{x}_1$;
and (iii) the terminal velocity encodes the attribute, $\bm{v}_1 = \dot{\bm{x}}(1) \propto \bm{n}_1$.
Conditions (i) and (ii) leave the \emph{shape} of the path largely unconstrained; by slightly bending the path away from a straight line to enforce (iii), we turn the terminal tangent---an otherwise free degree of freedom in the linear path---into a structured carrier of geometric information (see Figure~\ref{fig:att_illustration} and Figure~\ref{fig:atv_illustration} for an illustration).

In this section, we first introduce quadratic Hermite probability paths in Section~\ref{subsec:quadratic_hermite_path}. 
We then choose an arc-length-aware terminal velocity to stabilize time sampling in Section~\ref{subsec:atv}, extend canonicalization to the joint position-normal endpoints via joint canonicalization in Section~\ref{subsec:joint_canon}, and finally characterize the marginal terminal velocity at $t=1$ as a normal predictor together with its training and inference usage in Section~\ref{subsec:marginal_vel} and Section~\ref{subsec:implication}.


\begin{figure}[!htb]
    \centering
    \includegraphics[width=0.5\textwidth]{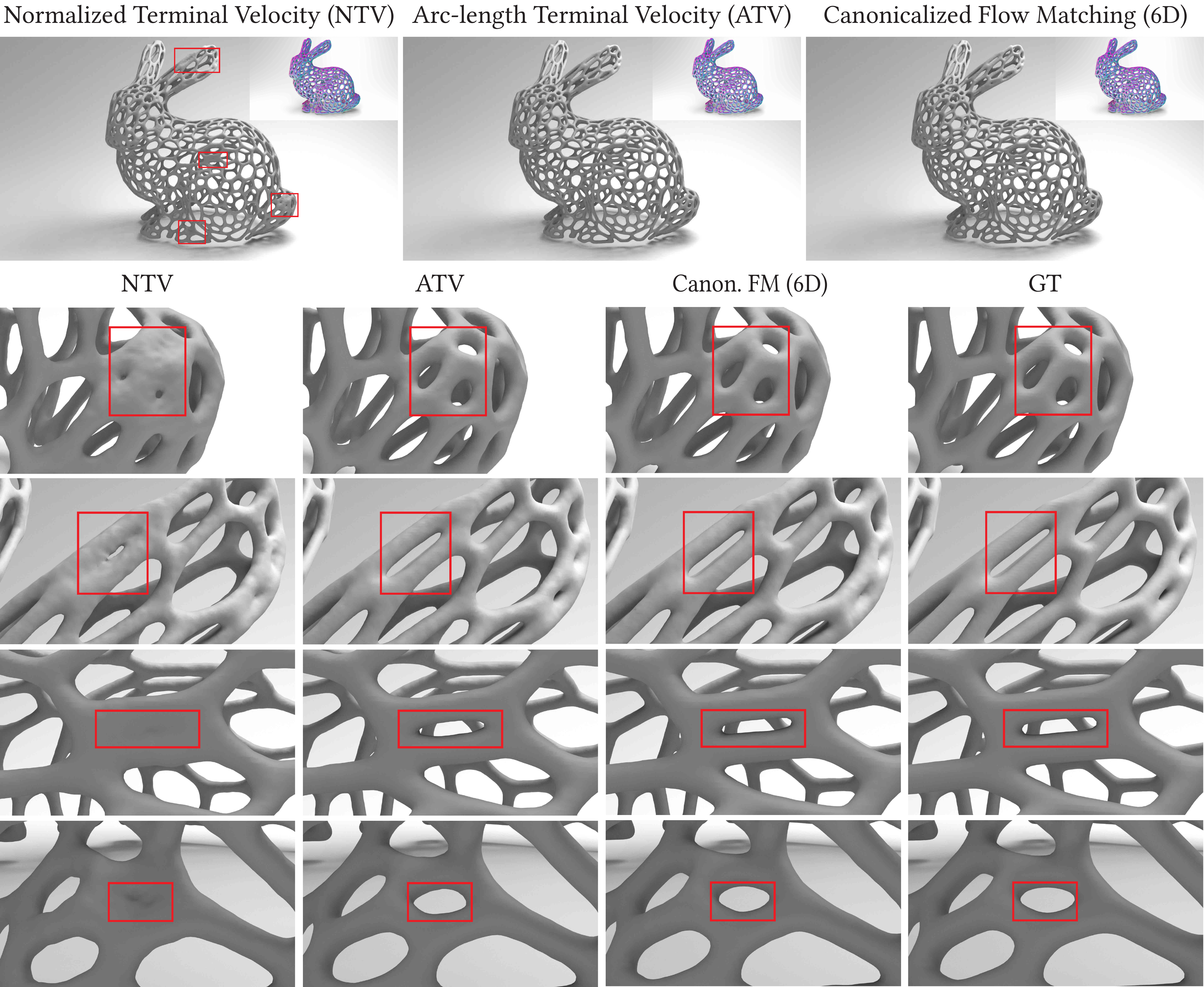}
    \caption{\rev{\textbf{Ablation on normal encoding strategies.} Top row: Screened Poisson reconstructions from NTV, ATV, and canonicalized 6D flow matching (Canon.\ FM 6D), with normal-colored point clouds inset. Bottom rows: zoomed-in comparisons against the ground truth (GT). ATV and Canon. FM (6D) achieve comparable quality and accurately reconstruct small Voronoi cells and thin structures that NTV fails to capture.}}
    \label{fig:surface_bunny_comparison}
\end{figure}

\subsection{Quadratic Hermite Curves}
\label{subsec:quadratic_hermite_path}

We construct the probability paths using a quadratic Hermite curve that satisfies the boundary conditions above.
We define the curve \(\gamma(t)\) as:
\begin{equation}
\label{eq:hermite_path}
\gamma(t) = \bm{x}_0 + \alpha(t) \cdot (\bm{x}_1 - \bm{x}_0) + \beta(t) \cdot \bm{v}_1,
\end{equation}
where the basis functions are
\(\alpha(t) = 2t - t^2\), \(\beta(t) = t^2 - t\),
and \(\bm{v}_1\) denotes the terminal tangent that we assign at \(t = 1\).


\paragraph{Conditional velocity field.}
Differentiating Eq.~\eqref{eq:hermite_path}, we obtain the conditional velocity field along the curve:
\begin{equation}
\label{eq:cond_vf_hermite}
\bm{u}_t^\mathrm{ref}(\bm{x}_t | \bm{z}) = \frac{2}{1-t}(\bm{x}_1 - \bm{x}_t) - \bm{v}_1.
\end{equation}

\begin{remark}
The quadratic Hermite path is the simplest polynomial path satisfying our three boundary conditions.
A cubic Hermite spline would introduce an additional degree of freedom (the tangent at $t=0$), which is unnecessary for our purpose.
We verify in our ablation study (Table~\ref{tab:ablation_study_geometric_path}) that the quadratic path achieves the best performance.    
\end{remark}

\subsection{Arc-Length Terminal Velocity (ATV)} 
\label{subsec:atv}

\begin{figure}
    \centering
    \includegraphics[width=0.45\textwidth]{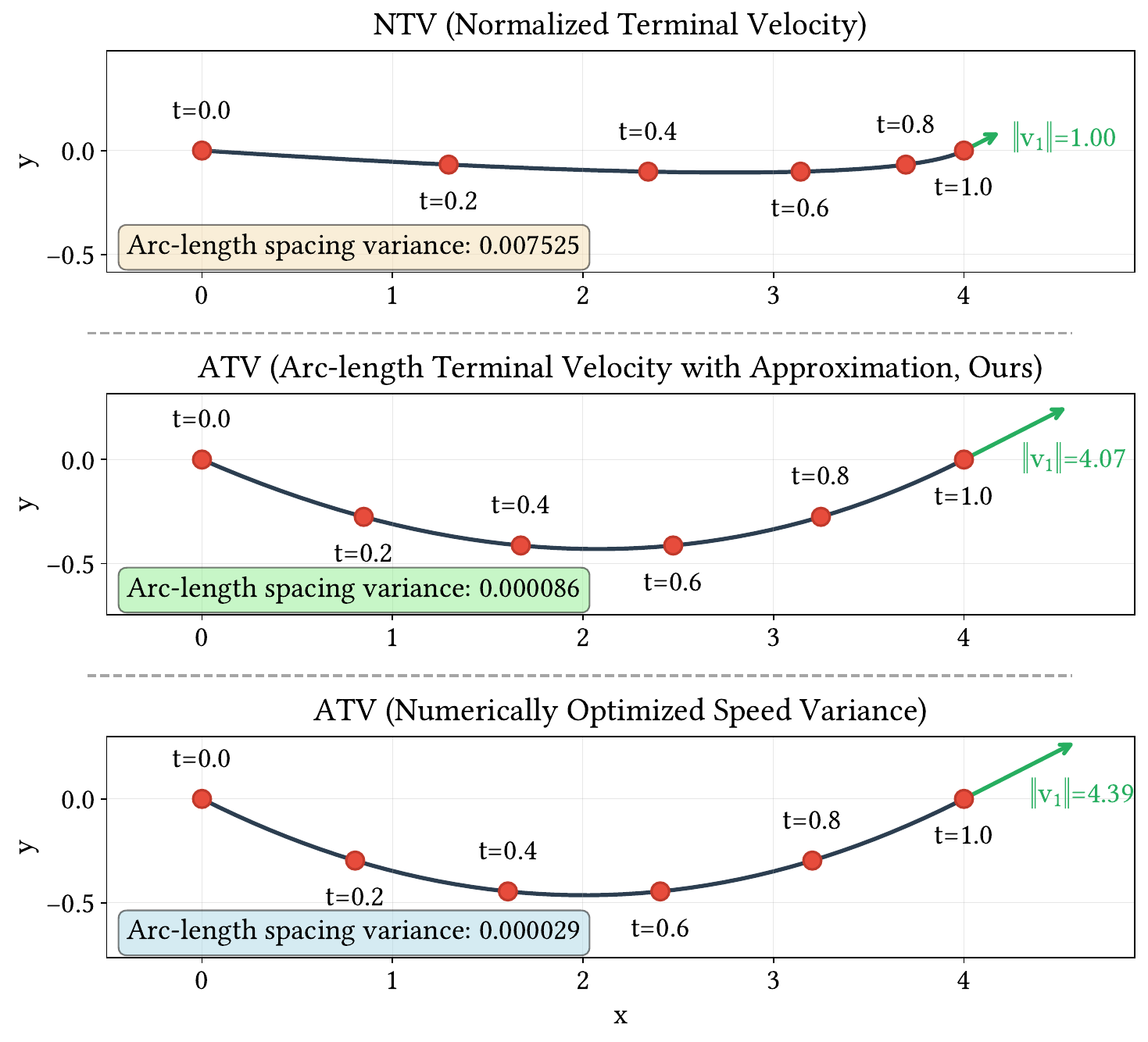}
    \caption{Comparison of terminal velocity magnitude choices. Red dots indicate uniform time samples $t \in \{0, 0.2, 0.4, 0.6, 0.8, 1.0\}$; green arrows show $\bm{v}_1$. \textbf{Top:} NTV sets $\|\bm{v}_1\| = 1$, yielding nonuniform arc-length spacing. \textbf{Middle:} Our ATV approximation sets $\|\bm{v}_1\| = D(1 + \lambda(1 - S))$, achieving near-uniform spacing with negligible overhead. \textbf{Bottom:} Numerically optimized ATV chooses $\|\bm{v}_1\|$ to minimize speed variance along the curve, giving optimal uniformity but requiring numerical optimization.}
    \label{fig:atv_illustration}
\end{figure}

\rev{For surface normals, only the \emph{direction} of $\bm{v}_1$ is constrained; its magnitude is a free parameter.
We exploit this freedom to achieve approximately uniform speed profiles along each trajectory, so that uniform time sampling $t \sim \mathrm{Uniform}(0,1)$ correlates well with uniform sampling along the curve.}
\begin{figure*}[!htb]
    \centering
    \includegraphics[width=1.0\textwidth]{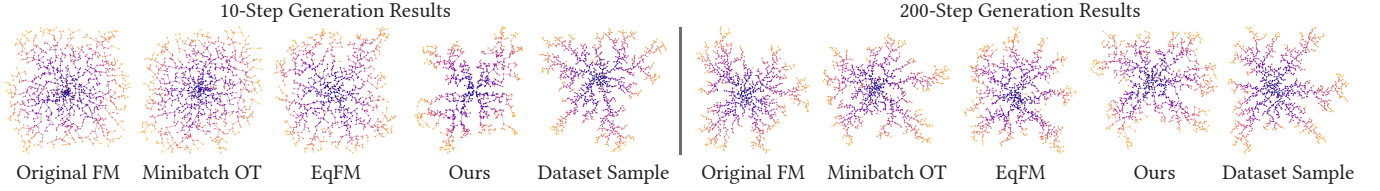}
    \caption{\textbf{DLA generation comparison.} 10-step (left) and 200-step (right) generation results. At 10 steps, baselines produce scattered, non-fractal structures, while ours exhibits realistic dendritic branching. At 200 steps, all methods improve; \rev{ours appears closest to} the ground-truth fractal morphology. Color encodes particle attachment order (early: dark, late: light).}
    \label{fig:dla_comparison}
\end{figure*}

\rev{Concretely, for each particle, Algorithm~\ref{alg:ogpp_training} (lines~5--13) computes the chord length $D = \|\bm{x}_1 - \bm{x}_0\|$ and the alignment
$S = \hat{\bm{d}} \cdot \hat{\bm{n}}_1$ between chord direction and normal, and sets the terminal velocity to
\begin{equation}
\label{eq:atv}
L_{\mathrm{arc}} \;=\; D \bigl(1 + \lambda (1 - S)\bigr),
\qquad
\bm{v}_1 \;=\; L_{\mathrm{arc}} \,\hat{\bm{n}}_1.
\end{equation}
The scaling $L_{\mathrm{arc}}$ adapts the terminal speed to the chord length and the angle between the chord and the normal: when the normal is aligned with the chord ($S \approx 1$), the path is nearly straight and $\|\bm{v}_1\| \approx D$; when they are misaligned ($S \ll 1$), the path bends more and a larger $\|\bm{v}_1\|$ compensates to maintain uniform speed. This computation is inexpensive (only norms and dot products) and empirically produces trajectories whose speed variation over $t$ is much smaller than the naive unit-norm baseline (normalized terminal velocity, NTV, which sets $\|\bm{v}_1\|=1$). As shown in Figure~\ref{fig:atv_illustration}, our ATV approximation closely matches a numerically optimized solution that directly minimizes speed variance. A detailed discussion of why NTV leads to nonuniform speed profiles is provided in Appendix~\ref{app:atv-detail}.
We additionally compare NTV and ATV in Figure~\ref{fig:surface_bunny_comparison} (see experimental details in Section~\ref{ablation:arc}).}

\subsection{Joint Canonicalization for Attribute-Encoded Paths}
\label{subsec:joint_canon}

The conditional-covariance analysis in
Section~\ref{subsec:per_particle_cov} was derived for the linear probability
path, where the regression target
$Y = (X_1 - X_t)/(1-t)$
depends only on the endpoint position $X_1$.
Under the geometric probability path Eq.~\eqref{eq:hermite_path}, differentiating
yields the regression target
\begin{equation}
\label{eq:hermite_YI_main}
Y_t
=
\frac{2}{1-t}\bigl(X_1 - X_t\bigr) - V_1,
\end{equation}
where $V_1$ is the stacked terminal velocity defined by Eq.~\eqref{eq:atv}. For each $X_t = \bm{x}$, the randomness now comes from the joint
endpoint
$Z := (X_1, V_1) \in (\mathbb{R}^6)^N$,
not just from $X_1$.

Under the same orbit-symmetry factorization as
Section~\ref{subsec:per_particle_cov}, extended to the joint endpoint
$Z$, orbit-space canonicalization of $Z$ reduces the
conditional covariance
$\mathrm{Cov}(Z \mid X_t = \bm{x})$;
since $Y_t$ is an affine function of $Z$, this decreases the
conditional covariance of $Y_t$ as well.
In particular, joint canonicalization lowers the irreducible MSE of the
velocity predictor, making the geometric-path
regression problem strictly easier to learn.

\paragraph{Implementation via 6D Hilbert curve.}
In practice, we concatenate the position $\bm{x}_1$ and attribute $\bm{n}_1$ for each particle into a six-dimensional vector $(\bm{x}_1, \bm{n}_1) \in \mathbb{R}^6$, and apply a N-dimensional Hilbert-curve \cite{skilling2004programming} in this joint space.
We compare 6D Hilbert ordering (position + normal) against 3D Hilbert ordering (position only) in Table~\ref{tab:ablation_study_geometric_path}: the joint canonicalization improves normal estimation (average cosine similarity from 0.91 to 0.92, standard deviation from 0.21 to 0.19) and generation quality (1-NNA accuracy from 0.78 to 0.61).

\subsection{Marginal Velocity at Terminal Time}
\label{subsec:marginal_vel}
The key theoretical question is: \emph{what does the marginal velocity field $\bm{u}_1^\mathrm{ref}(\bm{x})$ represent at the terminal time?}
By the marginalization trick (Theorem~\ref{thm:marginalization}), the trained network learns to approximate the marginal velocity, not the conditional one.
We now show that at $t=1$, the marginal velocity is precisely the expected attribute given the position.

Intuitively, at $t=1$, the conditional path collapses to a point mass $p_1(\cdot|\bm{z}) = \delta_{\bm{x}_1}$, so only conditioning variables with $\bm{x}_1 = \bm{x}$ contribute to the marginalization integral.
Since $\bm{u}_1^\mathrm{ref}(\bm{x}|\bm{z}) = N_1$, the marginal velocity becomes:
\begin{equation}
\bm{u}_1^\mathrm{ref}(\bm{x}) = \int N_1 \, p(N_1 | X_1 = \bm{x}) \, \mathrm{d}N_1 = \mathbb{E}\left[ N_1 \,\middle|\, X_1 = \bm{x} \right].
\end{equation}
See \rev{the supplement} for the full proof. 

\subsection{Implications for Training and Inference}
\label{subsec:implication}
The preceding analysis has direct practical implications.
During training, the same network $\bm{u}_t^\theta$ learns:
\begin{itemize}[leftmargin=*]
    \item the transport velocity for $t \in [0, 1)$, and
    \item at $t = 1$, the conditional expectation $\mathbb{E}[N_1 \mid X_1 = \bm{x}]$.
\end{itemize}
No separate network or training procedure is needed for normal estimation. Therefore, during inference:
\begin{enumerate}[leftmargin=*]
    \item first integrate the ODE from $t=0$ to $t=1$ to obtain the generated position $\bm{x}_1$;
    \item then evaluate $\bm{u}_1^\theta(\bm{x}_1)$ to get the predicted attribute $N_1$.
\end{enumerate}
The generated point cloud comes equipped with surface normals as a \emph{byproduct} of the flow, at no additional computational cost. Note that individual ODE trajectories are governed by the learned marginal velocity field rather than any single conditional Hermite path.


\section{\rev{Algorithm Overview}}
\label{sec:algorithm_overview}
\begin{algorithm}[!htb]
\caption{OGPP Training}
\label{alg:ogpp_training}
\begin{algorithmic}[1]
\Require Dataset $\mathcal{D} = \{(\bm{x}_1^{(j)}, \bm{n}_1^{(j)})\}_{j=1}^M$, number of particles per sample $N$, canonicalization map $C(\cdot)$, hyperparameter $\lambda$
\Ensure Trained velocity network $\bm{u}_t^\theta$
\Repeat
    \State Sample $\bm{x}_0^{(i)} \sim \mathrm{Uniform}([-1,1]^N)$  \Comment{Noise}
    \State Sample $\bm{z}_1^{(i)} = (\bm{x}_1^{(i)}, \bm{n}_1^{(i)})$ from $\mathcal{D}$ \Comment{Data with attributes}
    \State $(\bm{x}_1^{(i)}, \bm{n}_1^{(i)}) \gets C(\bm{x}_1^{(i)}, \bm{n}_1^{(i)})$ \Comment{Joint canon. (Sec.~\ref{sec:ogpp}, Sec.~\ref{subsec:joint_canon})}
    \For{each particle $k = 1, \ldots, N$ \textbf{in parallel}}
        \State $D^{(i),k} \gets \|\bm{x}_1^{(i),k} - \bm{x}_0^{(i),k}\|$ \Comment{Chord length}
        \State $\hat{\bm{d}}^{(i),k} \gets (\bm{x}_1^{(i),k} - \bm{x}_0^{(i),k}) / D^{(i),k}$ \Comment{Chord direction}
        \State $\hat{\bm{n}}_1^{(i),k} \gets \bm{n}_1^{(i),k} / \|\bm{n}_1^{(i),k}\|$ \Comment{Unit normal}
        \State $S^{(i),k} \gets \hat{\bm{d}}^{(i),k} \cdot \hat{\bm{n}}_1^{(i),k}$ \Comment{Directional alignment}
        \State $L_{\mathrm{arc}}^{(i),k} \gets D^{(i),k} \cdot (1 + \lambda(1 - S^{(i),k}))$ \Comment{Arc-length}
        \State $\bm{v}_1^{(i),k} \gets L_{\mathrm{arc}}^{(i),k} \cdot \hat{\bm{n}}_1^{(i),k}$ \Comment{ATV (Eq.~\ref{eq:atv}, Sec.~\ref{subsec:atv})}
        \State Construct $\gamma^{(i),k}(t)$ from $\bm{x}_0^{(i),k}$, $\bm{x}_1^{(i),k}$, $\bm{v}_1^{(i),k}$ \Comment{Sec.~\ref{subsec:quadratic_hermite_path}}
    \EndFor
    \State Sample $t \sim \mathrm{Uniform}([0, 1])$
    \For{each particle $k = 1, \ldots, N$ \textbf{in parallel}}
        \State $\bm{x}_t^{(i),k} \gets \gamma^{(i),k}(t)$ \Comment{Interpolated position}
        \State $\bm{v}_t^{(i),k} \gets \dot{\gamma}^{(i),k}(t)$ \Comment{Reference velocity (Eq.~\ref{eq:cond_vf_hermite})}
    \EndFor
    \State $\mathcal{L}^{(i)} \gets \frac{1}{N}\sum_{k=1}^{N} \|\bm{u}_t^{\theta,k}(\bm{x}_t^{(i)}) - \bm{v}_t^{(i),k}\|^2$ \Comment{MSE loss}
    \State Update $\theta$ via gradient descent on $\mathcal{L}^{(i)}$
\Until{converged}
\end{algorithmic}
\end{algorithm}

\begin{algorithm}[!htb]
\caption{\rev{OGPP Inference}}
\label{alg:ogpp_inference}
\begin{algorithmic}[1]
\Require Trained velocity network $\bm{u}_t^\theta$, number of particles per sample $N$, number of steps $K$
\Ensure Generated particles $\{\bm{x}_1^{i}\}_{i=1}^N$ with normals $\{\hat{\bm{n}}^{i}\}_{i=1}^N$
\State Sample $\bm{x}_0 \sim \mathrm{Uniform}([-1,1]^N)$ \Comment{Initial noise}
\State $\Delta t \gets 1/K$ \Comment{Step size}
\For{$k = 0, \ldots, K-1$}
    \State $t \gets k \cdot \Delta t$
    \State $\bm{u}_t^\theta(\bm{x}_t) \gets \text{evaluate NN at } (\bm{x}_t, t)$
    \For{each particle $i = 1, \ldots, N$ \textbf{in parallel}}
        \State $\bm{x}_{t+\Delta t}^{i} \gets \textsc{Step}(\bm{x}_t^{i},\; \bm{u}_t^{\theta, i}(\bm{x}_t),\; \Delta t)$ \Comment{ODE integration}
    \EndFor
\EndFor
\State $\bm{v}_1^{i} \gets \bm{u}_1^{\theta, i}(\bm{x}_1)$ for $i = 1, \ldots, N$ \Comment{Terminal velocity}
\State $\hat{\bm{n}}^{i} \gets \bm{v}_1^{i} / \|\bm{v}_1^{i}\|$ for $i = 1, \ldots, N$ \Comment{ Unit normal}
\State \Return $\{\bm{x}_1^{i},\; \hat{\bm{n}}^{i}\}_{i=1}^N$
\end{algorithmic}
\end{algorithm}

\rev{We summarize our training and inference procedures in Algorithm~\ref{alg:ogpp_training} and Algorithm~\ref{alg:ogpp_inference}, assuming a batch size of 1.} Training integrates three key components, as illustrated in Figure~\ref{fig:workflow}. First, \emph{orbit-space canonicalization} (Section~\ref{sec:ogpp}) canonicalizes only the terminal endpoint $X_1$ to reduce conditional covariance and straighten flows (lines 2--4). Second, \emph{particle index embedding} (Section~\ref{subsec:model_arch}) allows each index to specialize to its canonical role, turning the regression problem from a noisy mixture into well-separated families of trajectories. Third, \emph{geometric probability paths} (Section~\ref{sec:att_prob_path}) replace linear paths with quadratic Hermite paths, \revvvv{efficiently} encoding surface normals in the terminal tangent with arc-length-aware velocity (lines 6--12).

\rev{At inference time (Algorithm~\ref{alg:ogpp_inference}), we draw noise from the same prior and integrate the learned velocity field $\bm{u}_t^\theta$ forward from $t{=}0$ to $t{=}1$ using a standard ODE solver. The final positions $\bm{x}_1$ give the generated particle locations, and the terminal velocity $\bm{u}_1$ yields the surface normals after normalization.}

\section{Experiments}
\label{sec:validation}
\begin{figure*}[!htb]
    \centering
    \includegraphics[width=1.0\textwidth]{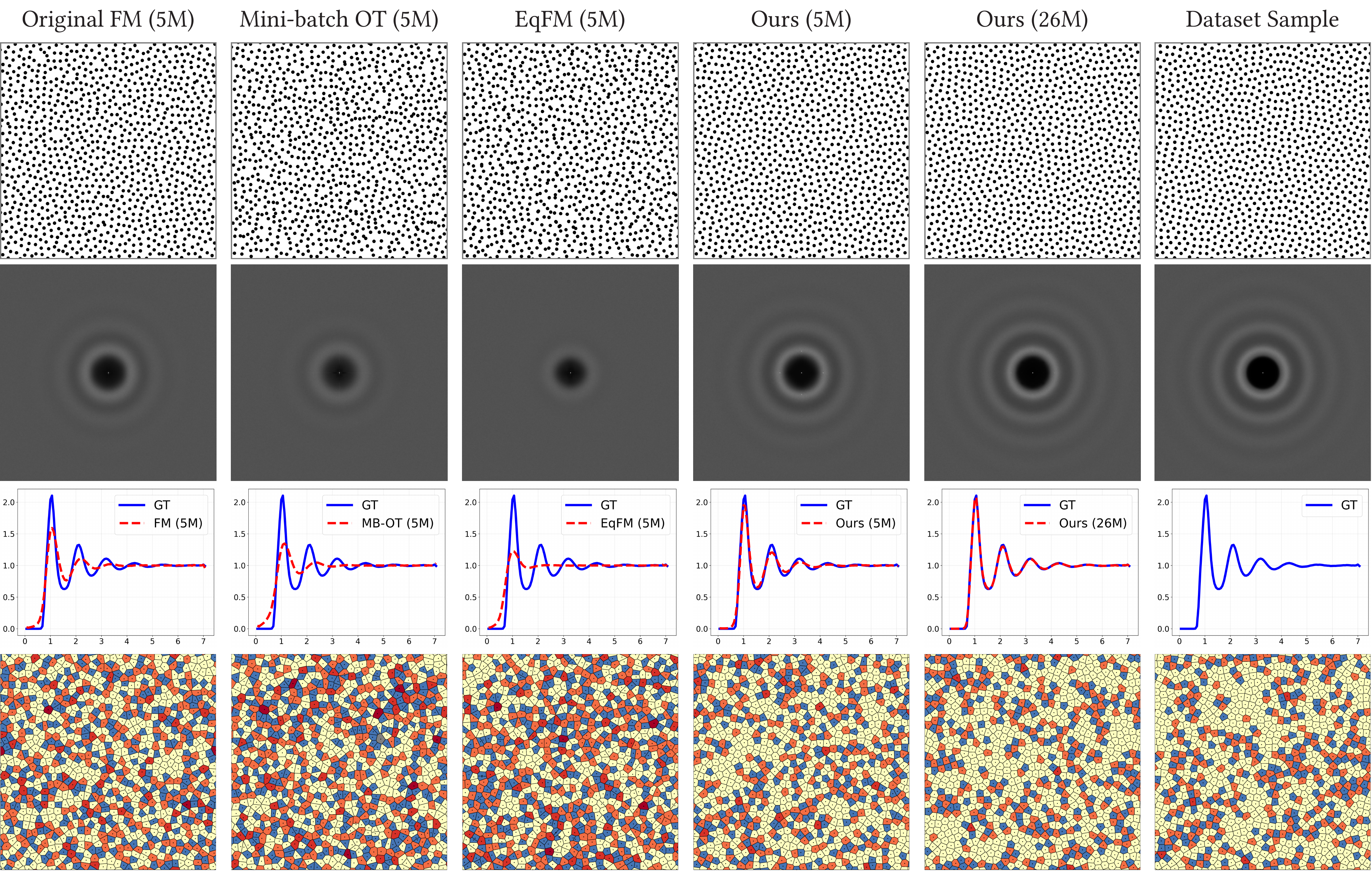}
    \caption{\textbf{Uniform blue-noise generation.}
    Comparison of flow-matching variants for 1024-point uniform blue-noise generation.
    \textit{Row 1:} One generated point set.
    \textit{Row 2:} 2D power spectrum averaged over 1K generated samples.
    \textit{Row 3:} Radial power spectrum averaged over 1K generated samples.
    \textit{Row 4:} Delaunay triangulation valence (color indicates neighbor count).
    Our method (5M and 26M) produces the sharpest spectral ring, and the results closely match the ground truth.}
    \label{fig:uniform_bn_comparison}
\end{figure*}
We evaluate our framework on two groups of tasks: energy-driven particle generation and 3D shape generation. By energy-driven, we refer to particle generation problems whose targets are equilibrium configurations of explicit physical or geometric energy functionals.
Such tasks include blue-noise sampling, minimal surfaces, diffusion-limited aggregation (DLA), and the multilayer Thomson problem. 3D shape and geometry tasks include point cloud generation on ShapeNet and single-shape encoding  \cite{zhang2025geometry} on complex meshes.

\paragraph{Model, Training, and Dataset Setup}

For model architecture, we adopt a plain transformer (See Section~\ref{subsec:model_arch}) with 5M or 26M parameters depending on task complexity, trained on NVIDIA RTX 4090 or H200 SXM GPUs. For canonicalization strategies, we use Hilbert curve sorting as our standard strategy across all experiments, except for minimal surface generation where we use counterclockwise polygon ordering to respect boundary structure. We train for 3K epochs on energy-driven tasks (6K for 26M blue noise), 8K epochs on CelebA, and 50K epochs on ShapeNet for ours and comparison models. Our method and baselines (Original FM, Minibatch OT) use batch size 200 for most tasks, with 256 for 5M blue noise and 3360 for 26M blue noise; Similar to \cite{hui2025not}, we use batch size 8 for EqFM due to its $\mathcal{O}(B^2 N^3)$ OT coupling cost, but we train it for the same wall-clock time as the other methods to ensure a fair comparison. Full training configurations are provided in Table~A-1 in the appendix. For energy-driven tasks, we generate training data using domain-specific algorithms and solvers. We use the CelebA dataset \cite{liu2015deep} for adaptive blue noise generation, ShapeNet \cite{chang2015shapenet} for point cloud generation, and Thingi10k \cite{zhou2016thingi10k} for single-shape encoding. Evaluation metrics and baselines specific to each task are described in the corresponding subsections. 

\subsection{Energy-driven Particle Generation}
We evaluate our approach on four energy-driven particle generation tasks, where equilibrium configurations arise as minimizers of physical or geometric objectives. For each task, we assess quality using intrinsic metrics that are aligned with the underlying energy functional. We first introduce the task background, then describe the dataset construction and evaluation metrics, and finally compare our method against competing baselines.

\begin{figure}[t]
    \centering
    \begin{subfigure}{\columnwidth}
        \centering
        \includegraphics[width=0.325\columnwidth]{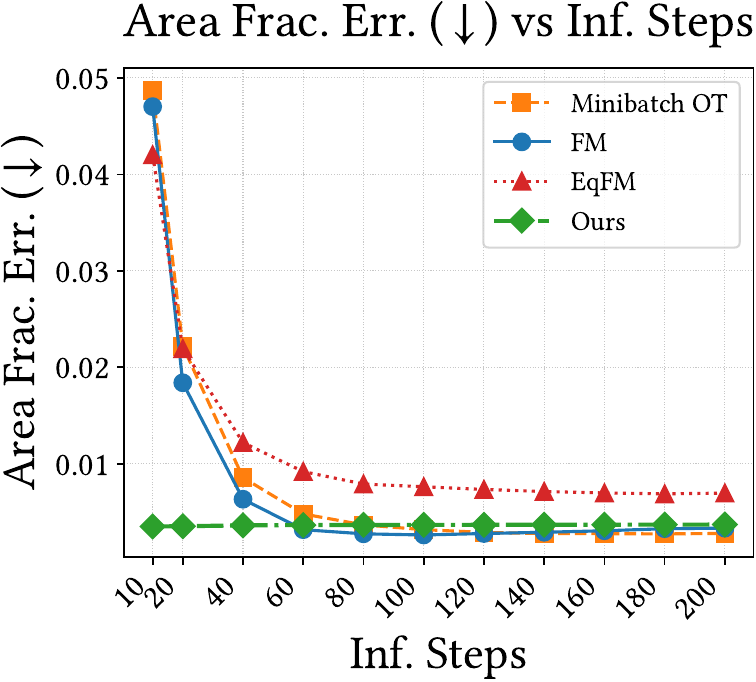}
        \hfill
        \includegraphics[width=0.325\columnwidth]{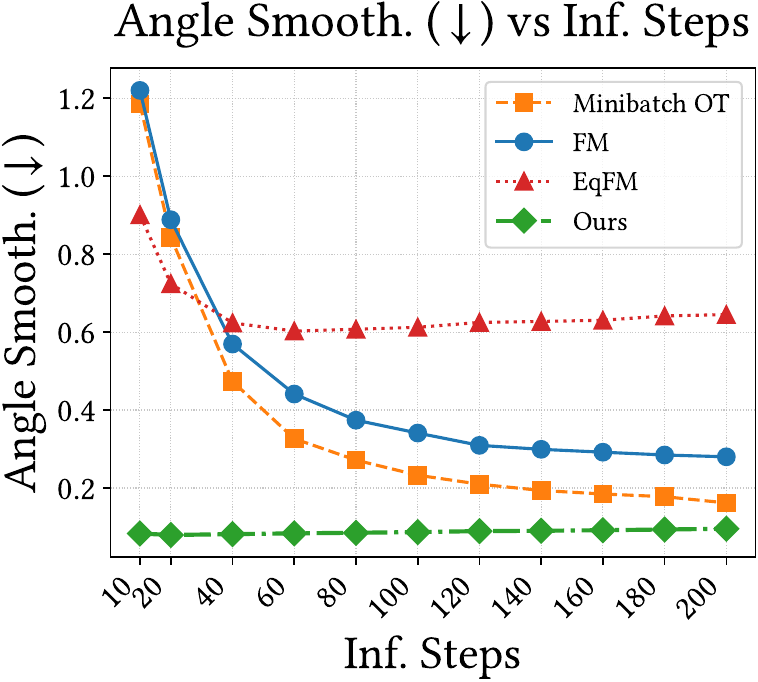}
        \hfill
        \includegraphics[width=0.325\columnwidth]{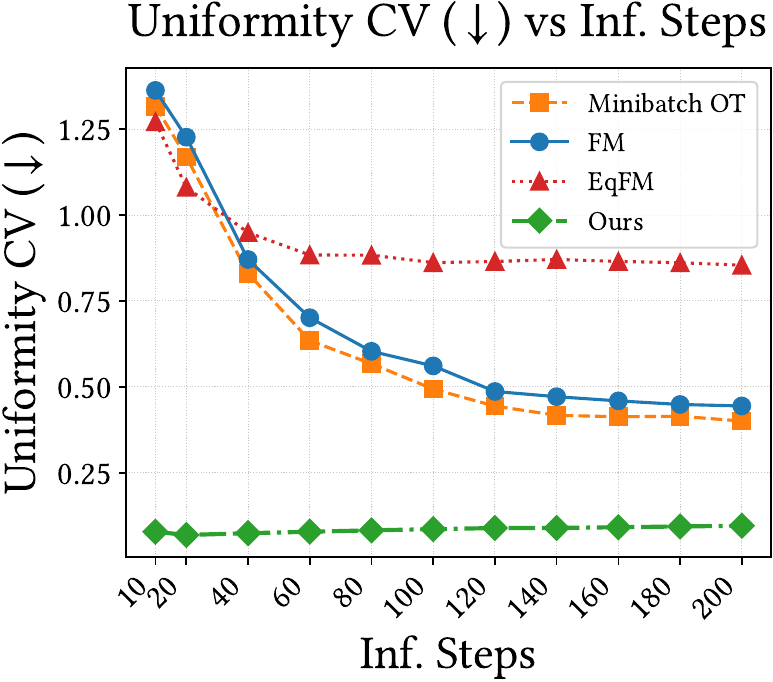}
        \caption{\rev{Minimal surface: area fraction error, angle smoothness, and uniformity CV (all lower is better). Circled points highlight representative steps.}}
        \label{fig:minimal_metrics}
    \end{subfigure}


    \begin{subfigure}{\columnwidth}
        \centering
        \begin{minipage}{0.48\columnwidth}
            \centering
            \includegraphics[width=\linewidth]{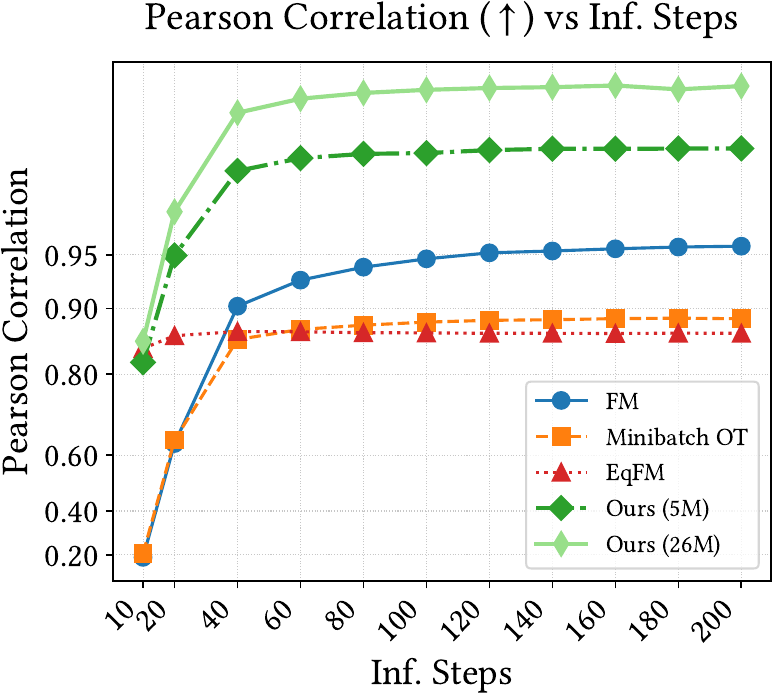}
        \end{minipage}
        \hfill
        \begin{minipage}{0.48\columnwidth}
            \centering
            \includegraphics[width=\linewidth]{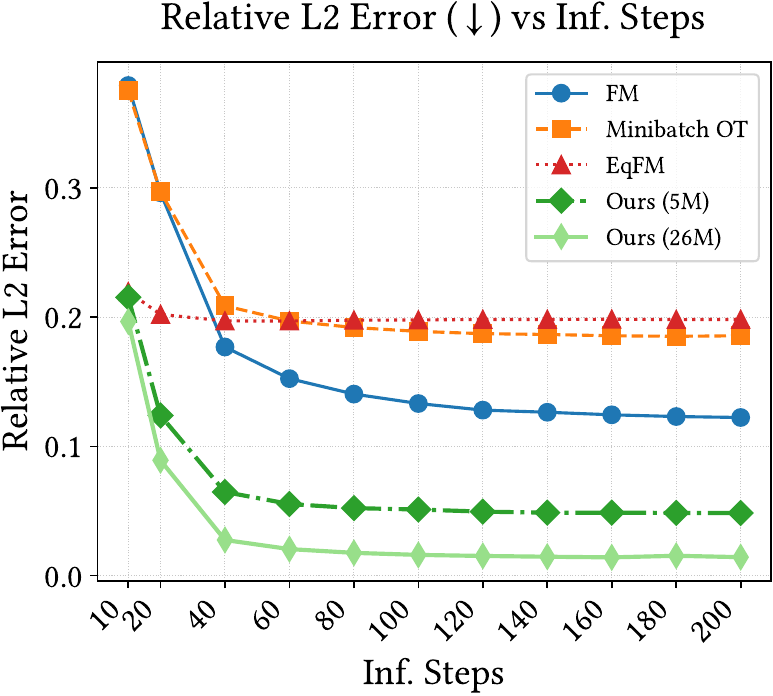}
        \end{minipage}
        \caption{\rev{Blue noise: Pearson correlation (higher is better) and relative $L_2$ error (lower is better).}}
        \label{fig:bn_metrics_steps}
    \end{subfigure}


    \begin{subfigure}{\columnwidth}
        \centering
        \begin{minipage}{0.48\columnwidth}
            \centering
            \includegraphics[width=\linewidth]{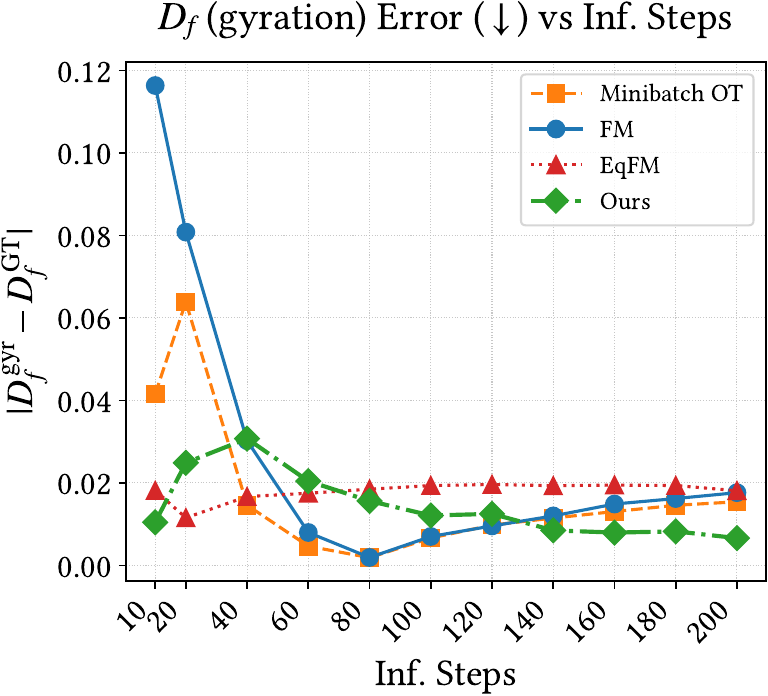}
        \end{minipage}
        \hfill
        \begin{minipage}{0.48\columnwidth}
            \centering
            \includegraphics[width=\linewidth]{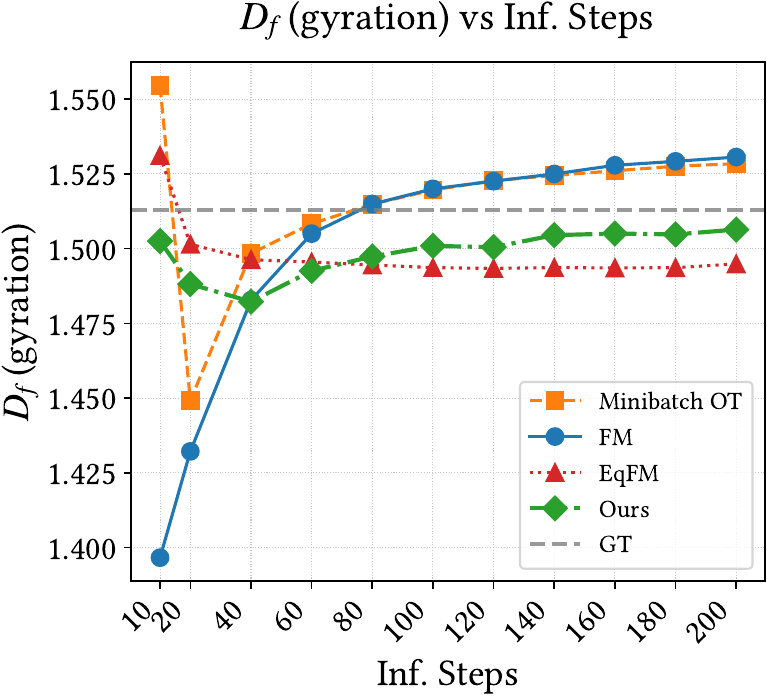}
        \end{minipage}
        \caption{\rev{DLA: absolute error $|D_f^{\text{gen}} - D_f^{\text{GT}}|$ and estimated fractal dimension $D_f$.}}
        \label{fig:dla_metrics}
    \end{subfigure}


    \begin{subfigure}{\columnwidth}
        \centering
        \begin{minipage}{0.48\columnwidth}
            \centering
            \includegraphics[width=\linewidth]{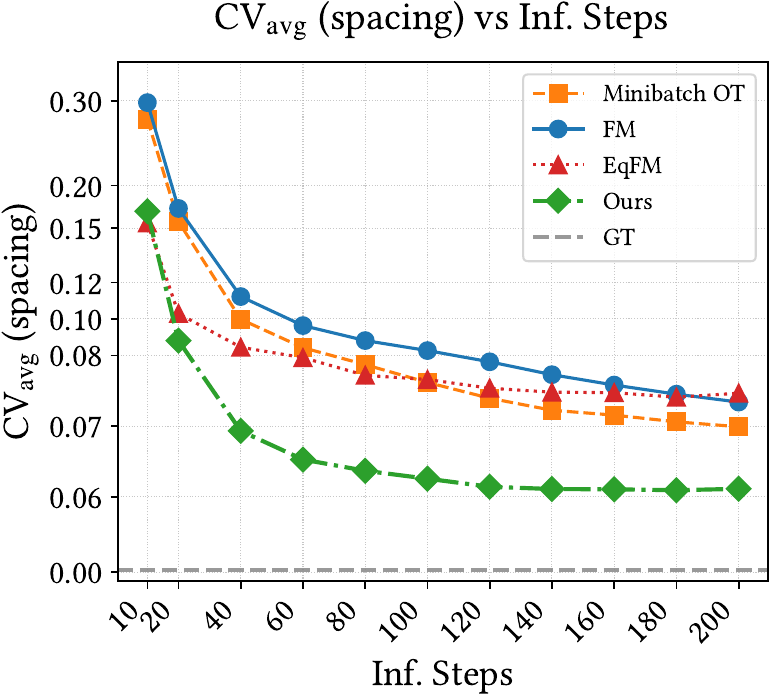}
        \end{minipage}
        \hfill
        \begin{minipage}{0.48\columnwidth}
            \centering
            \includegraphics[width=\linewidth]{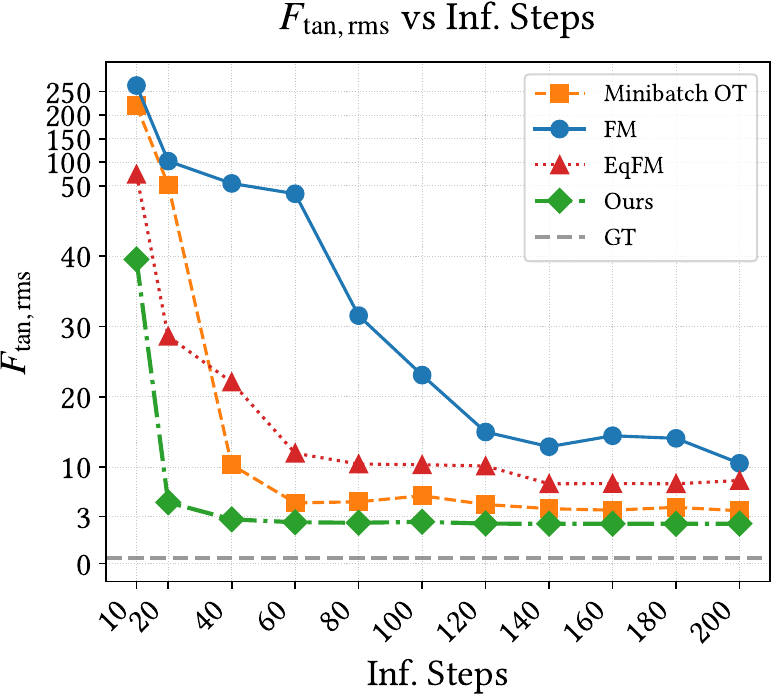}
        \end{minipage}
        \caption{\rev{Multilayer Thomson: CV average (spatial uniformity) and tangential force RMS (equilibrium deviation).}}
        \label{fig:thomson_metrics}
    \end{subfigure}

    \caption{\rev{\textbf{Quantitative metrics vs.\ inference steps for all energy-driven tasks.} Our method (green) achieves low error from early steps and remains stable, while baselines converge slower and plateau at higher error levels.}}
    \label{fig:all_metrics_steps}
\end{figure}

The plots that compare metrics against baselines as a function of inference steps (Figure~\ref{fig:all_metrics_steps}) demonstrate that our method yields straighter, higher-quality flows: it achieves \revvvv{substantially} better metric values with \revvvv{far} fewer steps and typically converges earlier than the baselines.

\subsubsection{Blue-Noise Generation}\label{sec:blue_noise}

Blue-noise distributions are fundamental to rendering and scientific computing, characterized by suppressed low-frequency content and isotropy that maximize sampling efficiency while minimizing aliasing artifacts \cite{ulichney1988dithering,cook1986stochastic}. Yellott \cite{yellott1983spectral} demonstrated that primate photoreceptor arrangements exhibit blue-noise characteristics, suggesting evolutionary optimization for visual sampling. 


\paragraph{Dataset Construction and Evaluation Metrics.} We use the state-of-the-art Gaussian Blue Noise (GBN) \cite{ahmed2022gaussian} to generate the uniform blue-noise dataset and its serial variant \cite{ahmed2024serial} to generate the adaptive blue-noise dataset.
We evaluate generation quality visually using the radial power spectrum and the valence of the Delaunay triangulation, and quantitatively using two metrics: Pearson correlation and relative $L_2$ error against the ground-truth spectral profile.

\paragraph{Uniform Blue Noise Generation Results.}
We generate 400K uniform blue-noise point sets of $N = 1024$ points using a constant density field as input to GBN. We train both the baseline methods and our approach on a subset of 50K point sets for unconditional generation, and additionally train a large variant of our model on the full set of 400K point sets. We compare against Original Flow Matching \cite{lipman2022flow}, Minibatch OT \cite{tong2023improving,pooladian2023multisample}, and Equivariant Flow Matching (EqFM) \cite{klein2023equivariant,song2023equivariant}, all trained with 5M parameters. To demonstrate scalability, we additionally train a 26M-parameter variant of our model on the full dataset.

\begin{table}[!htb]
    \centering
    \small
    \caption{Quantitative comparison on uniform blue-noise generation. Pearson correlation (higher is better) and relative $L_2$ error (lower is better) are computed against the ground-truth radial power spectrum over 1000 generated samples.}
    \label{tab:bn_metrics}
    \begin{tabular}{lcc}
        \toprule
        \textbf{Method} & \textbf{Pearson} $\uparrow$ & \textbf{$L_2$ Error} $\downarrow$ \\
        \midrule
        Original FM (5M) & 0.956 & 0.122 \\
        Minibatch OT (5M) & 0.888 & 0.185 \\
        EqFM (5M) & 0.867 & 0.198 \\
        \midrule
        Ours (5M) & 0.994 & 0.049 \\
        Ours (26M) & \textbf{0.999} & \textbf{0.014} \\
        \bottomrule
    \end{tabular}
\end{table}

Figure~\ref{fig:uniform_bn_comparison} shows qualitative and quantitative comparisons. The first row displays generated point samples; the second row shows the 2D power spectrum averaged over 1000 generated samples; the third row plots the radial power spectrum, computed by azimuthally averaging the 2D power spectrum; and the fourth row visualizes the valence of Delaunay triangulation, where each Voronoi cell is colored by its number of neighbors, more uniform coloring indicates better spatial regularity.
\rev{Among the methods compared, our method produces the closest spectral match to the ground-truth profile. Quantitative results are summarized in Table~\ref{tab:bn_metrics}: our 5M model outperforms the tested baselines, and our 26M model achieves Pearson correlation $0.999$ and $L_2$ error $0.014$.} Figure~\ref{fig:bn_metrics_steps} shows metric evolution across integration steps: \rev{our method reaches high Pearson correlation and low $L_2$ error from early steps, while the baselines require more steps to converge and plateau at higher error levels.} We also conduct an ablation study on canonicalization strategies for this task, detailed in Section~\ref{subsubsec:canonical}.

\begin{figure}
    \centering
    \includegraphics[width=0.5\textwidth]{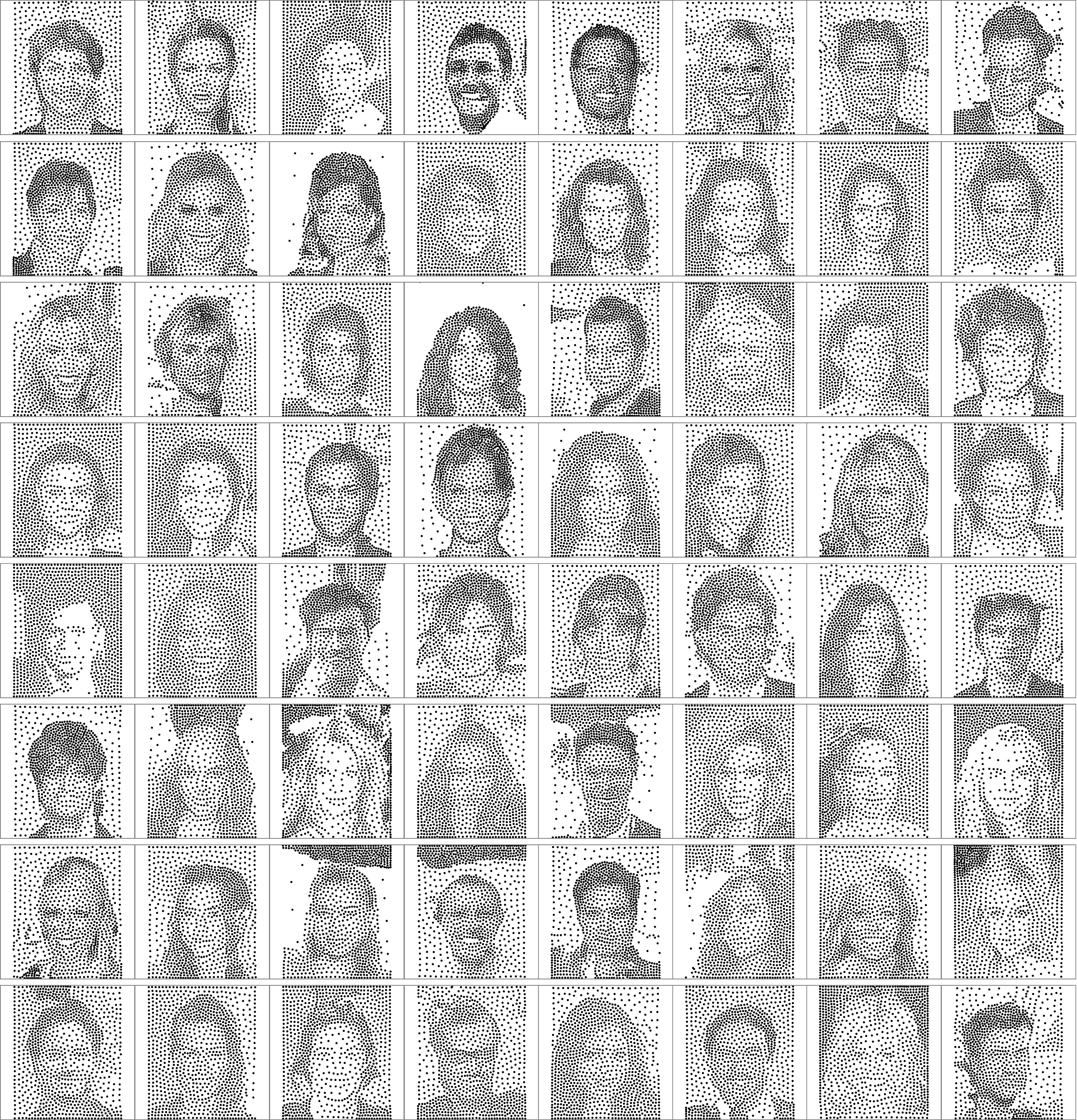}
    \caption{\textbf{Adaptive blue-noise generation on CelebA.} Unconditionally generated face distributions using our 26M model trained on 200K adaptive blue-noise samples. Point density varies with image intensity, revealing facial features---eyes, nose, mouth, and hair contours---while maintaining local blue-noise spectral characteristics throughout.}
    \label{fig:celeba_bn}
\end{figure}
\paragraph{Adaptive Blue Noise Generation Results.}
We extend our approach to adaptive blue-noise sampling, where point density varies spatially according to image intensity. Using the CelebA dataset \cite{liu2015deep}, we generate 200K adaptive blue-noise samples by applying serial GBN to each face image, then train our 26M model for unconditional generation. As shown in Figure~\ref{fig:celeba_bn}, the model successfully learns the joint distribution of facial geometries and the underlying blue-noise characteristics, generating diverse, stylized faces. 

\subsubsection{Minimal Surfaces (Area-Constrained)}

Minimal surfaces are surfaces that locally minimize area under given boundary constraints, arising naturally in soap films, biological membranes, and architectural structures \cite{plateau1873statique,isenberg1992science}. Given a set of anchor points defining the boundary, the minimal surface satisfies the Laplace equation with zero mean curvature. Classical computational methods solve this as a boundary-value problem through iterative optimization \cite{brakke1992surface,pinkall1993computing}. We reformulate this as a conditional generation task: given anchor points, directly generate boundary points that lie on the corresponding minimal surface.

\paragraph{Dataset Construction and Evaluation Metrics.} We sample random anchor configurations on the domain boundary of a $256 \times 256$ grid and compute minimal surface boundaries using an approximate method \cite{israelachvili2011intermolecular} with target area fraction $0.7$. Each sample consists of anchor positions as conditioning input and $256$ boundary points as the target output. We conduct two experiments: (i) fixed 3-anchor configurations, and (ii) variable 3--8 anchor configurations.
We evaluate generation quality using three metrics averaged over 100 samples: \textit{area fraction error} measures deviation from the target enclosed area; \textit{angle smoothness} quantifies boundary curve regularity via angular variation; and \textit{uniformity CV} (coefficient of variation) assesses the evenness of point spacing along the boundary. Lower values indicate better quality for all metrics.

\paragraph{Fixed Anchor Count Results.}
Figure~\ref{fig:minimal_surface} compares 1-step and 10-step generation results for configurations with 3 anchors at random positions. Here, all methods (Original FM, Minibatch OT, EqFM, and ours) use 5M parameters. \rev{Our method produces visually accurate minimal surface boundaries in a single inference step, while the baselines tested here fail to form coherent shapes. With 10 steps, baseline methods still exhibit noticeable artifacts: scattered points, irregular spacing, and boundary distortions.} Quantitative results in Table~\ref{tab:minimal_surface_metrics} confirm this observation. Specifically, with a single inference step, our method achieves area fraction error of $0.004$, angle smoothness of $0.33$, and uniformity CV of $0.34$, while baselines show area errors exceeding $0.69$ (Original FM, Minibatch OT) or poor smoothness and uniformity (EqFM). With 10 inference steps, our method further improves to area error $0.004$, angle smoothness $0.08$, and uniformity CV $0.08$, representing an order-of-magnitude improvement over all baselines. Figure~\ref{fig:minimal_metrics} shows metric evolution across inference steps: our method achieves low error from the first step and remains stable, whereas baselines improve slowly and plateau at substantially higher error levels even with 200 inference steps.

\begin{figure*}[t]
    \centering
    \includegraphics[width=1.0\textwidth]{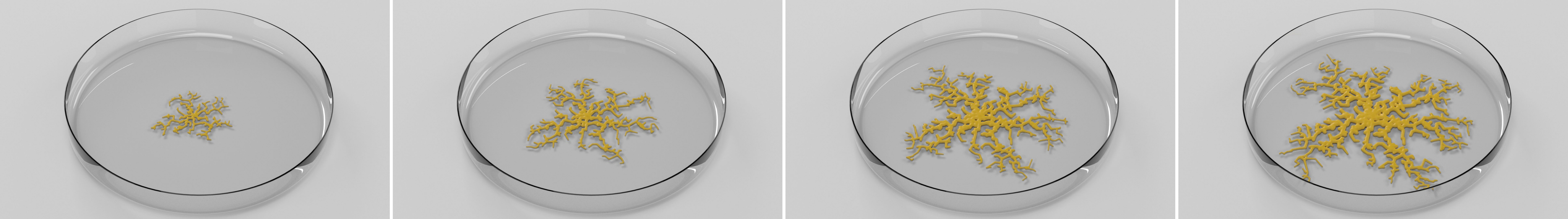}
    \caption{Our generated DLA growth process, rendered as a growing bacterial colony in a Petri dish.}
    \label{fig:dla_process}
\end{figure*}
\begin{table}[t]
    \centering
    \small
    \caption{Quantitative comparison on minimal surface generation (3 anchors, random positions). All metrics are lower-is-better, averaged over 100 samples.}
    \label{tab:minimal_surface_metrics}
    \begin{tabular}{lccc}
        \toprule
        \textbf{Method} & \textbf{Area Err.} $\downarrow$ & \textbf{Angle Smooth.} $\downarrow$ & \textbf{Unif. CV} $\downarrow$ \\
        \midrule
        \multicolumn{4}{l}{\textit{1-step generation}} \\
        Original FM & 0.700 & 1.974 & 1.123 \\
        Minibatch OT & 0.689 & 2.011 & 0.906 \\
        EqFM & 0.040 & 1.444 & 1.304 \\
        Ours & \textbf{0.004} & \textbf{0.330} & \textbf{0.343} \\
        \midrule
        \multicolumn{4}{l}{\textit{10-step generation}} \\
        Original FM & 0.047 & 1.220 & 1.363 \\
        Minibatch OT & 0.049 & 1.185 & 1.317 \\
        EqFM & 0.042 & 0.901 & 1.272 \\
        Ours & \textbf{0.004} & \textbf{0.083} & \textbf{0.078} \\
        \bottomrule
    \end{tabular}
\end{table}

\paragraph{Variable Anchor Count Results.}
We further evaluate our method under varying anchor counts (3--8) at random boundary positions, using a conditional model architecture that generalizes across different configurations (see Section~\ref{subsec:model_arch}).
As shown in Figure~\ref{fig:minimal_surface_multi}, our method produces smooth and accurate minimal-surface boundaries with \revvvv{only} 3 inference steps across all anchor counts and positions.

\subsubsection{Diffusion-Limited Aggregation}
\label{sec:dla_validation}

Diffusion-limited aggregation (DLA) models fractal growth through Brownian-motion particle attachment, producing dendritic structures observed in electrodeposition, mineral formation, and biological branching \cite{witten1981diffusion,meakin1983formation}. A key characteristic of DLA clusters is their fractal dimension $D_f$, which approaches $1.71 \pm 0.01$ in 2D as $N \to \infty$ \cite{witten1981diffusion,meakin1983diffusion}. We formulate DLA generation as an unconditional task where the model learns to produce realistic fractal clusters.
\paragraph{Dataset Construction and Evaluation Metrics.} We run standard DLA simulations with $N = 1024$ particles on a $256 \times 256$ grid using a circular seed, generating 50K samples. For each sample, we record per-particle positions and attachment times as triplets $(x, y, t)$, where $t$ is the time step at which the particle first appears in the cluster. To visualize the DLA growth process, we first generate these $(x, y, t)$ triplets and sort the particles by their time coordinate $t$. In this example, our canonicalization consists of a Hilbert sort applied to the $(x, y, t)$ triplets.
We evaluate using the fractal dimension $D_f$ computed via the gyration method: the radius of gyration scales as $R_g(N) \sim N^{1/D_f}$, and $D_f$ is obtained by fitting $\log R_g$ versus $\log N$. For finite $N = 1024$, the expected fractal dimension is $D_f \approx 1.58$ because of finite-size scaling effects \cite{meakin1983diffusion,tolman1989off}, rather than the asymptotic value $1.71$. Because of a slightly different dataset construction procedure and the finite-sample estimation based on 100 dataset samples, our simulated dataset has a smaller ground-truth fractal dimension of $D_f^{\text{GT}} \approx 1.51$. We report the absolute error $|D_f^{\text{gen}} - D_f^{\text{GT}}|$ averaged over 100 generated samples.

\begin{table}[t]
    \centering
    \small
    \caption{Fractal dimension error on DLA generation. $|D_f^{\text{gen}} - D_f^{\text{GT}}|$ computed via gyration method, averaged over 100 samples (lower is better).}
    \label{tab:dla_metrics}
    \begin{tabular}{lcc}
        \toprule
        \textbf{Method} & \textbf{10-step} $\downarrow$ & \textbf{200-step} $\downarrow$ \\
        \midrule
        Original FM & 0.116 & 0.018 \\
        Minibatch OT & 0.042 & 0.015 \\
        EqFM & 0.018 & 0.018 \\
        Ours & \textbf{0.011} & \textbf{0.007} \\
        \bottomrule
    \end{tabular}
\end{table}

\paragraph{Generation Results.} Table~\ref{tab:dla_metrics} summarizes the results. Our method achieves the lowest fractal dimension error at both 10 and 200 inference steps. Figure~\ref{fig:dla_metrics} shows $D_f$ and its error across varying inference steps: while all methods exhibit some oscillation, ours remains the most stable and converges to the lowest error. Figure~\ref{fig:dla_comparison} provides a qualitative comparison at 10 and 200 inference steps. For a fair visual comparison, we unconditionally generate 400 samples with each trained model and, for a chosen dataset example, retrieve the closest generated sample from each method using the Chamfer Distance (CD). At 10 steps, baseline methods produce scattered, non-fractal structures lacking the characteristic dendritic branching of DLA, while our method \revvvv{already} exhibits realistic fractal morphology comparable to a dataset sample. At 200 steps, all methods improve, but ours maintains the closest resemblance to the dataset sample in terms of branching density and radial structure. \revvvv{This visual comparison aligns with the quantitative metrics: our method achieves lower fractal dimension error at both step counts, indicating better capture of the underlying DLA physics}. Furthermore, Figure~\ref{fig:dla_process} visualizes the temporal evolution of our generated DLA clusters rendered as a growing bacterial colony.

\subsubsection{Multilayer Thomson Problem}

The Thomson problem seeks minimum-energy configurations of $N$ electrons on a sphere under Coulomb repulsion \cite{thomson1904xxiv,smale1998mathematical}. We extend this to a multilayer setting: particles are distributed across concentric spherical shells, interacting via pairwise Coulomb repulsion $E_{\text{coul}} = \sum_{i<j} 1/|\mathbf{x}_i - \mathbf{x}_j|$, while being constrained to their respective shells by a radial potential. This models atomic shell structures and provides a challenging 3D equilibrium problem with both intra-layer and inter-layer interactions.
\begin{figure*}[t]
    \centering
    \includegraphics[width=1.0\textwidth]{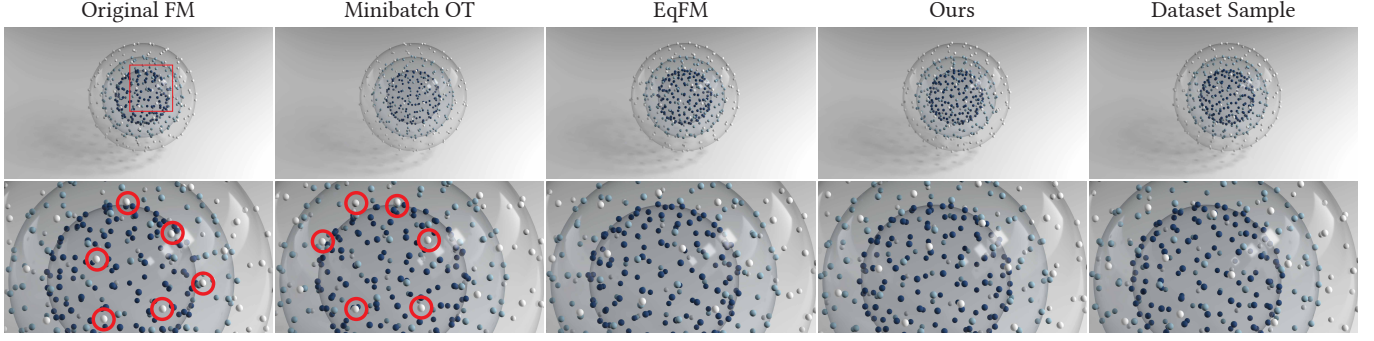}
\caption{\textbf{Multilayer Thomson problem generation.} Comparison of generated three-shell electron configurations (128 particles per shell). The top row shows the full configuration, and the bottom row zooms into the region indicated by the red box. Red circles mark irregular particle-spacing artifacts that remain in Original FM and Minibatch OT, while EqFM and our method produce Poisson-disk-like particle distributions on each shell that closely match the ground-truth equilibrium structure.}
    \label{fig:thomson_comparison}
\end{figure*}

\paragraph{Dataset Construction and Evaluation Metrics.} We simulate 3 concentric shells with 128 particles each (384 total), using gradient-based optimization with Coulomb forces and shell-confining springs until convergence. We generate 20K equilibrium configurations as training data. We evaluate generation quality using two metrics averaged over 100 samples: \textit{CV average} (coefficient of variation of nearest-neighbor distances) measures spatial uniformity on each shell, and \textit{$F_{\tan}$ RMS} (root-mean-square tangential force) quantifies deviation from force equilibrium. At a true minimum, tangential forces vanish.

\begin{table}[t]
    \centering
    \small
    \caption{Quantitative comparison on multilayer Thomson problem (3 shells $\times$ 128 particles). CV average and tangential force RMS, averaged over 100 samples (lower is better).}
    \label{tab:thomson_metrics}
    \begin{tabular}{lcccc}
        \toprule
        \multirow{2}{*}{\textbf{Method}} & \multicolumn{2}{c}{\textbf{20-step}} & \multicolumn{2}{c}{\textbf{200-step}} \\
        \cmidrule(lr){2-3} \cmidrule(lr){4-5}
        & CV $\downarrow$ & $F_{\tan}$ $\downarrow$ & CV $\downarrow$ & $F_{\tan}$ $\downarrow$ \\
        \midrule
        Original FM & 0.173 & 102.4 & 0.073 & 10.55 \\
        Minibatch OT & 0.158 & 52.16 & 0.070 & 3.80 \\
        EqFM & 0.103 & 28.61 & 0.075 & 8.11 \\
        Ours & \textbf{0.088} & \textbf{4.99} & \textbf{0.061} & \textbf{2.54} \\
        \bottomrule
    \end{tabular}
\end{table}

\paragraph{Generation Results.} Table~\ref{tab:thomson_metrics} summarizes the results.
Our method achieves the best performance on both metrics at 20 and 200 inference steps. \rev{At 20 steps, the tangential force RMS is roughly an order of magnitude lower than Original FM ($4.99$ vs.\ $102.4$), suggesting that the generated configurations lie closer to energy minima.} Figure~\ref{fig:thomson_metrics} shows metric evolution across inference steps: our method converges faster and achieves lower error throughout. Figure~\ref{fig:thomson_comparison} provides qualitative comparison, where our generated configurations exhibit uniform particle spacing within each shell and proper inter-shell separation, closely matching ground-truth equilibrium structures.



\subsection{3D Shape Generation}

We evaluate our framework on 3D shape generation tasks, \rev{testing both} our canonicalization strategy for position-only generation and our geometric probability paths for joint position-normal generation. All experiments in this subsection use point clouds with $N = 2048$ points and 26M-parameter models. 



\subsubsection{ShapeNet Point-Cloud Generation}
\label{sec:shapenet_pcd}
Following prior work, we evaluate on three ShapeNet~\cite{chang2015shapenet} categories: airplane, chair, and car. We compare against the same baselines (Original FM, Minibatch OT, EqFM), all trained under identical settings for fair comparison.
Using the evaluation protocol of Yang et al.~\cite{yang2019pointflow}, we report 1-NNA accuracy under both Chamfer Distance (CD) and Earth Mover's Distance (EMD), where values closer to $50\%$ indicate better generation quality.

\paragraph{Position-Only ShapeNet Generation.}
\begin{figure*}[t]
    \centering
    \includegraphics[width=1.0\textwidth]{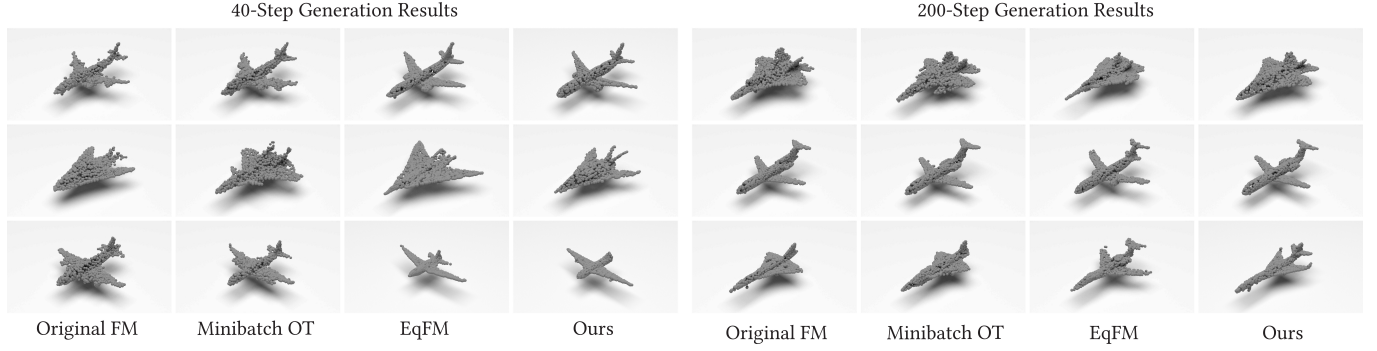}
    \caption{\textbf{ShapeNet airplane generation.} Comparison at 40-step and 200-step inference.}
    \label{fig:shapenet_airplane}
\end{figure*}

\begin{figure*}[t]
    \centering
    \includegraphics[width=1.0\textwidth]{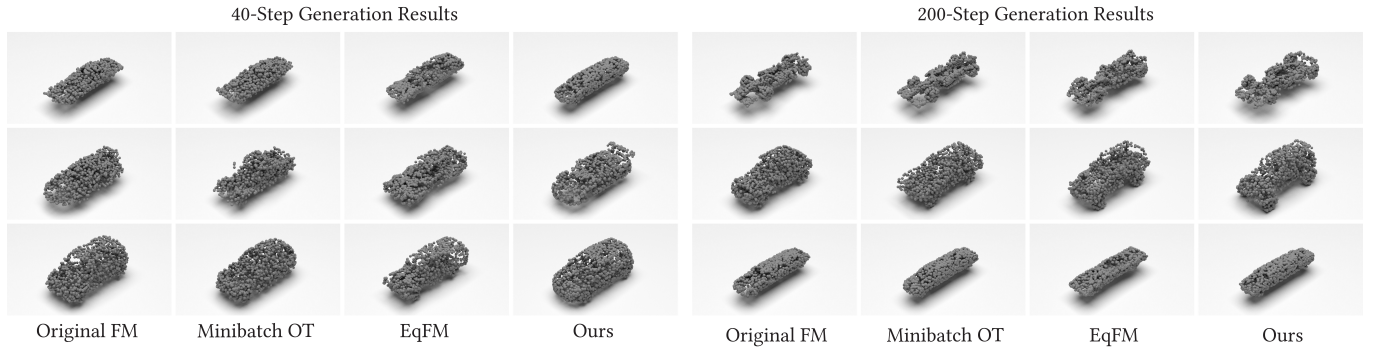}
    \caption{\textbf{ShapeNet car generation.} Comparison at 40-step and 200-step inference.}
    \label{fig:shapenet_car}
\end{figure*}

\begin{figure*}[t]
    \centering
    \includegraphics[width=1.0\textwidth]{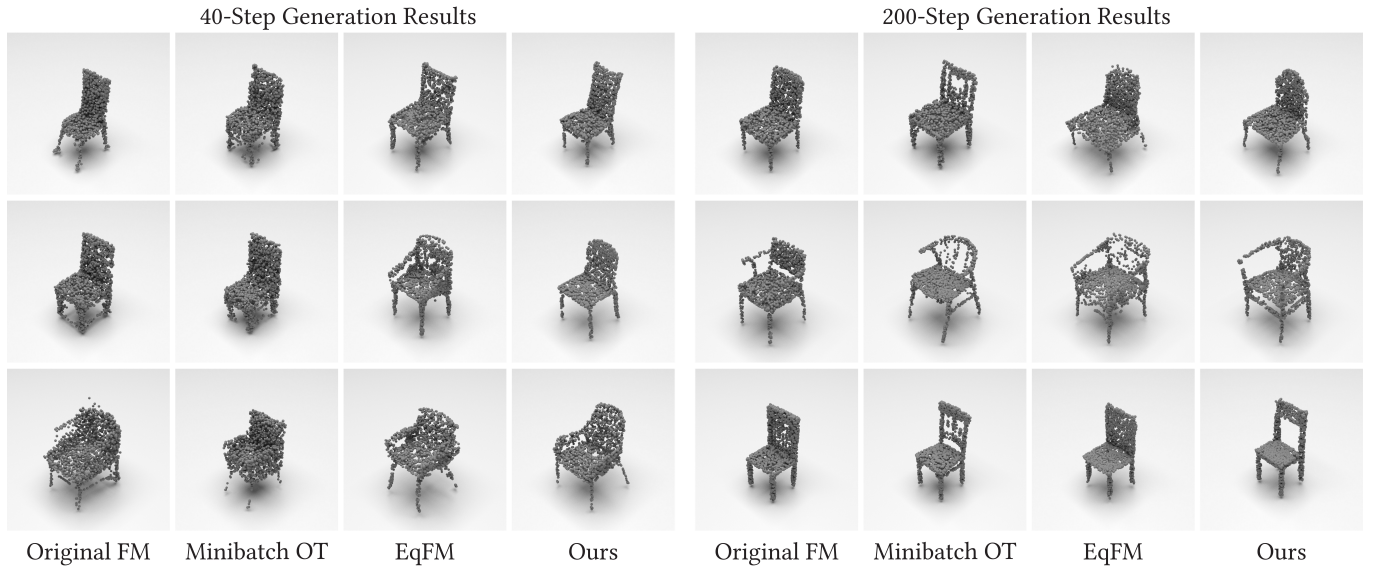}
    \caption{\textbf{ShapeNet chair generation.} Comparison at 40-step and 200-step inference.}
    \label{fig:shapenet_chair}
\end{figure*}
\begin{table*}[t]
    \centering
    \small
    \caption{Quantitative comparison on ShapeNet. 1-NNA accuracy (\%) with Chamfer Distance (CD) and Earth Mover's Distance (EMD); closer to 50\% is better. Two-stage latent methods first train a VAE to compress point clouds into a latent space, then train generative models in that space. $^\dagger$: trained by us for fair comparison; others from original papers.}
    \label{tab:shapenet_comparison}
    \begin{tabular}{c|c|ccc|cc|cc|cc}
        \toprule
        \multirow{2}{*}{\textbf{Model}} & \multirow{2}{*}{\textbf{Method}} & \textbf{\# Params} & \textbf{Infer} & \textbf{Two-stage} & \multicolumn{2}{c|}{\textbf{Airplane}} & \multicolumn{2}{c|}{\textbf{Chair}} & \multicolumn{2}{c}{\textbf{Car}} \\
        & & \textbf{(M)} & \textbf{Steps} & \textbf{Latent} & CD $\downarrow$ & EMD $\downarrow$ & CD $\downarrow$ & EMD $\downarrow$ & CD $\downarrow$ & EMD $\downarrow$ \\
        \midrule
        \multirow{8}{*}{\shortstack{FM}}
        & PVD-DDIM \cite{zhou20213d} & 28 & 100 & \ding{55} & 76.21 & 69.84 & 61.54 &  57.73 & 60.95 &  59.35 \\
        & Original FM$^\dagger$ \cite{lipman2022flow} & 25 & 200 & \ding{55} & 81.98 & 66.29 & 66.77 & 65.03 & 75.57 & 60.94 \\
        & Minibatch OT$^\dagger$ \cite{tong2023improving}
        & 25 & 200 & \ding{55} & 80.12 & 67.90 & 71.68 & 69.49 & 71.31 & 61.22 \\
        & Equivariant FM$^\dagger$ \cite{klein2023equivariant} & 25 & 200 & \ding{55} & 91.85 & 85.93 & 74.62 & 70.32 & 92.90 & 79.69 \\
        & NSOT \cite{hui2025not} & - & 1000 & \ding{55} & \textbf{68.64} & 61.85 & \textbf{55.51} & 57.63 & 59.66 & 53.55 \\
        & \textbf{Ours} & 26 & 200 & \ding{55} & 69.38 & \textbf{58.77} & 61.93 & 58.38 & 60.65 & 55.39 \\
        & \textbf{Ours (1000)} & 26 & 1000 & \ding{55} & 71.35 & 62.96 & 58.23 & \textbf{55.14} & \textbf{59.54} & \textbf{53.26} \\

        \cmidrule{1-11}
        \multirow{9}{*}{\shortstack{Diffusion}}
            & DPM \cite{luo2021diffusion} & 3.9 & 100 & \ding{55} & 76.42 & 86.91 & 60.05 & 74.77 & 68.89 & 79.97 \\
        & PVD \cite{zhou20213d} & 28 & 1000 & \ding{55} & 73.82 & 64.81 & 56.26 & 53.32 & 54.55 & 53.83 \\
        & LION \cite{vahdat2022lion} & 111 & 1000 & \ding{51} & 67.41 & 61.23 & 53.70 & 52.34 & 53.41 & 51.14 \\
        & FrePoLad \cite{zhou2024frepolad} & - & 1000 & \ding{51} & \textbf{65.25} & 62.10 & 52.35 & 53.23 & 51.89 & 50.26 \\
        & NWD \cite{hui2022neural} & 31 & 100 & \ding{55} & \textbf{59.78} & \textbf{53.84} & 56.35 & 57.98 & 61.75 & 58.54 \\
        & 3DShape2VecSet \cite{zhang20233dshape2vecset} & 270 & 18 & \ding{51} & 62.75 & 61.01 & 54.06 & 56.79 & 86.85 & 80.91 \\
        & DiT-3D (S) \cite{mo2023dit} & 33 & 1000 & \ding{55} & - & - & 60.72 &  56.04 & - & -\\
        & DiT-3D (XL) \cite{mo2023dit} & 675 & 1000 & \ding{55} & 62.35 & 58.67 & \textbf{49.11} & \textbf{50.73} & \textbf{48.24} & \textbf{49.35} \\

        \cmidrule{1-11}
        \multirow{6}{*}{Others}
            & l-GAN \cite{achlioptas2018learning} &  1.9 & 1 & \ding{51} & 87.30 & 93.95 & 68.58 & 83.84 & 66.49 & 88.78 \\
        & PointFlow \cite{yang2019pointflow} & 1.6 & var. & \ding{51} & 75.68 & 70.74 & 62.84 & 60.57 & \textbf{58.10} & 56.25 \\
        & DPF-Net \cite{klokov2020discrete} & 3.8 & var. & \ding{51} & \textbf{75.18} & \textbf{65.55} & 62.00 & \textbf{58.53} & 62.35 & \textbf{54.48} \\
        & SoftFlow \cite{kim2020softflow} & - & var. & \ding{51} & 76.05 & 65.80 & 59.21 & 60.05 & 64.77 & 60.09 \\
        & SetVAE \cite{kim2021setvae} & 0.7 & 1 & \ding{51} & 75.31 & 77.65 & \textbf{58.76} & 61.48 & 59.66 & 61.48 \\
        & ShapeGF \cite{cai2020learning} & 5.3 & 100 & \ding{51} & 80.00 & 76.17 & 68.96 & 65.48 & 63.20 & 56.53 \\
        \bottomrule
        
    \end{tabular}
\end{table*}

Table~\ref{tab:shapenet_comparison} compares our method against flow-matching baselines and prior work. Results for Original FM, Minibatch OT, EqFM, and our method are from models we trained; other results are taken from respective papers with the same setting. Our method achieves the best EMD scores among flow-matching approaches across all categories \rev{(EMD is generally considered a more informative metric for global shape distribution quality than CD)}. Notably, we match the performance of NSOT \cite{hui2025not} with approximately $5\times$ fewer inference steps (200 vs.\ 1000), and achieve airplane EMD comparable to DiT-3D (XL) \cite{mo2023dit} using roughly $26\times$ fewer parameters (26M vs.\ 675M) and $5\times$ fewer inference steps. Figures~\ref{fig:shapenet_airplane}, \ref{fig:shapenet_car}, and \ref{fig:shapenet_chair} show qualitative comparisons at 40 and 200 inference steps. Following Hui et al.~\cite{hui2025not}, we unconditionally generate hundreds of shapes with each method and take samples generated by our model as references and, for each such sample, retrieve from every baseline the generated point cloud with the smallest Chamfer Distance (CD) to that reference, yielding shape-wise aligned comparisons. At 40 inference steps, competing baselines \rev{tend to produce less coherent} shapes, whereas our model \revvvv{already} generates diverse, realistic geometries; \revvvv{even} at 200 steps, some of their samples remain \revvvv{noticeably} less detailed and faithful than ours.

\rev{Following \cite{zhang20233dshape2vecset}, we further evaluate generation quality via Rendering-FID and Rendering-KID on the ShapeNet airplane category. Each generated and training   shape is rendered from 10 viewpoints, and both metrics are computed between the generated and training rendering sets using Clean-FID \cite{parmar2022aliased}. As shown in Table~\ref{tab:shapenet_fid_kid}, our method achieves the lowest FID and KID among the flow-matching baselines.}

\begin{table}[t]
    \centering
    \small
    \caption{\rev{Rendering-FID and Rendering-KID ($\times 10^3$) on ShapeNet airplane (lower is better).}}
    \label{tab:shapenet_fid_kid}
    \rev{\begin{tabular}{lcc}
        \toprule
        \textbf{Method} & \textbf{FID} $\downarrow$ & \textbf{KID} ($\times 10^3$) $\downarrow$ \\
        \midrule
        Original FM & 7.659 & 3.582 \\
        EqFM & 11.586 & 6.990 \\
        Ours & \textbf{6.693} & \textbf{2.708} \\
        \bottomrule
    \end{tabular}}
\end{table}

\begin{figure*}[t]
    \centering
    \includegraphics[width=1.0\textwidth]{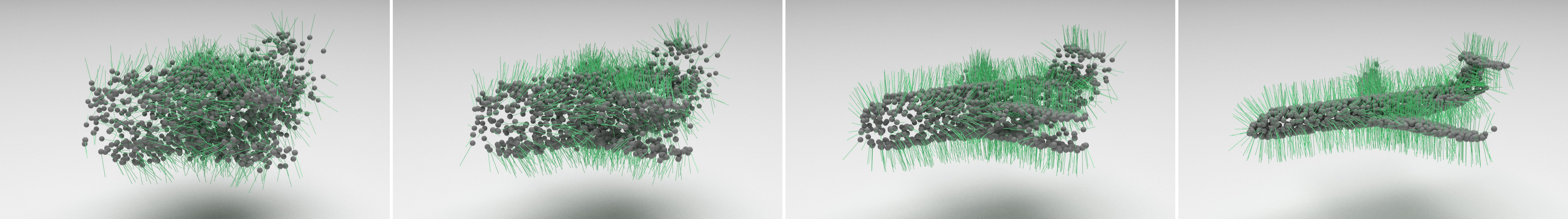}
    \caption{\textbf{3D generation with encoded normals on ShapeNet airplane.} Green line segments show velocity directions during generation, which converge to surface normals at the terminal frame.}
    \label{fig:shapenet_normal_traj1}
\end{figure*}

\begin{figure*}[t]
    \centering
    \includegraphics[width=1.0\textwidth]{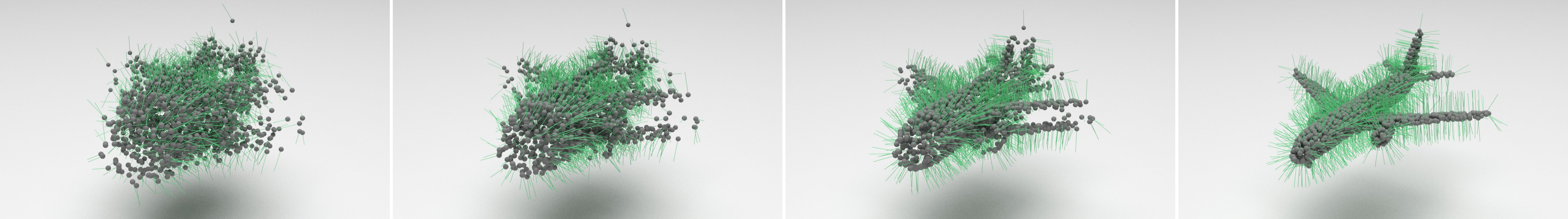}
    \caption{\textbf{3D generation with encoded normals on ShapeNet airplane.} Additional samples demonstrating consistent normal generation across diverse airplane geometries.}
    \label{fig:shapenet_normal_traj2}
\end{figure*}

\begin{figure}[t]
    \centering
    \includegraphics[width=0.5\textwidth]{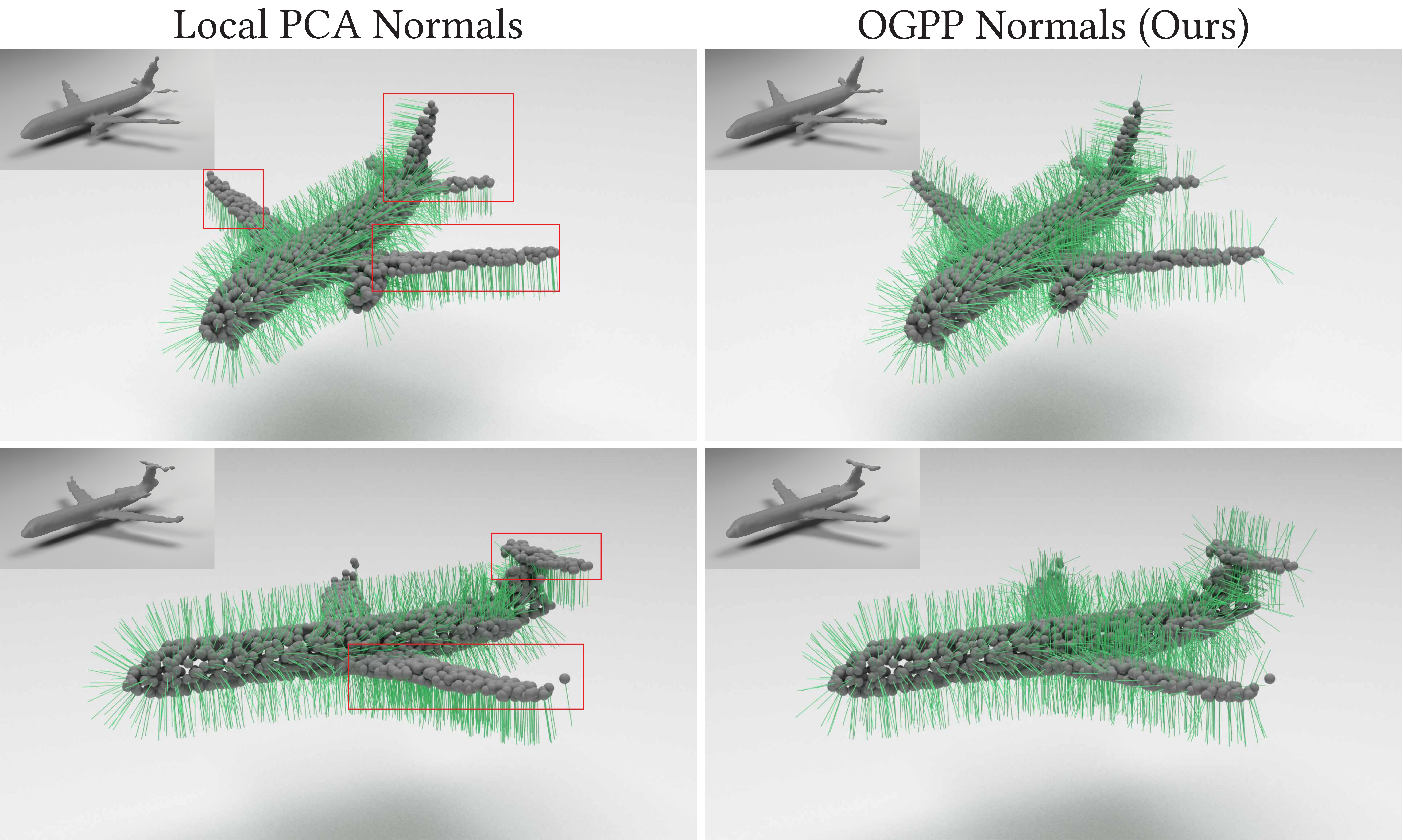}
    \caption{\textbf{Normal comparison on ShapeNet airplane.} Top-left: Poisson reconstruction from our generated normals. Our method produces consistent, accurate normals compared to PCA-estimated normals.}
    \label{fig:shapenet_normal_compare}
\end{figure}

\paragraph{ShapeNet Point-Cloud Generation with Encoded Normals}
\rev{We further evaluate unconditional joint position-normal generation on the ShapeNet airplane category. A key advantage of our geometric probability paths is that they produce \emph{consistently oriented} surface normals as a zero-cost byproduct of the flow, without requiring any additional network output or post-processing.} For ShapeNet dataset, we first apply marching cubes to the voxelized shapes to obtain watertight meshes, so that we have consistently oriented ground-truth surface normals. Figures~\ref{fig:shapenet_normal_traj1} and \ref{fig:shapenet_normal_traj2} visualize two unconditional generation processes using $200$ inference steps with green line segments indicating velocity directions, which converge to surface normals at the terminal time. \revvvv{The consistent normal alignment across diverse airplane geometries demonstrates the robustness of our approach.} Figure~\ref{fig:shapenet_normal_compare} compares our generated normals against PCA-estimated normals on the same point cloud. \rev{While PCA can recover approximate normal directions, it cannot determine consistent orientations, leading to failures at thin structures like wings and tail fins.} The top-left part additionally shows Screened Poisson reconstruction \cite{kazhdan2013screened,kazhdan2006poisson} comparison, where PCA-based reconstruction exhibits artifacts due to these inconsistent normal orientations. \rev{We further quantify normal accuracy via unoriented angular deviation in Section~\ref{ablation:arc}, confirming that our method also achieves lower angular error than PCA estimation.} Together, these results demonstrate that our geometric probability paths produce accurate, consistently oriented normals, enabling high-quality surface reconstruction.

\subsubsection{Single-Shape Encoding}

We evaluate on the single-shape encoding task proposed in Geometry Distributions \cite{zhang2025geometry}, where a generative model encodes a single geometry. Following this setup, we train per-shape models with our geometric probability paths on complex meshes from Thingi10k \cite{zhou2016thingi10k}. At  inference time, we generate 256 batches of $N = 2048$ points, yielding about 500K points with normals using $300$ inference steps, which we feed into Screened Poisson reconstruction \cite{kazhdan2013screened,kazhdan2006poisson}.





\begin{figure}[t]
    \centering
    \includegraphics[width=0.5\textwidth]{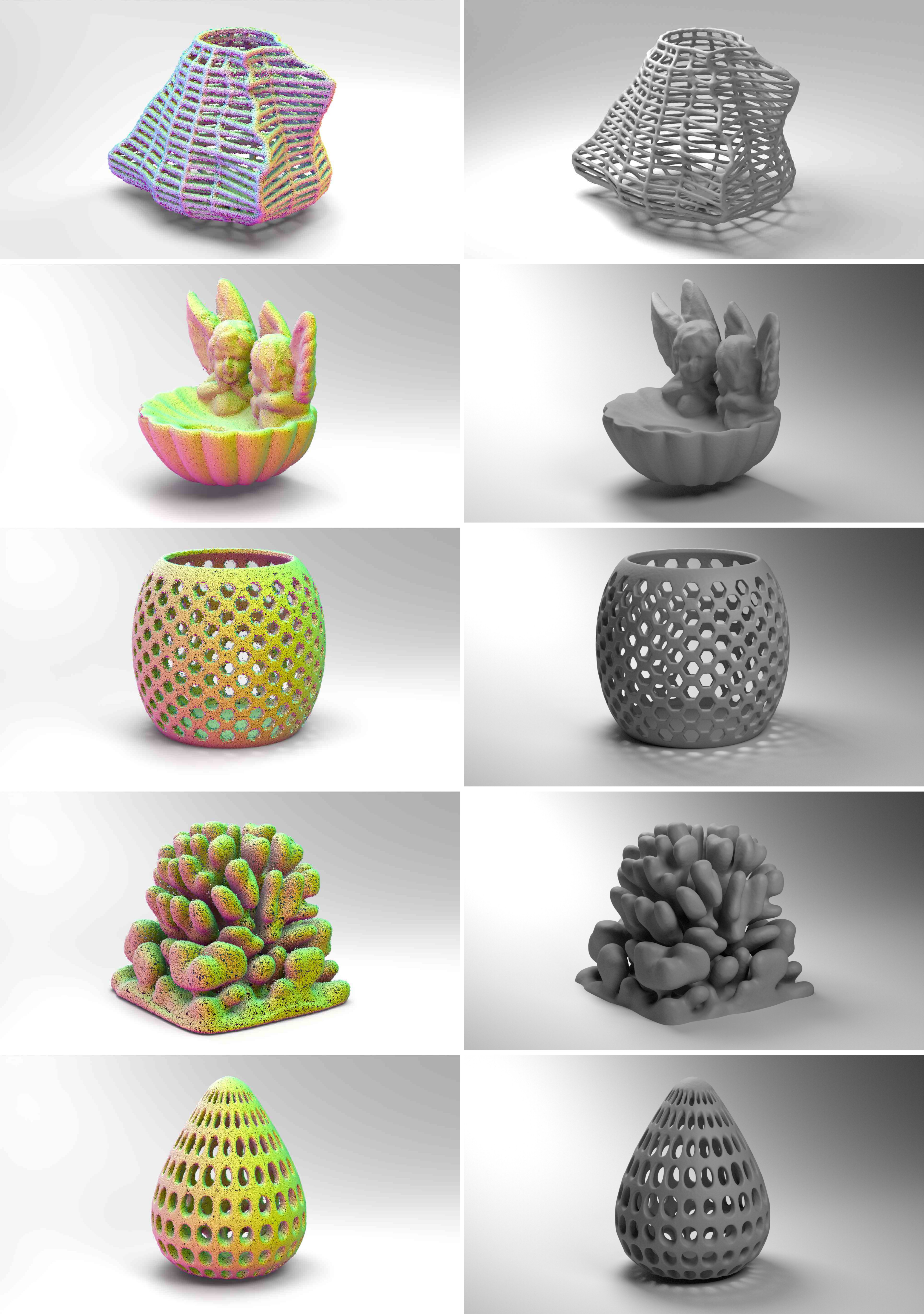}
    \caption{\rev{\textbf{Single shape encoding.} Left: point clouds colored by predicted normals. Right: reconstructed mesh.}}
    \label{fig:single_shape_encoding}
\end{figure}

\begin{figure*}[t]
    \centering
    \includegraphics[width=1.0\textwidth]{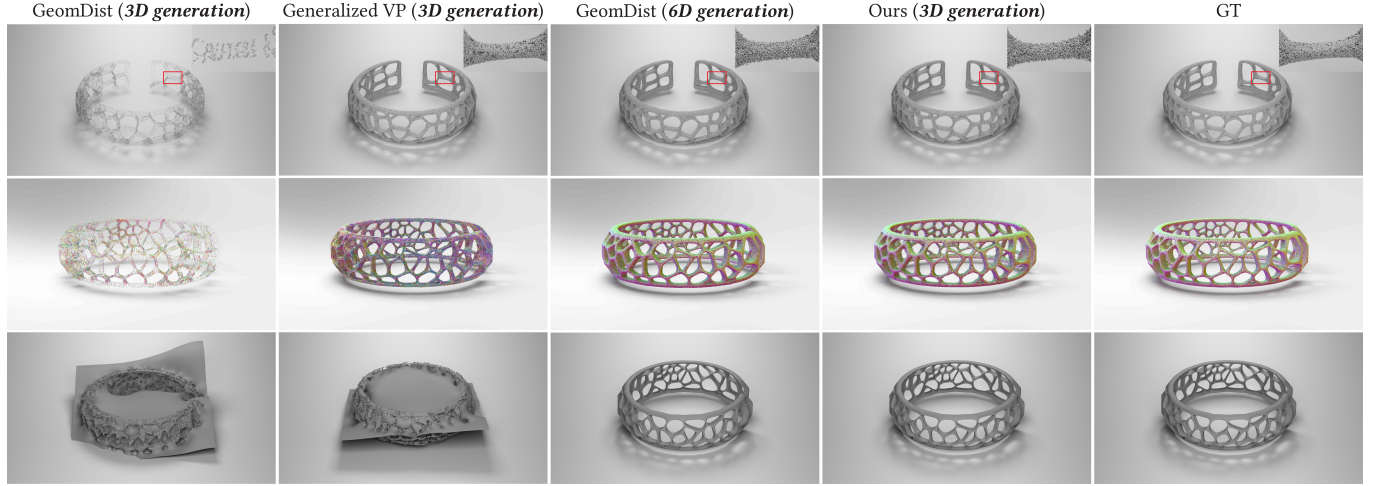}
    \caption{\textbf{Single-shape encoding comparison on Coral Cuff.}
    Row 1: generated point clouds with zoomed-in details.
    Row 2: generated point clouds colored by normal direction.
    Row 3: meshes reconstructed via Screened Poisson.
    Geometry Distributions (3D) produces sparse, clustered points, while Generalized VP yields noisy normals on thin structures.
    Geometry Distributions (6D) further requires 6D outputs and higher computation.
    Our 3D geometric probability paths achieve quality comparable to 6D methods while maintaining the efficiency of a purely 3D generation process.}
    \label{fig:coral_comparison}
\end{figure*}

We compare against three baselines: (1) Geometry Distributions (3D) \cite{zhang2025geometry}, which generates positions only and estimates normals via PCA;
(2) Generalized Variance Preserving (gVP) Path (3D)~\cite{chang20243d,albergo2022building,ma2024sit}, which, similar to our approach, interprets terminal velocities as normals but strongly relying on the assumption that the learned density collapses to a near-delta distribution around the surface; and 
(3) Geometry Distributions (6D), which explicitly generates 6D position-normal vectors. Figure~\ref{fig:coral_comparison} shows comparison on a challenging coral cuff mesh with thin structures. Geometry Distributions (3D) produces sparse, clustered point distributions, resulting in poor mesh quality with PCA-estimated normals. Generalized VP recovers normals from terminal velocities, but the predicted normals are noisy and often misaligned, so the reconstructed mesh exhibits pronounced artifacts. Geometry Distributions (6D) achieves results comparable to ours but requires generating 6D outputs. \revvvv{Our method, using 3D geometric probability paths, produces accurate normals as a byproduct of the path geometry, enabling high-fidelity reconstruction of complex thin structures without increasing output dimensionality.}

\rev{Figure~\ref{fig:single_shape_encoding}} shows additional results on diverse Thingi10k meshes, including thin structures (single tear), solid objects (dendrite, angel), and shapes with complex topology (alien egg, honeycomb jar). Left columns show generated point clouds colored by predicted normals, while right columns show meshes reconstructed with Screened Poisson. Our method consistently recovers accurate normals and supports high-fidelity reconstructions across this range of geometric complexity.


\begin{figure*}[t]
    \centering
    \includegraphics[width=1.0\textwidth]{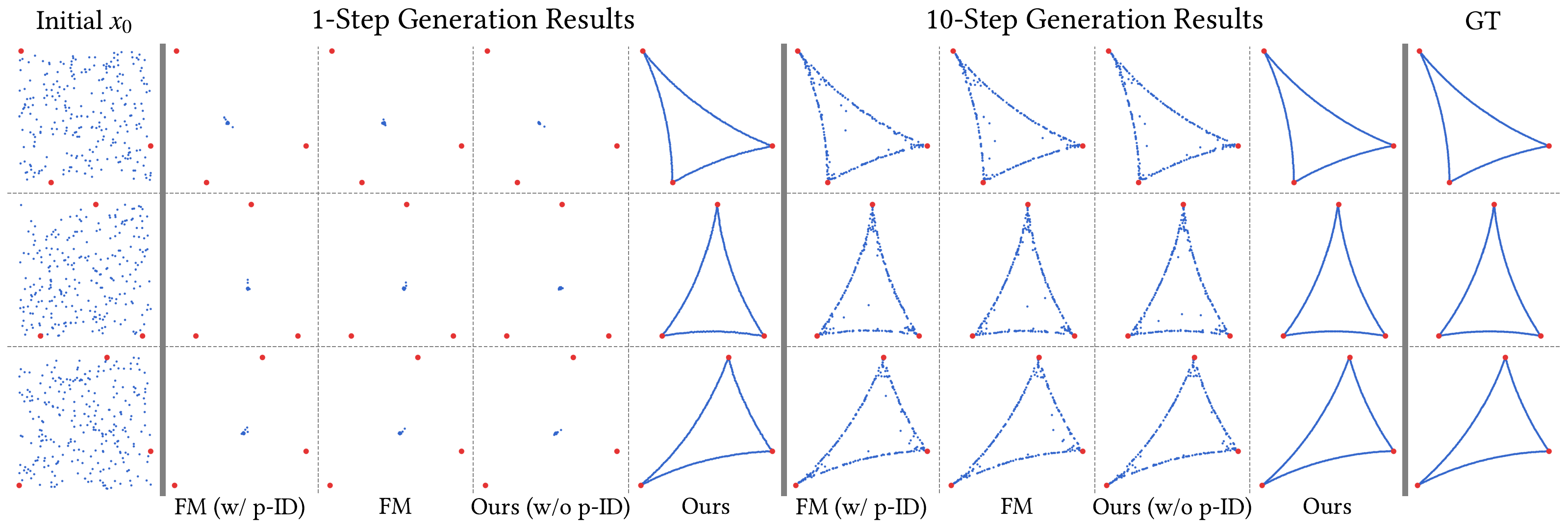}
    \caption{\textbf{Minimal surface (area-constrained) generation ablation study (3 anchors).} Comparison of 1-step and 10-step generations; red dots indicate anchor points (conditioning locations), and ground truth (GT) is shown on the right.
Without per-particle index (identity) embeddings, our method has only similar expressive power to vanilla Flow Matching (Eulerian view), while equipping vanilla Flow Matching with particle identities alone still fails to produce high-quality minimal surfaces.}
    \label{fig:minimal_surface_ablation}
\end{figure*}

\subsection{Ablation Studies}
\label{sec:ablation}
We conduct ablation studies to analyze three key design choices in our framework: (1) orbit-space canonicalization strategy and initial noise distribution, (2) particle index  embeddings, and (3) geometric probability paths for normal generation.

\subsubsection{Canonicalization Strategy and Initial Noise Distributions}
\label{subsubsec:canonical}
\begin{figure*}[t]
    \centering
    \includegraphics[width=1.0\textwidth]{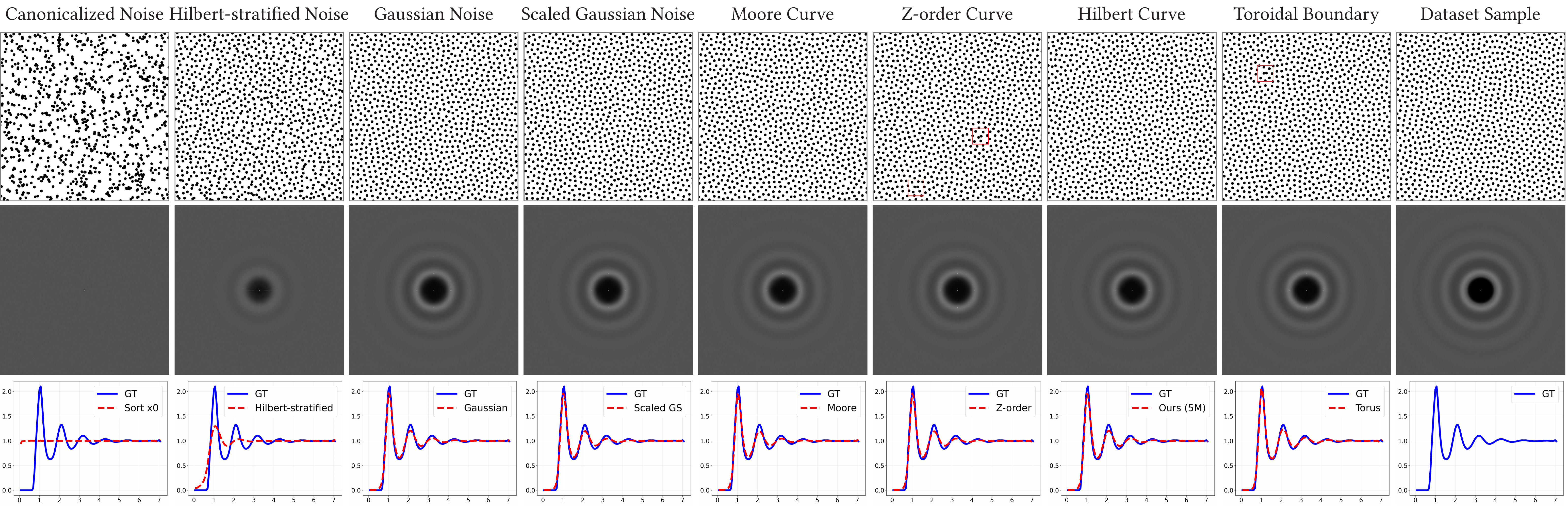}
    \caption{\textbf{Ablation study on canonicalization strategies and initial noise distributions.} Row 1: generated point sets. Row 2: averaged 2D power spectrum. Row 3: radial power spectrum compared to ground truth. Red boxes highlight point pairs that are too close to each other.}
    \label{fig:ablation_bn}
\end{figure*}

We evaluate different canonicalization strategies and initial noise distributions on the uniform blue-noise task (Section~\ref{sec:blue_noise}) using the same 5M model and 50K training samples. Table~\ref{tab:ablation_sort} and Figure~\ref{fig:ablation_bn} summarize the results.

\begin{table}[t]
    \centering
    \small
    \caption{Ablation study on canonicalization strategies and initial noise distributions for uniform blue-noise generation. Pearson correlation (higher is better) and $L_2$ error (lower is better) against ground-truth radial power spectrum.}
    \label{tab:ablation_sort}
    \begin{tabular}{lcc}
        \toprule
        \textbf{Method} & \textbf{Pearson} $\uparrow$ & \textbf{$L_2$ Error} $\downarrow$ \\
        \midrule
        Canonicalized Noise & 0.210 & 0.383 \\
        Hilbert-stratified Noise & 0.896 & 0.184 \\
        \midrule
        Gaussian Noise & 0.994 & 0.048 \\
        Scaled Gaussian Noise & 0.994 & 0.049 \\
        Moore Curve & 0.993 & 0.053 \\
        Z-order Curve & 0.993 & 0.050 \\
        Hilbert Curve (Ours) & 0.994 & 0.049 \\
        Toroidal Boundary & \textbf{0.994} & \textbf{0.046} \\
        \bottomrule
    \end{tabular}
\end{table}

Several observations emerge from Table~\ref{tab:ablation_sort} and Figure~\ref{fig:ablation_bn}. 
First, canonicalizing only the noise endpoint $X_0$ without sorting $X_1$ fails to capture blue-noise structure (Pearson 0.21), confirming our theoretical analysis that canonicalization on the $X_1$ side is essential. 
Second, Hilbert-stratified noise (second column) implements a two-sided canonicalization strategy: we sample $X_0$ by placing one particle at each grid-cell center, adding small jitter, and then sorting these points by their Hilbert indices, while $X_1$ is Hilbert-sorted in the usual way. This construction slightly improves over canonicalized noise alone (first column) but still performs markedly worse than the other configurations.
Third, using Gaussian and scaled Gaussian noise (third and fourth columns) yields nearly identical power-spectrum metrics, indicating that modest changes in the noise variance have little effect on performance in this setting. 
Fourth, among space-filling curve orderings (Moore, Z-order, Hilbert; fifth--seventh columns), all achieve comparable performance, but we occasionally observe that samples generated with Z-order curves contain points that are too close to each other. We therefore adopt Hilbert ordering as our default due to its stronger locality-preserving properties. 
Fifth, employing a toroidal probability path (last column), which effectively imposes periodic boundary conditions so that trajectories can wrap across the domain boundary, leads to a slight additional improvement. However, this gain is marginal, we also occasionally observe point pairs that are too close, and such a path may not be equally beneficial for competing methods, so we retain the simpler linear path when comparing across baselines in Section~\ref{sec:blue_noise}.

Overall, we adopt Hilbert curve sorting with uniform noise as our standard configuration, as it offers a simple, well-standardized choice with consistently strong performance.

\subsubsection{Particle Index Embedding}\label{ablation:index}
As shown in Figure~\ref{fig:minimal_surface_ablation}, we conduct an ablation on area-constrained minimal surface generation with three anchors to disentangle the roles of orbit-space path design and per-particle identity embeddings (Figure~\ref{fig:minimal_surface_ablation}). We compare four variants: vanilla flow matching in the Eulerian view, with and without index (identity) embeddings, and our OGPP framework, also with and without identity embeddings. Equipping vanilla flow matching with particle identities alone fails to recover high-quality minimal surfaces, even with more inference steps. Without per-particle identities, our method produces shapes that are only on par with vanilla flow matching. Only the full OGPP model, which combines orbit-space canonicalization with Lagrangian identity-conditioned trajectories, yields smooth, well-formed minimal surfaces in as few as one ODE step, matching the intuition from our introduction that both components are necessary to untangle mixed particle roles and straighten the learned flows.

\subsubsection{Geometric Probability Path Design}

We ablate geometric probability path configurations on the ShapeNet airplane category with joint position-normal generation. Our geometric probability paths involve several design choices: (1) \textit{canonicalization dimension}, i.e., whether to sort particles using Hilbert curves in 3D position space or 6D position-normal space (joint canonicalization); (2) \textit{Hermite degree}, where quadratic interpolation encodes only the terminal tangent (normal), while cubic interpolation additionally specifies the initial tangent $\bm{n}_0$; (3) \textit{initial tangent $\bm{n}_0$}, which for cubic paths can be set to zero or aligned with the displacement direction $(\bm{x}_1 - \bm{x}_0)/\|\bm{x}_1 - \bm{x}_0\|$; and (4) \textit{noise shape}, i.e., the distribution of initial points $\bm{x}_0$, which can be box (uniform in $[-1,1]^3$), sphere (uniform on the unit sphere), or shell (uniform in a spherical shell). Table~\ref{tab:ablation_study_geometric_path} summarizes the results, evaluated by average cosine similarity between generated normals and dataset normals, its standard deviation, and joint position-normal 1-NNA accuracy.

\begin{table}[t]
    \centering
    \small
    \caption{\textbf{Ablation study on geometric probability path design for position-normal generation}. Avg.\ Cos.\ Sim.: average cosine similarity between generated and dataset normals (higher is better). Std.\ Cos.\ Sim.: standard deviation (lower is better). 1-NNA Acc.: joint position-normal 1-NNA accuracy (closer to 50\% is better).}
    \label{tab:ablation_study_geometric_path}
    \resizebox{\columnwidth}{!}{%
    \begin{tabular}{cccccc}
        \toprule
        \textbf{Canon.} & \textbf{Hermite} & $\bm{n}_0$ & \textbf{Noise} & \textbf{Avg. Cos.} & \textbf{1-NNA} \\
        & \textbf{Degree} & & \textbf{Shape} & \textbf{Sim.} $\uparrow$ & \textbf{Acc.} $\downarrow$ \\
        \midrule
         Hilbert 6D & Cubic & $\frac{\bm{x}_1 - \bm{x}_0}{\|\bm{x}_1 - \bm{x}_0\|}$ & Box & 0.89 & 0.82 \\
         Hilbert 6D & Cubic & $\bm{0}$ & Box & 0.90 & 0.74 \\
         Hilbert 3D & Quadratic & N/A & Box & 0.91 & 0.78 \\
         Hilbert 6D & Quadratic & N/A & Sphere & 0.83 & 0.99 \\
         Hilbert 6D & Quadratic & N/A & Shell & 0.91 & 0.65 \\
         Hilbert 6D & Quadratic & N/A & Box & \textbf{0.92} & \textbf{0.61} \\
        \bottomrule
    \end{tabular}%
    }
\end{table}

Quadratic interpolation consistently outperforms cubic (rows 1--2 vs.\ 3--6), suggesting that encoding only the terminal tangent is sufficient and that additionally constraining the initial tangent over-specifies the path. Joint canonicalization, i.e., sorting in 6D position-normal space, improves over 3D sorting (row 3 vs.\ 6), which is consistent with our earlier analysis in Section~\ref{subsec:joint_canon}. Noise shape has a substantial impact: sphere noise yields poor results (0.83 cosine similarity, 0.99 1-NNA), likely because particles start on a lower-dimensional manifold, whereas box noise provides full-dimensional support and achieves the best performance. For cubic paths, setting $\bm{n}_0 = \bm{0}$ outperforms aligning it with the displacement direction, indicating that simpler initial conditions aid optimization. 

Based on these results, we adopt joint canonicalization with 6D Hilbert-curve sorting, quadratic Hermite probability paths, and box noise as our default configuration.

\subsubsection{Arc-length Terminal Velocity}\label{ablation:arc}

We ablate the effect of terminal velocity magnitude on surface generation quality. As described in Section~\ref{subsec:atv}, for normal encoding, only the direction of the terminal velocity $\bm{v}_1$ is constrained, so its magnitude is a free parameter. Normalized Terminal Velocity (NTV) sets $\|\bm{v}_1\| = 1$ for all particles, while our Arc-length Terminal Velocity (ATV) scales $\|\bm{v}_1\|$ based on chord length and normal alignment (Eq.~\ref{eq:atv}) to achieve more uniform speed profiles along trajectories.

Figure~\ref{fig:surface_bunny_comparison} compares NTV and ATV on the Voronoi bunny mesh. ATV reconstructs more accurate geometry, particularly visible in the zoomed-in regions (red boxes): small Voronoi cells and thin hole boundaries are clearly preserved with ATV, while NTV fails to capture these fine details. \rev{We additionally evaluate normal accuracy via the unoriented angular deviation between predicted and GT normals at the closest projected surface points on $100$K generated points. The median unoriented angular error is $7.6^{\circ}$ (ATV) vs.\ $12.4^{\circ}$ (PCA-ATV), and $10.9^{\circ}$ (NTV) vs.\ $17.3^{\circ}$ (PCA-NTV). The PCA baselines differ because the two methods generate different point distributions, changing the local neighborhoods used for covariance estimation.}

\rev{\subsubsection{Comparison with Direct 6D Generation}\label{ablation:6d}

We also compare our path-based normal encoding against directly generating 6D position-normal pairs via canonicalized flow matching (Canon.\ FM 6D). As shown in Figure~\ref{fig:surface_bunny_comparison}, Canon.\ FM (6D) achieves comparable reconstruction quality to ATV, confirming that our geometric probability path encodes normals as effectively as explicit 6D generation. The advantage of our approach is representational economy: the flow transports only 3D positions, while normals are recovered from the terminal velocity at no extra cost.}

\begin{figure}[t]
    \centering
    \includegraphics[width=\columnwidth]{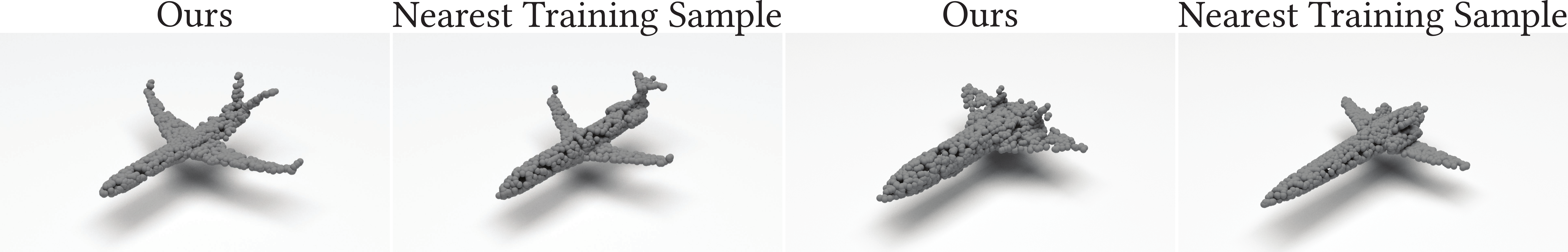}
    \caption{\rev{Generated airplanes (left of each pair) and their nearest training samples under Chamfer Distance (right of each pair). The generated shapes are visually distinct from their closest training neighbors, indicating that the model produces novel geometry rather than memorizing training data.}}
    \label{fig:nearest_neighbor}
\end{figure}

\rev{\subsubsection{Generalization: Nearest-Neighbor Analysis}\label{ablation:generalization}

To assess overfitting risks, we retrieve the nearest training neighbor under Chamfer Distance for each generated airplane. As shown in Figure~\ref{fig:nearest_neighbor}, generated samples differ visibly from their closest matches, suggesting novel geometry synthesis rather than memorizing.
}

\subsubsection{Permutation Equivariance and Inference Efficiency}

Table~\ref{tab:inference_time} compares inference performance between our Plain Transformer backbone and PVCNN. Our plain transformer architecture achieves significantly higher throughput, \revvvv{particularly} benefiting from the efficiency of attention-based computation on modern GPUs.

\begin{table}[t]
    \centering
    \caption{Inference time benchmark comparing Plain Transformer and PVCNN. Measured on a single NVIDIA H100 SXM GPU with BF16, batch size 256, averaged over 100 runs.}
    \label{tab:inference_time}
    \resizebox{\columnwidth}{!}{%
    \begin{tabular}{llccccccc}
        \toprule
        \textbf{Model} & \textbf{Size} & \textbf{$N$} & \textbf{Dim} & \textbf{Params} & \textbf{Particle ID} & \textbf{Perm.-eq.} & \textbf{ms/samp.} & \textbf{samp./s} \\
        \midrule
        \multirow{8}{*}{\shortstack[l]{Plain\\Trans.}}
        & default & 1024 & 2 & 5M  
        & \checkmark & \ding{55} & 0.121 & 8299 \\
        & default & 2048 & 3 & 5M  
        & \checkmark & \ding{55} & 0.278 & 3595 \\
        & large   & 1024 & 2 & 26M 
        & \checkmark & \ding{55} & 0.324 & 3089 \\
        & large   & 2048 & 3 & 26M 
        & \checkmark & \ding{55} & 0.755 & 1324 \\
        \cmidrule{2-9}
        & default & 1024 & 2 & 5M  
        & \ding{55} & \checkmark & 0.120 & 8306 \\
        & default & 2048 & 3 & 5M  
        & \ding{55} & \checkmark & 0.278 & 3597 \\
        & large   & 1024 & 2 & 25M 
        & \ding{55} & \checkmark & 0.323 & 3092 \\
        & large   & 2048 & 3 & 25M 
        & \ding{55} & \checkmark & 0.754 & 1326 \\
        \midrule
        \multirow{2}{*}{PVCNN}
        & --      & 1024 & 3 & 28M 
        & \ding{55} & \checkmark & 16.952 & 59 \\
        & --      & 2048 & 3 & 28M 
        & \ding{55} & \checkmark & 17.008 & 59 \\
        \bottomrule
    \end{tabular}%
    }
\end{table}

\subsection{Mid-time analysis for Lipschitz ratio and directional cancellation}
\label{subsec:mid_time_analysis}

\begin{figure*}[t]
    \centering
    \begin{minipage}{0.245\textwidth}
        \centering
        \includegraphics[width=\linewidth]{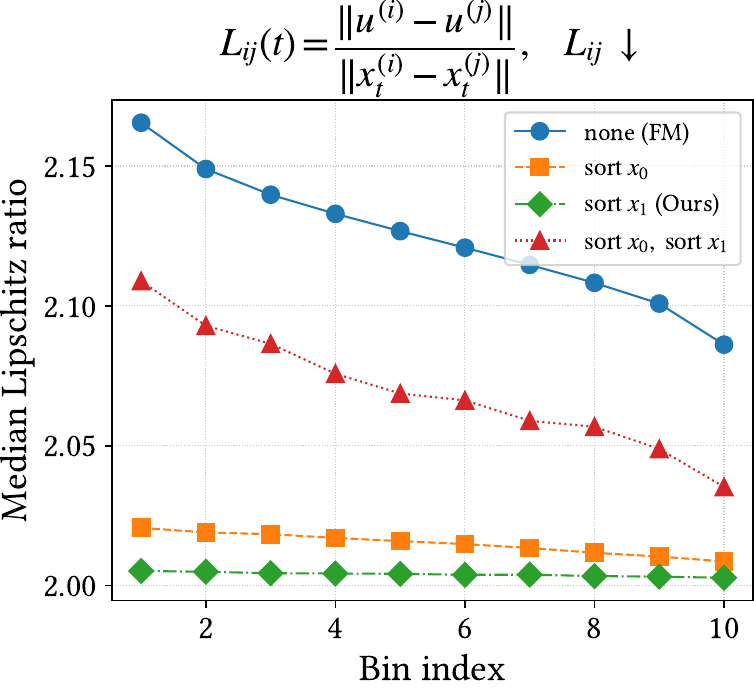}
    \end{minipage}
    \begin{minipage}{0.245\textwidth}
        \centering
        \includegraphics[width=\linewidth]{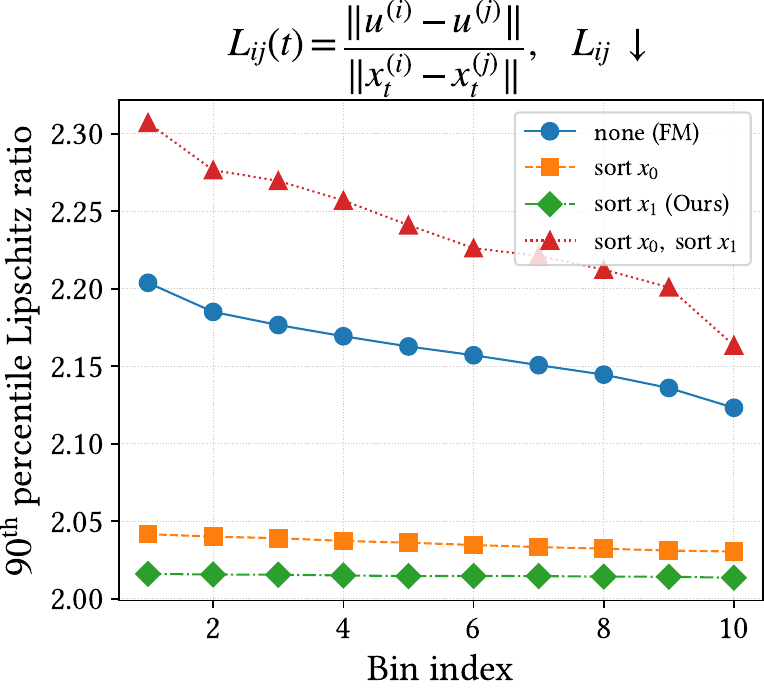}
    \end{minipage}
    \begin{minipage}{0.245\textwidth}
        \centering
        \includegraphics[width=\linewidth]{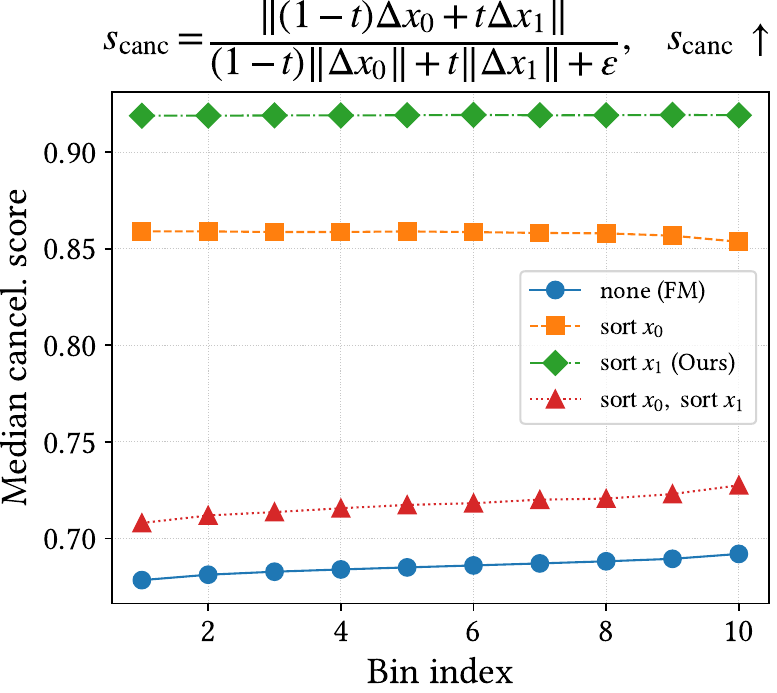}
    \end{minipage}
    \begin{minipage}{0.245\textwidth}
        \centering
        \includegraphics[width=\linewidth]{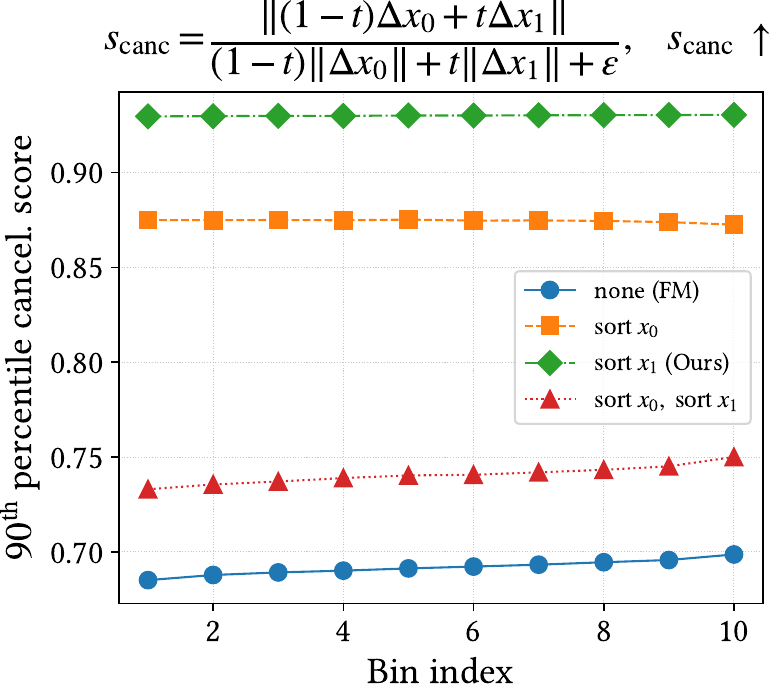}
    \end{minipage}
\caption{\textbf{Mid-time analysis of Lipschitz ratios and directional cancellation.}
From left to right: median and 90th percentile Lipschitz ratio, median and 90th percentile cancellation score at $t = 1/2$ over $k$-NN edges (bins ordered by distance).
Lower $L_{ij}$ and higher $s_{\mathrm{canc}}$ are better.
Canonicalizing $X_1$ only (``sort $x_1$'', Ours) yields the lowest Lipschitz ratios and highest cancellation scores, matching our directional-cancellation analysis. See more experimental details in Section~\ref{subsec:mid_time_analysis}.}
    \label{fig:midtime-analysis}
\end{figure*}

Here we empirically evaluate the Lipschitz ratio and directional cancellation introduced in Section~\ref{subsec:lipschitz_view} in a realistic training setting on our uniform blue-noise dataset.

\paragraph{Experimental setup.}
We sample $N = 500{,}000$ pairs $(\bm{x}_t^{(i)}, \bm{u}_t^{(i)})$ at the midpoint $t = 0.5$ from our uniform blue noise dataset (See Section~\ref{sec:blue_noise}), randomly select $A = 4{,}000$ anchors, and build a $k$-NN graph with $K = 32$ neighbors per anchor.
We partition anchor--neighbor pairs into $B = 10$ equal-frequency distance bins, where bin 1 contains the closest pairs and bin 10 the most distant.
Within each bin, we report summary statistics (median and 90th percentile); 
Details of the quantile-bin construction are given in \rev{the supplement}. 

At the midpoint \(t = \tfrac{1}{2}\), the squared Lipschitz ratio is:

\begin{equation}
\label{eq:L-mid}
L_{ij}\!\left(\tfrac{1}{2}\right)^2
=
4\,
\frac{\|\Delta_1^{(ij)} - \Delta_0^{(ij)}\|^2}
     {\|\Delta_1^{(ij)} + \Delta_0^{(ij)}\|^2}.
\end{equation}
Large values of \(L_{ij}(1/2)\) are driven by near-cancellation in the denominator,
i.e.,\ by configurations where
\(\Delta_0^{(ij)} \approx -\Delta_1^{(ij)}.\)

\paragraph{Cancellation score.}
To quantify the degree of directional cancellation, we define the \emph{cancellation score} for each $k$-NN edge $(i,j)$:
\[
s_{\mathrm{canc}}^{(ij)} := \frac{\|(1-t)\Delta_0^{(ij)} + t\,\Delta_1^{(ij)}\|}{(1-t)\|\Delta_0^{(ij)}\| + t\|\Delta_1^{(ij)}\| + \varepsilon},
\]
where $\varepsilon > 0$ is a small constant for numerical stability.
A score close to $1$ indicates that $\Delta_0^{(ij)}$ and $\Delta_1^{(ij)}$ are roughly aligned,
while a score close to $0$ indicates near-perfect cancellation ($\Delta_0^{(ij)} \approx -\frac{t}{1-t}\Delta_1^{(ij)}$).

\rev{Figure~\ref{fig:midtime-analysis} reports both quantities at $t=1/2$ across four configurations (no canonicalization, canonicalize $X_0$ only, canonicalize $X_1$ only, and both). The results strongly support our theoretical analysis: one-sided canonicalization of $X_1$ (Ours) achieves the lowest and most stable Lipschitz ratios (median $\approx 2.00$, P90 $\approx 2.02$) and the highest cancellation scores ($\approx 0.92$), indicating that most $k$-NN edges correspond to genuinely close pairs with minimal spurious cancellation. In contrast, two-sided canonicalization produces the highest Lipschitz ratios (P90 up to $2.30$) and the lowest cancellation scores ($\approx 0.72$), confirming that canonicalizing both endpoints contracts $\Delta_0^{(ij)}$ to the same small scale as $\Delta_1^{(ij)}$, thereby drastically increasing the frequency of directional cancellation events. 
}

\section{\rev{Discussion}}
\label{sec:discussion}

\rev{
\paragraph{Why canonicalization helps.}
Canonicalization fundamentally works by exploiting \emph{structural commonalities} shared across samples. For 3D shapes, there is typically a point that is relatively ``bottom-left'' of the configuration; consistently assigning index~0 to that point concentrates the positional range covered by that index.

\paragraph{Conditions for reduced benefit.}
Canonicalization provides smaller gains when samples share less common structure or when the chosen ordering captures it less effectively. In the extreme case where the data has no exploitable common structure, OGPP gracefully degrades to standard flow matching. This is consistent with our experiments: on synthetic benchmarks such as minimal surfaces, where target configurations are relatively simple, canonicalization yields large improvements, while on complex real-world shapes (e.g., diverse ShapeNet categories) the gains are more moderate.

\paragraph{Domain-specific alternatives.}
When the Euclidean space-filling curve is insufficient, a more domain-appropriate canonicalization can be substituted. We already demonstrate this: for minimal surfaces (Figures~\ref{fig:minimal_surface_multi} and~\ref{fig:minimal_surface}), we use counterclockwise polygon ordering instead of Hilbert sorting. For articulated shapes, a promising direction is sorting in the spectral domain. More generally, domain knowledge about expected commonalities can be translated into a canonicalization strategy, making OGPP adaptable to diverse tasks.
}

\section{Conclusion}
In this work, we introduced \emph{Orbit-Space Geometric Probability Paths (OGPP)}, \revv{a flow-matching framework designed for generative modeling of particle systems.} While most modern generative models in graphics adopt a grid view, OGPP treats particles as persistent entities evolving through physical space, with identities, trajectories, and geometry-aware dynamics. \revv{We explored whether explicitly respecting permutation symmetry and physical semantics in the probability-path design can improve the learning problem for particle generation.}

\revv{Concretely, OGPP addresses permutation symmetry through terminal canonicalization, which our analysis and experiments suggest reduces per-particle ambiguity and helps each particle assume a more consistent role. Particle index embeddings further introduce identity-aware conditioning, aiming to disentangle mixed regression targets. Our geometric probability paths show that the terminal velocity in flow matching can serve as a carrier for per-particle attributes such as surface normals. Together, these components are designed to make particle flow matching a more structured learning problem, and our experiments on the tested benchmarks indicate straighter flows and reduced inference cost.} More broadly, this formulation shows that flow-based generative modeling can be designed natively for particles, and opens up new opportunities for graphics generative systems that tightly integrate sampling, geometry, and physics with particle representations. \revv{We view this work as an initial investigation into particle-centric probability-path design for flow matching, and hope it motivates further study in this direction.}

\paragraph{Limitations.}
Our approach has several limitations that  suggest promising directions for future research.
First, our current framework operates on a fixed number of particles and relies on full attention, whose quadratic cost in particle count limits scalability to larger particle systems.
Second, our geometric probability paths do not correspond to Wasserstein-2 optimal transport; lacking the geodesic property of W$_2$ displacement interpolation, they may induce slightly more curved probability flows.
Third, orbit-space canonicalization introduces an additional design degree of freedom and can induce useful structure in the canonical indexing (e.g., locality or ordered semantics). However, our current framework does not explicitly leverage this induced structure to encode extra information, leaving the canonicalization choice underutilized.
\rev{Fourth, the benefit of canonicalization depends on how much common structure the data exhibits; on datasets with high variability, the improvements are more moderate than on structurally regular benchmarks.}

\paragraph{Future Work.}
Motivated by these limitations, future work will explore sparse, hierarchical, and locality-aware architectures inspired by physical interactions, where only nearby particles exert significant influence. We also plan to extend OGPP to variable particle counts.
On the probability-path side, a natural direction is to explore a richer family of geometric probability paths, e.g., higher-order or piecewise-smooth constructions that could better trade off geometric attribute encoding and transport optimality, potentially approaching W$_2$-consistent behavior when desired.
Finally, we will investigate canonicalization as an explicit information channel by designing canonicalizers whose index order aligns with task-relevant semantics (e.g., temporal order), enabling such signals to be encoded implicitly through indices rather than introducing additional generation dimensions.

\begin{acks}
We sincerely thank the anonymous reviewers for their valuable feedback. Georgia Tech authors acknowledge NSF CAREER \#2420319, IIS \#2433307, OISE \#2433313, IIS \#2433322, ECCS \#2318814 for funding support. We credit the Houdini education license for video animations.
\end{acks}

\bibliography{ot, flow_matching, spatial_partition, particle_dist, geometry, pcd_gen, PhysicsGen}
\bibliographystyle{ACM-Reference-Format}
\appendix


\rev{\section{Flow Matching Details}
\label{sec:appendix_flow_matching}

For more detailed expositions of the flow matching framework, see~\cite{lipman2022flow,lipman2024flow,holderrieth2025introduction}.

\paragraph{Probability paths and velocity fields.}
To define suitable training targets, flow matching specifies, for each data point $\bm{x}_1 \sim p_\mathrm{data}$, a \emph{conditional probability path} $p_t(\cdot \mid \bm{x}_1)$, $t \in [0,1]$, which starts from $p_\mathrm{init}$ at $t = 0$ and collapses to a point mass at $\bm{x}_1$ at $t = 1$.
Intuitively, $p_t(\cdot \mid \bm{x}_1)$ describes how noise samples are transported toward the terminal location $\bm{x}_1$.
Each such path is realized by a reference \emph{conditional velocity field} $\bm{u}_t^\mathrm{ref}(\cdot \mid \bm{x}_1)$ such that the solution $X_t$ of the induced ODE satisfies $X_t \sim p_t(\cdot \mid \bm{x}_1)$.
By averaging $p_t(\cdot \mid \bm{x}_1)$ over $\bm{x}_1 \sim p_\mathrm{data}$, one obtains a \emph{marginal probability path} $p_t$ that interpolates between $p_\mathrm{init}$ and $p_\mathrm{data}$.

\paragraph{Marginalization trick.}
A central tool in flow matching is the \emph{marginalization trick}, which expresses the marginal velocity field as a posterior-weighted average of conditional velocities.
Let $p_t$ be the marginal probability path induced by the conditional paths $p_t(\cdot \mid \bm{x}_1)$.
Then the marginal velocity field can be written as Eq.~\eqref{eq:marginal_vf},
where the weighting factor is exactly the posterior of $\bm{x}_1$ given $X_t = \bm{x}$.
With this choice, the ODE Eq.~\eqref{eq:flow_ode} driven by $\bm{u}_t^\mathrm{ref}$ transports $p_\mathrm{init}$ along $p_t$ and reaches $p_\mathrm{data}$ at $t=1$.

\paragraph{Flow matching training.}
In practice, the marginal velocity field Eq.~\eqref{eq:marginal_vf} is intractable to evaluate directly.
Instead, flow matching trains $\bm{u}_t^\theta$ to regress onto the conditional reference velocity along the probability paths via the conditional flow matching loss Eq.~\eqref{eq:cfm_loss}.
This objective is equivalent, up to a constant, to a marginal regression loss that matches $\bm{u}_t^\theta$ to the marginal velocity $\bm{u}_t^\mathrm{ref}(\bm{x})$.}

\rev{\section{Group Theory Details}
\label{sec:appendix_group_theory}

\paragraph{Groups and group actions.}
A \emph{group} $(G, \cdot)$ is a set $G$ equipped with a binary operation $\cdot$ satisfying associativity, the existence of an identity element, and the existence of inverses.
A group $G$ \emph{acts} on a set $X$ if there is a map $G \times X \to X$, written $(g, x) \mapsto g \cdot x$, such that $e \cdot x = x$ for the identity $e \in G$ and $(g_1 \cdot g_2) \cdot x = g_1 \cdot (g_2 \cdot x)$ for all $g_1, g_2 \in G$ and $x \in X$.

\paragraph{Rigid motions as preprocessing.}
Physically, particle configurations are defined only up to global rigid motions
(translations and rotations) and permutations.
In all our experiments we first normalize away global pose by recentering each
configuration and aligning a canonical frame (e.g., via PCA), so that the
remaining symmetry is purely combinatorial: permutations of particle indices.
We note that PCA-based alignment cannot resolve axis sign flips and may become
ambiguous when the inertia tensor has degenerate eigenvalues (e.g., for
near-isotropic shapes); we address sign ambiguity with a fixed sign convention
and did not observe a noticeable impact on generation quality in our experiments.

\paragraph{Orthogonal representations.}
In our setting, we consider groups acting on Euclidean spaces via
\emph{orthogonal representations}.
An orthogonal representation is a group homomorphism
$\rho: G \to O(d)$, where $O(d)$ denotes the orthogonal group of
$d \times d$ matrices $R$ satisfying $R^\top R = I$.
This means each group element $g \in G$ is represented by an orthogonal
matrix $\rho(g)$, and the group action on $\mathbb{R}^d$ is given by
$g \cdot x = \rho(g) x$.

\paragraph{Orbits and invariant maps.}
The \emph{orbit} of a configuration $x \in \mathbb{R}^d$ under the group action is the set of all configurations reachable from $x$ by group transformations:
$\mathrm{Orb}(x) := \{ \rho(g) x : g \in G \}$.
Configurations in the same orbit represent the same underlying object under symmetry transformations (here, permutations of particle indices after pose normalization).
A function $f: \mathbb{R}^d \to Y$ is called \emph{$G$-invariant} if $f(\rho(g) x) = f(x)$ for all $g \in G$ and $x \in \mathbb{R}^d$; that is, $f$ is constant on each orbit.

\paragraph{Canonicalization (extended).}
A \emph{canonicalization map} $C: \mathbb{R}^d \to \mathbb{R}^d$ selects a representative from each orbit in a $G$-invariant way.
Formally, we require:
\begin{enumerate}
    \item $C(\rho(g) x) = C(x)$ for all $g \in G$ and $x \in \mathbb{R}^d$ ($G$-invariance);
    \item $C(x) \in \mathrm{Orb}(x)$ for all $x \in \mathbb{R}^d$ (the output lies in the orbit of $x$).
\end{enumerate}
The image $C(x)$ is called the \emph{canonical representative} of $x$.
Together, these conditions imply that $C(x_1) = C(x_2)$ if and only if $x_1$ and $x_2$ lie in the same orbit, so $C$ induces a bijection between orbits and their canonical representatives.}

\rev{\section{Conditional Covariance Reduction via Orbit-Space Canonicalization: Detailed Derivation}
\label{sec:appendix_cov_derivation}

This section provides the full derivation for the conditional covariance reduction result in Section~\ref{subsec:per_particle_cov}.

\paragraph{Setup.}
We consider the full configuration
$\bm{X}_t = (\bm{X}_t^1,\dots,\bm{X}_t^N) \in (\mathbb{R}^D)^N$,
and a network that predicts a velocity field
$\bm{u}^{\theta}(\bm{X}_t, t) \in (\mathbb{R}^D)^N$
for the entire configuration.
For the linear path, the regression target is
\begin{equation}
\label{eq:whole-target}
Y
:=
\frac{\bm{X}_1 - \bm{X}_t}{1-t}
=
\bm{X}_1 - \bm{X}_0.
\end{equation}

\paragraph{Covariance of the regression target.}
Combining Eq.~\eqref{eq:whole-target} with the Bayes-optimal velocity Eq.~\eqref{eq:bayes_opt}, the only source of randomness in $Y$ given $X_t=\bm{x}$ is the endpoint $X_1$, and
\begin{equation}
\label{eq:whole-cov-Y}
\mathrm{Cov}\!\left( Y \mid X_t = \bm{x} \right)
=
\frac{1}{(1-t)^2}\,
\mathrm{Cov}\!\left( X_1 \mid X_t = \bm{x} \right).
\end{equation}
A smaller conditional covariance thus directly lowers the irreducible MSE.

\paragraph{Orbit-factorization details.}
Under the orbit-symmetry factorization Eq.~\eqref{eq:orbit-factorization},
\begin{equation}
X_1 \mid (X_t = \bm{x})
\;\overset{d}{=}\;
\rho(G)\,\zeta_{\bm{x}},
\end{equation}
where $G \in S_N$ is a random permutation and $\rho(G)$ is its permutation representation on $(\mathbb{R}^D)^N$.
Let $C : (\mathbb{R}^D)^N \to (\mathbb{R}^D)^N$ be a $G$-invariant canonicalization map (Section~\ref{def:canonical_map}) and define
$\widetilde{X}_1 := C(X_1)$.
The $G$-invariance of $C$ implies
\begin{equation}
\widetilde{X}_1 \mid (X_t = \bm{x}, G = g)
\;\overset{d}{=}\;
\widetilde{X}_1 \mid (X_t = \bm{x}),
\label{eq:whole-drop_G}
\end{equation}
so the conditional law of $\widetilde{X}_1$ no longer depends on $G$.

\paragraph{Conditional covariance decomposition.}
Applying the conditional law of total covariance to $X_1$ with respect to $G$ gives
\begin{equation}
\label{eq:whole-cov-decomp}
\begin{aligned}
\mathrm{Cov}( X_1 \mid X_t\!=\!\bm{x} )
&=
\mathbb{E}_{G}\bigl[
  \mathrm{Cov}(
    X_1 \mid X_t\!=\!\bm{x}, G
  )
\bigr]
\\
&\quad+
\mathrm{Cov} \bigl(
  \mathbb{E}[
    X_1 \mid X_t\!=\!\bm{x}, G
  ]
  \bigm|
  X_t\!=\!\bm{x}
\bigr).
\end{aligned}
\end{equation}
The first term is the intrinsic variability under a fixed permutation, averaged over $G$.
The second term is a positive semidefinite covariance capturing additional variability from random $G$.

Applying the same decomposition to $\widetilde{X}_1$ yields
\begin{equation}
\label{eq:whole-cov-tilde}
\begin{aligned}
\mathrm{Cov}(
  \widetilde{X}_1 \mid X_t\!=\!\bm{x}
)
&=
\mathbb{E}_{G}\bigl[
  \mathrm{Cov}(
    \widetilde{X}_1 \mid X_t\!=\!\bm{x}, G
  )
\bigr]
\\
&\quad+
\underbrace{\mathrm{Cov} \bigl(
  \mathbb{E}[
    \widetilde{X}_1 \mid X_t\!=\!\bm{x}, G
  ]
  \bigm|
  X_t\!=\!\bm{x}
\bigr)}_{=\,0}.
\end{aligned}
\end{equation}
By Eq.~\eqref{eq:whole-drop_G}, the inner conditional expectation does not depend on $G$, so the second term vanishes.

\paragraph{Trace comparison.}
For each fixed $G=g$, the action $\rho(g)$ is a permutation matrix on the full configuration space and is therefore orthogonal, and orthogonal transformations preserve covariance trace:
\begin{equation}
\mathrm{tr}\,
  \mathrm{Cov}\!\left(
    X_1 \mid X_t = \bm{x}, G = g
  \right)
=
\mathrm{tr}\,
  \mathrm{Cov}\!\left(
    \widetilde{X}_1 \mid X_t = \bm{x}, G = g
  \right).
\end{equation}
Taking traces in Eq.~\eqref{eq:whole-cov-decomp} and Eq.~\eqref{eq:whole-cov-tilde}, and using linearity of trace and expectation, we obtain
\begin{equation}
\label{eq:whole-trace-ineq-x}
\begin{aligned}
\mathrm{tr}\,
\mathrm{Cov}( X_1 \mid X_t\!=\!\bm{x} )
&=
\mathrm{tr}\,
\mathrm{Cov}( \widetilde{X}_1 \mid X_t\!=\!\bm{x} )
\\
&\quad+
\mathrm{tr}\,
\mathrm{Cov} \bigl(
  \mathbb{E}[
    X_1 \mid X_t\!=\!\bm{x}, G
  ]
  \bigm|
  X_t\!=\!\bm{x}
\bigr)
\\
&\ge
\mathrm{tr}\,
\mathrm{Cov}( \widetilde{X}_1 \mid X_t\!=\!\bm{x} ).
\end{aligned}
\end{equation}
Combining Eq.~\eqref{eq:whole-cov-Y} and Eq.~\eqref{eq:whole-trace-ineq-x} yields the main result Eq.~\eqref{eq:per-particle-trace-ineq-Y}.}

\rev{\section{Lipschitz Ratio Analysis: Canonicalizing \(X_0\)}
\label{sec:appendix_lipschitz}

This section provides the detailed derivation for Section~\ref{subsec:lipschitz_view}, showing why two-sided canonicalization inflates the nearest-neighbor Lipschitz ratios of the velocity field.

\paragraph{Setup.}
We view each configuration
$\bm{x}_t^{(i)} = (\bm{x}_t^{(i),1},\dots,\bm{x}_t^{(i),N}) \in (\mathbb{R}^3)^N$
as a stacked vector in $\mathbb{R}^{3N}$.

\paragraph{Nearest-neighbor Lipschitz ratios.}
Draw i.i.d.\ pairs
$\{(\bm{x}_0^{(i)}, \bm{x}_1^{(i)})\}_{i=1}^M$ with
$\bm{x}_0^{(i)}, \bm{x}_1^{(i)} \in (\mathbb{R}^3)^N$,
fix $t \in (0,1)$, and form the interpolants
\[
\bm{x}_t^{(i)}
=
(1-t)\,\bm{x}_0^{(i)} + t\,\bm{x}_1^{(i)},
\qquad
\bm{u}^{(i)}
=
\bm{u}\bigl(\bm{x}_t^{(i)}, t\bigr)
=
\frac{\bm{x}_1^{(i)} - \bm{x}_t^{(i)}}{1-t}.
\]
A short calculation shows that
$\bm{u}^{(i)} = \bm{x}_1^{(i)} - \bm{x}_0^{(i)}$.
To quantify how $\bm{u}$ varies under small perturbations, we build a $k$-NN graph on $\{\bm{x}_t^{(i)}\}_{i=1}^M$, and for each
edge $(i,j)$ define the Lipschitz ratio as
\[
L_{ij}(t)
:=
\frac{\bigl\|\bm{u}^{(i)} - \bm{u}^{(j)}\bigr\|}
     {\bigl\|\bm{x}_t^{(i)} - \bm{x}_t^{(j)}\bigr\|},
\]
where the norms are in $\mathbb{R}^{3N}$.
With the per-edge differences
$\Delta_0^{(ij)} := \bm{x}_0^{(i)} - \bm{x}_0^{(j)}$,
$\Delta_1^{(ij)} := \bm{x}_1^{(i)} - \bm{x}_1^{(j)}$, and
$\Delta_t^{(ij)} := \bm{x}_t^{(i)} - \bm{x}_t^{(j)}$,
the Lipschitz ratio takes the form of Eq.~\eqref{eq:Lij-deltas}.

\paragraph{One-sided vs.\ two-sided canonicalization.}
Since we focus on canonicalization at $X_0$, we assume $X_1$ has already been canonicalized so that $\Delta_1^{(ij)}$ is typically small.
We compare two regimes:
\begin{itemize}[leftmargin=*]
    \item \textbf{One-sided canonicalization (ours).}
    Canonicalize $X_1$ only.
    Endpoint differences are $(\Delta_0^{(ij)}, \Delta_1^{(ij)})$, and
    $L_{ij}(t)^2$ is given by Eq.~\eqref{eq:Lij-deltas}.

    \item \textbf{Two-sided canonicalization.}
    Also canonicalize $X_0$ via the same map $C$.
    Write $\widetilde{\bm{x}}_0^{(i)} := C(\bm{x}_0^{(i)})$ and
    $\widetilde{\Delta}_0^{(ij)} := \widetilde{\bm{x}}_0^{(i)} - \widetilde{\bm{x}}_0^{(j)}$.
\end{itemize}

By construction, canonicalization contracts the pairwise dispersion: for some $0 < \alpha_0 \le 1$,
\[
\mathbb{E}\bigl[\|\widetilde{\Delta}_0^{(ij)}\|^2\bigr]
=
\alpha_0^2\,
\mathbb{E}\bigl[\|\Delta_0^{(ij)}\|^2\bigr],
\]
with $\alpha_0 < 1$ whenever the group symmetry is non-trivial.

\paragraph{Directional cancellation.}
The key phenomenon is \emph{directional cancellation} in the denominator of Eq.~\eqref{eq:Lij-deltas}.
When $\Delta_0^{(ij)}$ and $\Delta_1^{(ij)}$ point in approximately opposite directions and have comparable magnitudes, the denominator $(1-t)\Delta_0^{(ij)} + t\Delta_1^{(ij)}$ becomes small while the numerator $\Delta_1^{(ij)} - \Delta_0^{(ij)}$ remains large.

Since the $k$-NN graph is built from small values of $\|\Delta_t^{(ij)}\|$, nearest-neighbor edges are \emph{biased} toward such cancellation events.
Canonicalizing $X_1$ already makes $\Delta_1^{(ij)}$ small, so the denominator becomes sensitive to $\Delta_0^{(ij)}$:
\begin{itemize}[leftmargin=*]
    \item If we \emph{keep $X_0$ uncanonicalized}, $\Delta_0^{(ij)}$ has a relatively large spread.
    It is statistically unlikely that
    $\Delta_0^{(ij)} \approx -\tfrac{t}{1-t}\,\Delta_1^{(ij)}$,
    so most nearest-neighbor edges correspond to genuinely close configurations and the denominator does not become spuriously small.

    \item If we also \emph{canonicalize $X_0$},
    $\widetilde{\Delta}_0^{(ij)}$ reaches a similar scale to $\Delta_1^{(ij)}$.
    It becomes much easier for the two small vectors to nearly cancel in
    $(1-t)\,\widetilde{\Delta}_0^{(ij)} + t\,\Delta_1^{(ij)}$, while the
    numerator $\Delta_1^{(ij)} - \widetilde{\Delta}_0^{(ij)}$ stays comparable.
    The $k$-NN construction then selects many edges with tiny denominators but non-tiny numerators, yielding large $L_{ij}(t)$ and a less smooth velocity field.
\end{itemize}}

\rev{\section{Orbit-Continuous Canonicalization and Straight Flows: Detailed Derivation}
\label{sec:appendix_orbit_continuous}

This section provides the detailed derivation for the orbit-continuous canonicalization analysis in Section~\ref{subsec:orbit_straightness}.

\paragraph{Smoothness of endpoint distributions over the orbit space.}
We assume that the endpoint distribution
$X_1 \mid X_t = \bm{x}$ varies smoothly over the orbit space
$\mathcal{O}$, in the sense that nearby orbits
$\mathrm{Orb}(\bm{x})$ and $\mathrm{Orb}(\bm{x}')$ induce nearby terminal
endpoint distributions.
Without canonicalization, this smoothness naturally lives at the level of
orbits; however, a poorly behaved canonicalization map $C$ could destroy it by
introducing abrupt representative changes between nearby orbits.
To avoid such pathologies, we require $C$ to be orbit-continuous in the sense
of Eq.~\eqref{eq:orbit-Lipschitz}.
Under these conditions, the canonical means $\bm{m}(\bm{x})$ inherit
orbit-Lipschitz regularity: nearby orbits induce nearby values of
$\bm{m}(\bm{x})$.

\paragraph{Lipschitz bound derivation.}
Combining this orbit-Lipschitz regularity of $\bm{m}$ with
Eq.~\eqref{eq:bayes-vel-canon}, we obtain the local Lipschitz bound Eq.~\eqref{eq:u-star-Lip},
where $L_{\mathrm{vel}}(t)$ is a time-dependent constant controlled by the
orbit-Lipschitz constant $L_{\mathrm{orb}}$ (through the choice of $C$) and the intrinsic smoothness of the
canonical means.

\paragraph{From orbit metric to Euclidean metric.}
In practice, $\bm{u}^*(\cdot,t)$ is defined on the Euclidean configuration space
$(\mathbb{R}^D)^N$.
When $d_{\mathcal{O}}$ is chosen as the standard orbit metric
\[
d_{\mathcal{O}}(\mathrm{Orb}(\bm{x}),\mathrm{Orb}(\bm{x}'))
=
\inf_{g \in G} \|\bm{x} - \rho(g)\bm{x}'\|,
\]
we have
\[
d_{\mathcal{O}}(\mathrm{Orb}(\bm{x}),\mathrm{Orb}(\bm{x}'))
\le
\|\bm{x} - \bm{x}'\|,
\]
so Eq.~\eqref{eq:u-star-Lip} also implies a corresponding local Lipschitz bound
with respect to the Euclidean distance.
Thus, the orbit-space analysis directly controls the regularity of the velocity
field in the actual input space seen by the network.}

\rev{\section{Arc-Length Terminal Velocity: Detailed Discussion}
\label{app:atv-detail}

This section provides a detailed discussion of the arc-length terminal velocity (ATV) design described in Section~\ref{subsec:atv}.

\paragraph{Why normalized terminal velocity (NTV) is suboptimal.}
A naive choice is to set $\|\bm{v}_1\| = 1$ for all particles (normalized terminal velocity, NTV).
However, under NTV, paths with very different chord lengths $\|\bm{x}_1 - \bm{x}_0\|$ share the same terminal speed.
Distant points must either move quickly at early times and then slow down, or nearby points must start slowly and then accelerate, so different trajectories exhibit very nonuniform speed profiles in $t$.
This makes $t$ a poor surrogate for progress along the curve (i.e., normalized arc length), so uniform sampling in $t$ no longer corresponds to approximately uniform sampling along the trajectory.
One might try to correct this with a fixed, hand-crafted non-uniform schedule in $t$, but any such schedule can only compensate for a particular family of acceleration patterns (e.g., accelerate-then-decelerate trajectories) and will necessarily be suboptimal for trajectories that accelerate and decelerate in the opposite order.

\paragraph{Optimal speed-variance minimization.}
For the quadratic Hermite path Eq.~\eqref{eq:hermite_path},
the speed $\|\dot{\gamma}(t)\|$ is the square root of a quadratic polynomial in $t$, and the arc length
\(
L(\bm{x}_0,\bm{x}_1,\bm{v}_1)
= \int_0^1 \|\dot{\gamma}(t)\|\,\mathrm{d}t
\)
admits a closed-form expression in terms of $\sqrt{\cdot}$ and $\log(\cdot)$ (see the additional supplementary material for the explicit formula and derivation).
Since only the \emph{direction} of $\bm{v}_1$ is fixed by the normal, we may write $\bm{v}_1 = \alpha \hat{\bm{n}}_1$ with a scalar $\alpha$, and choose $\alpha$ by solving a one-dimensional optimization problem that minimizes the variance of the speed profile,
\(
\alpha^\star
\;=\;
\arg\min_{\alpha \in [0.5,\,15.0]}
\operatorname{Var}_{t \in [0,1]}\bigl(\|\dot{\gamma}(t; \alpha)\|\bigr),
\)
yielding, for each $(\bm{x}_0,\bm{x}_1,\hat{\bm{n}}_1)$, a Hermite trajectory whose speed over $t\in[0,1]$ is as uniform as possible.

\paragraph{Cheap ATV approximation.}
In practice, optimizing the speed variance for every particle at every training step would introduce nontrivial overhead.
The ATV formula (Eq.~\eqref{eq:atv}) approximates the optimal solution using only the chord length $D$ and the chord-normal alignment $S$.
This approximation is inexpensive (only norms and dot products), yet empirically produces trajectories with much more uniform speed profiles in $t$ than under NTV.
}

\end{document}